\documentclass[fleqn,usenatbib]{mnras}

\usepackage{newtxtext,newtxmath}

\usepackage[T1]{fontenc}
\usepackage{ae,aecompl}
\usepackage{subfig}

\usepackage{graphicx}  
\usepackage{amsmath} 
\usepackage{amssymb} 
\usepackage{chemformula}

\title[The atmosphere of 55 Cnc e]{The PEPSI Exoplanet Transit Survey (PETS) I: 
\newline
Investigating the presence of a silicate atmosphere on the super-Earth 55 Cnc e}

\author[E. Keles et al.]{
Engin Keles,$^{1}$
Matthias Mallonn,$^{1}$
Daniel Kitzmann,$^{2}$
Katja Poppenhaeger,$^{1,4}$
\newauthor
H. Jens Hoeijmakers,$^{3}$
Ilya Ilyin,$^{1}$
Xanthippi Alexoudi,$^{1,4}$
Thorsten A. Carroll,$^{1}$
\newauthor
Julian Alvarado-Gomez,$^{1}$
Laura Ketzer,$^{1}$
Aldo S. Bonomo,$^{5}$ 
Francesco Borsa,$^{6}$
\newauthor
Scott Gaudi,$^{7}$
Thomas Henning,$^{8}$
Luca Malavolta,$^{9,10}$
Karan Molaverdikhani,$^{8,11,18,19}$
\newauthor
Valerio Nascimbeni,$^{10}$
Jennifer Patience,$^{12}$
Lorenzo Pino,$^{13}$
Gaetano Scandariato,$^{14}$
\newauthor
Everett Schlawin,$^{15}$
Evgenya Shkolnik,$^{12}$
Daniela Sicilia,$^{14}$
Alessandro Sozzetti,$^{5}$
\newauthor
Mary G. Foster,$^{1}$
Christian Veillet,$^{16}$
Ji Wang,$^{7}$
Fei Yan,$^{17}$
Klaus G. Strassmeier,$^{1,4}$
\\
\\
$^{1}$Leibniz-Institut f\"ur Astrophysik Potsdam (AIP), Potsdam, Germany\\
$^{2}$University of Bern, Center for Space and Habitability, Bern, Switzerland\\
$^{3}$Lund Observatory, Lund, Sweden \\
$^{4}$Institut f\"ur Physik \& Astronomy, University of Potsdam, Potsdam, Germany \\ 
$^{5}$INAF - Osservatorio Astrofisico di Torino, Torinese, Italy \\
$^{6}$INAF – Osservatorio Astronomico di Brera, Merate, Italy \\
$^{7}$The Ohio State University, Ohio, USA  \\
$^{8}$Max-Planck-Institute for Astronomy, Koenigstuhl 17, 69117 Heidelberg, Germany  \\
$^{9}$Dipartimento di Fisica e Astronomia ”Galileo Galilei”, Universita degli Studi di Padova, Padova, Italy \\
$^{10}$INAF – Osservatorio Astronomico di Padova, Padova, Italy   \\
$^{11}$Landessternwarte Heidelberg, Heidelberg, Germany  \\
$^{12}$School of Earth and Space Exploration, Arizona State University, Arizona, USA  \\
$^{13}$INAF - Osservatorio Astrofisico di Arcetri, Firenze, Italy \\
$^{14}$INAF - Osservatorio Astrofisico di Catania, Catania, Italy \\
$^{15}$Steward Observatory, University of Arizona, Arizona, USA  \\
$^{16}$Large Binocular Telescope Observatory, Arizona, USA \\
$^{17}$Institute of Astrophysics, University of Göttingen, Göttingen, Germany \\ 
$^{18}$LMU: Universitäts-Sternwarte, Ludwig-Maximilians-Universität München, Scheinerstrasse 1, D-81679 München, Germany \\
$^{19}$ORIGINS: Exzellenzcluster Origins, Boltzmannstraße 2, 85748 Garching, Germany \\ }

\date{Accepted XXX. Received YYY; in original form ZZZ}

\pubyear{2020}

\begin{document}
\label{firstpage}
\pagerange{\pageref{firstpage}--\pageref{lastpage}}
\maketitle

\begin{abstract}
The study of exoplanets and especially their atmospheres can reveal key insights on their evolution by identifying specific atmospheric species. For such atmospheric investigations, high-resolution transmission spectroscopy has shown great success, especially for Jupiter-type planets. Towards the atmospheric characterization of smaller planets, the super-Earth exoplanet 55 Cnc e is one of the most promising terrestrial exoplanets studied to date. Here, we present a high-resolution spectroscopic transit observation of this planet, acquired with the PEPSI instrument at the Large Binocular Telescope. Assuming the presence of Earth-like crust species on the surface of 55 Cnc e, from which a possible silicate-vapor atmosphere could have originated, we search in its transmission spectrum for absorption of various atomic and ionized species such as \ion{Fe}{}, \ion{Fe}{}\textsuperscript{+}, \ion{Ca}{}, \ion{Ca}{}\textsuperscript{+}, \ion{Mg}{} and \ion{K}{}, among others. Not finding absorption for any of the investigated species, we are able to set absorption limits with a median \mbox{value of 1.9 $\times$ R\textsubscript{P}}. In conclusion, we do not find evidence of a widely extended silicate envelope on this super-Earth reaching several planetary radii.
\end{abstract}




\section{Introduction}
\label{sec:Intro}
The detection of atmospheric fingerprints from terrestrial planets outside the Solar System is an important aim of exoplanet science. A suitable method for this is to analyze transmitted stellar light through the atmosphere of an exoplanet, a method called transmission spectroscopy \citep{SeagerSasslow2000}. The comparison of the atmospheric fingerprints of the solar system planets to those worlds, especially those with similar properties to Earth (see e.g. \citealt{Grenfell_2017,Keles2018,Wunderlich_2021}), will help to increase our knowledge about their evolution.

Terrestrial exoplanets could possess different atmospheres, such as a primordial atmosphere, being rich in hydrogen and helium, or a secondary atmosphere with a larger mean molecular weight and absence of hydrogen and helium \citep{Kite_2020}. The presence of a secondary type atmosphere on an exoplanet can be attributed to the evolution of a primordial atmosphere. Under the absence of a magnetic field, there is a possibility that a planet in a close-in orbit can lose its hydrogen-rich primordial atmosphere and remain with a rocky surface. Strong stellar irradiation can sputter particles from the exposed surface \citep{Schaefer_Fegley_2009} and can build up a partly ionized silicate atmosphere or exosphere \citep{Ridden_Harper_2016,Vidotto_2018}. Such silicate-vapor atmospheres might be built up also by outgassing processes from molten magma on the surface \citep{Miguel_2011,Ito_2015} and can contain species such as \ion{Na}, \ion{K}, \ion{Fe}, \ion{Si}, \ion{SiO}, O and O\textsubscript{2}, and probably also hydrogen due to resupply processes from magma outgassing \citep{Kite_2019}.

Among the close-in terrestrial exoplanets detected to date, the super-Earth 55 Cnc e \citep{Demory_2011,Winn_2011}, with M\textsubscript{P} = 8.0 $\pm$ 0.3 M\textsubscript{$\oplus$} and \mbox{R\textsubscript{P} = 1.88 $\pm$ 0.03 R\textsubscript{$\oplus$}}  \citep{Bourrier_2018b}, is one of the most studied targets, due to its short orbital period of P $\approx$ 0.66 days and bright host star with a V\textsubscript{mag} $\approx$ 6. However, the atmospheric conditions for this planet are still under discussion, due to the applicability of different interior composition scenarios estimated from the observed planetary radius and mass.

Under the assumption of a solid interior mainly composed of iron, carbon, and silicates, the planet might have no thick atmospheric envelope, i.e. exposes mostly a rocky surface \citep{Madhusudhan_2012}, enabling the evolution of silicate-species building up a thin envelope around the planet. On the other hand, considering a less solid rocky interior, the presence of a thick atmospheric envelope is necessary \citep{Bourrier_2018b}, e.g. an envelope rich in water \citep{Demory_2011,Gillon_2012} or in carbon dioxide and molecular nitrogen \citep{Angelo_2017}. A primordial gaseous envelope dominated by hydrogen-helium is mainly ruled out as it may evaporate in short timescales, which is also supported by the absence of hydrogen Ly$\alpha$ \citep{Ehrenreich_2012} and helium absorption \citep{Zhang_2021}.

Different investigations made efforts to confirm one of those scenarios, by searching for atomic and molecular features applying moderate and high-resolution transmission spectroscopy using ground-based facilities:

Assuming the presence of a rocky surface, \citet{Ridden_Harper_2016} investigated if the strong stellar irradiation ejected atoms from the surface by sputtering to the exosphere. In their work, a tentative absorption relative to the stellar spectrum of $\sim$2.3 × 10\textsuperscript{-3} in the Na-D-lines at a 3-$\sigma$ level and $\sim$7 × 10\textsuperscript{-2} in the ionized \ion{CaII}{H\&K} lines at a 4.1-$\sigma$ level is shown. However, the authors did not claim detection due to low significance of the Na absorption and variable Ca\textsuperscript{+} signal. Another study on the rocky surface scenario for 55 Cnc e is presented by \citet{Tabernero_2020} investigating the absorption in the Na-D-lines and H$\alpha$ line. The authors showed an upper limit of (3.4 $\pm$ 0.4) × 10\textsuperscript{-4} to the presence of the Na-D-lines and \mbox{(7 $\pm$ 1) × 10\textsuperscript{-4}} to the presence of the H$\alpha$ line, not detecting absorption of these species. The non-detection of hydrogen is in agreement with the findings shown by \citet{Ehrenreich_2012}, confirming the absence of a primordial atmosphere. Another study that supports this result is demonstrated by \citet{Zhang_2021}, who did not find evidence of helium absorption on this planet. 

Furthermore, different studies investigated molecular absorption features within the atmosphere of 55 Cnc e studying the presence of a secondary atmosphere. For instance, \citet{Esteves_2017} and \citet{Jindal_2020} investigated the water absorption in the optical wavelength range, not detecting water absorption signature. In addition, \citet{Jindal_2020} searched for the absorption signature of TiO also not finding any evidence for this molecule. Note that a water-rich steam atmosphere would dissociate at high altitudes, enriching the upper atmosphere with hydrogen \citep{Bourrier_2018b}. Thus, the non-detection of water is in agreement with the non-detection of hydrogen \citep{Ehrenreich_2012}, favoring a scenario with a dry atmosphere \citep{Bourrier_2018b}. Further molecular features were investigated by \citet{Deibert_2021}. The authors searched for absorption features from CO, CO\textsubscript{2}, H\textsubscript{2}O, HCN, NH\textsubscript{3} and C\textsubscript{2}H\textsubscript{2}. This study did not detect these features, but they were able to put upper volume mixing ratio limits on the last three of them.

Overall, the aforementioned studies remark that the results demonstrated are still consistent with a heavyweight atmosphere scenario as well as a rocky surface scenario, further to the possible presence of clouds which might mask absorption signature. This is a valid assumption for 55 Cnc e where the formation of mineral clouds is possible \citep{Mahapatra_2017}. Nevertheless, to date, no significant absorption signature for this planet, using high-resolution transit observations, has been reported.

In this study, we make use of one high-resolution transit observation of 55 Cnc e using a larger telescope than in previous studies to search for absorption by various atmospheric silicate species. This paper is structured as follows: In Section~\ref{sec:2} the observation and data reduction process is described. Section~\ref{sec:Data Analysis & Modelling} explains the methodology of this paper. The results are presented in Section~\ref{sec:Results} and discussed and concluded in Section~\ref{sec:Discussion}.
\begin{table*}
              \centering
              \caption{The shown atomic species are the major elements in Earth's crust with mixing ratios larger than 0.01\% \citep{Lide_book}. Not all species have absorption lines in the investigated wavelength range in this work. The species searched for in the atmosphere of 55 Cnc e are marked with a checkmark, the asterisk marks the species were also the singly ionized forms were investigated.}
              \label{tab:Table1}
              \setlength{\tabcolsep}{3.4pt}
              \begin{tabular}{cccccccccccccccccccccc}
              \hline
              Species & O          & Si      & Al        & Fe\textsuperscript{*}     & Ca\textsuperscript{*} & Na & Mg & K & Ti\textsuperscript{*} & H & P & Mn\textsuperscript{*} & F & Ba\textsuperscript{*} & Sr & S & C & Zr\textsuperscript{*} & Cl & V & Cr \\
              \hline
              mixing ratio [$\%$]   & 46.10 & 28.20 & 8.23 & 5.60 & 4.15 & 2.36 & 2.33 & 2.09 & 0.57 & 0.14 & 0.11 & 0.10 & 0.06 & 0.04 & 0.04 & 0.04 & 0.02 & 0.02 & 0.01 & 0.01 & 0.01  \\
              \vspace{0.5cm}
              Investigated & x & x & \checkmark & \checkmark & \checkmark & \checkmark & \checkmark & \checkmark & \checkmark & x & x & \checkmark & x & \checkmark & \checkmark & \checkmark & x & \checkmark & x & \checkmark & \checkmark \\
              \end{tabular}
\end{table*}

\section{Observation and Data Reduction}
\label{sec:2}

\subsection{The PEPSI Exoplanet Transit Survey}
\label{sec:PETS}
This observation is part of the PEPSI Exoplanet Transit Survey (PETS). PETS is a high-resolution spectroscopic survey of exoplanet transits, secondary eclipses, and host-star characterization of a large number of targets. Our goal is to characterize their planetary atmospheres in exquisite detail. The strengths and uniqueness of this project are the combination of the high-resolution spectrograph PEPSI (Potsdam Echelle Polarimetric and Spectroscopic Instrument) and throughput and the light-collecting power of the Large- Binocular- Telescope (LBT) in binocular mode with its 2 $\times$ 8.4m mirrors. 

\subsection{Observation}
\label{sec:Observation}
One transit of 55 Cnc e was observed on 2021 January 13 from 07:56 to 11:13 UT at the LBT operated in binocular mode with the fiber-fed spectrograph PEPSI \citep{Strassmeier2015,Strassmeier22018}. The spectrograph is located in a pressure-controlled chamber at a constant temperature and humidity to ensure that the refractive index of the air inside stays constant over a long-term period, providing a radial velocity stability of around 5\,m/s  \citep{Strassmeier2018}. 

PEPSI has a blue and a red arm with three cross-dispersers (CD) in each arm: only one CD per arm can be used at a time. The resolving power obtained from the ThAr frames is R $\sim$ 130\,000 in the CD selected in the blue arm (CD3: 4800-5441\AA) and R $\sim$ 115\,000 in the idem in the red arm (CD6: 7419--9067\AA). The ThAr wavelength solution is determined from about 3000 ThAr lines with an error of the fit at the image center of 4\,m/s. A CCD binning factor of two was used in wavelength direction. The light from the target and the sky from the two telescopes is rendered through four image slicers via respective octagonal fibers into the spectrograph, in which the echelle image is formed on two $10.5\times 10.5$k STA1600LN CCDs with $9\,\mu$m pixel size and 16 amplifiers. 

We obtained 83 spectra from CD3 and CD6, with 41 spectra acquired during transit and 42 spectra acquired out-of-transit. Due to the high velocity orbital semi-amplitude of K\textsubscript{P} $\approx$ 228 km/s of 55 Cnc e, a short exposure time of 90 seconds was used to prevent the smearing of possible atmospheric absorption signatures. The continuum signal-to-noise ratio (S/N) per pixel is $\sim$500 in CD3 and $\sim$700 in CD6.

\subsection{Data Reduction}
\label{sec:Reduction}
The data reduction was performed on a generic software platform written in C++, on Linux, and it is called the Spectroscopic Data Systems (SDS), which can work with different \'echelle spectrographs with the same tools. The image processing pipeline is specifically designed to handle PEPSI data calibration flow and image specific content (SDS4PEPSI based on \citet{Ilya_Thesis}). It provides an automated process from the beginning to the end product without human interaction and it relies on robust statistics. Further information for its application to PEPSI is provided in \citep{Strassmeier2018}.

The specific steps of the image processing include bias subtraction and variance estimation of the source images, super-master flat-field correction for the CCD spatial noise, echelle orders definition from the tracing flats, scattered light subtraction, wavelength solution for the ThAr images, optimal extraction of image slicers and cosmic spikes elimination of the target image, wavelength calibration and merging slices in each order, normalization to the master flat field spectrum to remove CCD fringes and blaze function, a global 2D fit to the continuum of the normalized image, and rectification of all spectral orders in the image to a 1D spectrum.

The continuum of the final spectra in the time series is rectified with the use of the mean spectrum that is calculated from weighted average of all the spectra. The mean spectrum is normalized to the continuum to eliminate any residual effects in the continuum. The ratio of each spectrum and the mean is then used to fit a smoothing spline which constitutes the improved continuum for the individual spectrum. The final spectra are then shifted to the stellar rest frame with the Earth-Sun and stellar radial velocities removed from the laboratory wavelength scale, and put on a uniform wavelength grid with 0.01\AA\space separation considering error propagation.

\section{Methods}
\label{sec:Data Analysis & Modelling}
In this work, we search for the presence of atomic species in the high spectral resolution transmission spectrum of the super-Earth 55 Cnc e. Due to the presence of strong telluric contamination of mostly saturated oxygen and water lines within the observed wavelength range, we exclude the wavelength ranges 7580\AA\space - 7680\AA, 8120\AA\space - 8350\AA\space and 8940\AA\space - 9007\AA\space from the analysis. We employ the pixel-by-pixel approach (see Section~\ref{sec:Absorption}) using a second-order polynomial function to create the combined transmission spectrum of 55 Cnc e. Here, we fit the normalized and wavelength calibrated flux value at a certain wavelength position over time and divide the data by the fit. We do this for all flux values at each wavelength position. Within the combined transmission spectrum we derive this way, we investigate the absorption by various neutral atomic and ionized species on 55 Cnc e, assumed to originate from a silicate-vapor atmosphere. In order to be able to make a preselection of the species that might build up the silicate atmosphere, we assume that the crust of 55 Cnc e contains similar species as the Earth's crust \citep{Lide_book}. Table~\ref{tab:Table1} shows the species with mixing ratios larger than 0.01\% in the Earth's crust composition. We inspect the transmission spectrum around the \ion{Mg}{I}-lines at 5167.3\AA, 5172.7\AA\space and 5183.6\AA, the \ion{K}{I}- line at 7699.0\AA\space, the H$\beta$-line at 4861.3\AA\space and the \ion{Ca}{II}- lines at 8498.0\AA, 8542.0\AA\space and 8662.1\AA. Furthermore, we calculate synthetic template spectra (see Section~\ref{sec:Templates}) and compare these with the combined transmission spectrum via the cross-correlation method where a large number of absorption lines are co-added \citep{Snellen2010,Brogi2016,Hoeijmakers2018}. We do this within a velocity range of $\pm$200 km/s in steps of 1 km/s and use the PyAstronomy\footnote{\url{https://github.com/sczesla/PyAstronomy}} tool \textit{pyasl.crosscorrRV}, expecting an upward peak at 0 km/s velocity for absorption. To calculate the significance of the detection, we divide the cross-correlation signal by the standard deviation of it excluding $\pm$20 km/s around the 0 km/s region \citep{Hoeijmakers2018}. Finally, we determine the upper limit of the atmospheric extension R\textsubscript{ext} for each species (see Section~\ref{sec:Extent}). Note, that we do not correct for the effects of centre-to-limb variation (see e.g. \citealt{Yan2017}) and the Rossiter–McLaughlin effect (see e.g. \citealt{Borsa_2021}) which are insignificant for this observation. The following provides further information regarding the data analysis.


\subsection{Synthetic species template absorption spectra}
\label{sec:Templates}
Using the cross-correlation technique we aim to ascertain the potential presence of different atoms and ions in the observed high-resolution spectrum of 55 Cnc e. To perform cross-correlation calculations, we create a set of spectral templates as described in e.g. \citet{Hoeijmakers2018} or \citet{Hoeijmakers2020}. Where available, we use the VALD line list data \citep{Ryabchikova2015PhyS...90e4005R, Pakhomov2019ARep...63.1010P} to compute the opacities of the atoms and ions. For the remaining cases, we employ the Kurucz line list database \citep{Kurucz2017CaJPh..95..825K}. The opacity calculations are done with the \texttt{HELIOS-k} code, described in \citet{Grimm2021ApJS..253...30G}.

We adopt the mass and radius values for 55 Cnc e from \citet{Bourrier_2018b} and use a surface pressure of 1.4 bar following the study of \citet{Angelo_2017}. We use element abundances that correspond to the composition of the Earth's mantle (see Table~\ref{tab:Table1}), assuming the potential atmosphere originates from evaporated mantle material. Due to the lack of constraints regarding the atmospheric structure of 55 Cnc e's terminator region, we assume the atmosphere to be isothermal during the calculations of the transmission spectra for simplicity and set the atmospheric temperature to \mbox{T = 2500 K}, which is close to the equilibrium temperature T\textsubscript{eq} $\approx$ 2350 K (inferred from the TEPCat\footnote{\url{https://www.astro.keele.ac.uk/jkt/tepcat/}} database \citep{Tepcat}). Due to the absence of absorption lines in the investigated wavelength range, some atomic species have been excluded from the analysis. For the species Fe, Ca, Ti, Mn, Ba and Zr, also the singly ionized forms are investigated. Table~\ref{tab:Table1} gives an overview of the selected species. As non-thermal processes (such as sputtering) could play a significant role to heat-up the near surface environment above the equilibrium temperature, we repeat the analysis accounting for hotter atmospheric temperature with T = 5000 K for the selected species. The abundances of the chemical species are determined by the chemical equilibrium code \textsc{FastChem 2.0}\footnote{\url{https://github.com/exoclime/FastChem}} \citep{Stock2018MNRAS.479..865S} for different values of the temperature and assumed element abundances. 

All template spectra are initially calculated with a constant step size in wavenumber space of 0.01 cm$^{-1}$ and converted via \mbox{1-R\textsubscript{P}\textsuperscript{2}/R\textsubscript{S}\textsuperscript{2}}. Finally, the template spectra are continuum corrected, convolved to the resolution of the PEPSI instrument and put on a uniform wavelength grid with 0.01\AA\space at the planetary rest-frame. The template spectra are shown in the Figures~\ref{fig:A1} - \ref{fig:A4}.

\subsection{Searching for atmospheric absorption feature}
\label{sec:Absorption}
We aim to probe the planetary atmosphere of 55 Cnc e for possible absorption signatures, which will manifest themselves on top of the stellar spectra observed within the in-transit phases. Due to the planetary motion, the wavelength position of possible absorption signature becomes Doppler-shifted at different \mbox{orbital phases $\phi$}, depending on the line-of-sight velocity of the atmosphere via \mbox{RV = K\textsubscript{P} $\times$ $\sin{(2\pi\phi)}$}. To extract the planetary absorption, it is necessary to eliminate the stellar lines from the observations by creating the transmission spectrum for each single exposed spectrum. Finally, the combined transmission spectrum can be created by shifting the in-transit phase transmission spectra to the planetary rest-frame and then by co-adding these. This procedure allows the detection of exoplanet atmospheres and is successfully demonstrated in different investigations where atmospheric absorption from atoms and molecules has been inferred especially for giant planets (see e.g. \citealt{Snellen2008,Casasayas_Barris_2019,Hoeijmakers2018,Keles2020,Cauley2021}).

\begin{figure}
    \includegraphics[width=1.0\columnwidth]{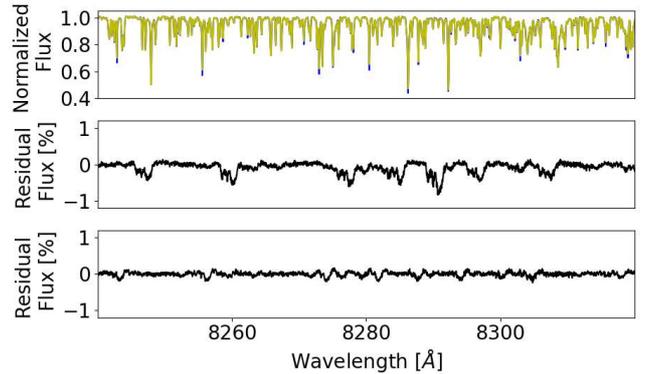}
    \caption{The top panel shows the spectrum observed at the lowest airmass (yellow) and highest airmass (blue) at a spectral region where telluric water lines are evident. The central panel shows the transmission spectrum derived by dividing the in-transit spectra by the mean out-of-transit spectrum. The bottom panel shows the pixel-by-pixel approach to derive the transmission spectrum.}
    \label{fig:func}
\end{figure}

Different possibilities exist to derive the transmission spectrum. One approach is to divide the spectra from the in-transit phases by the mean out-of-transit spectrum. This approach showed great success and was used in several investigations \citep{Wyttenbach2015,YanHenning2018,Casasayas_Barris_2019}. However, using this approach, one has to take care about possible telluric line contamination arising from Earth's atmosphere. As the depth of telluric lines varies overnight, the telluric lines need to be removed from the observations for instance by using telluric absorption correction tools such as Molecfit \citep{Smette_2015}, as else they would introduce spurious signature in the transmission spectra, if they arise in the wavelength region of interest. This is demonstrated in Figure~\ref{fig:func}, which shows in the top panel the spectrum observed at the lowest airmass (yellow) and highest airmass (blue) at a spectral region where telluric water lines are evident. As the entire observation is taken at excellent airmass conditions below 1.2, the telluric lines do not show strong variation. The central panel shows the combined transmission spectrum from the in-transit spectra, where the residuals from the telluric lines are visible. 
\begin{figure*}
    \includegraphics[width=1.00\textwidth]{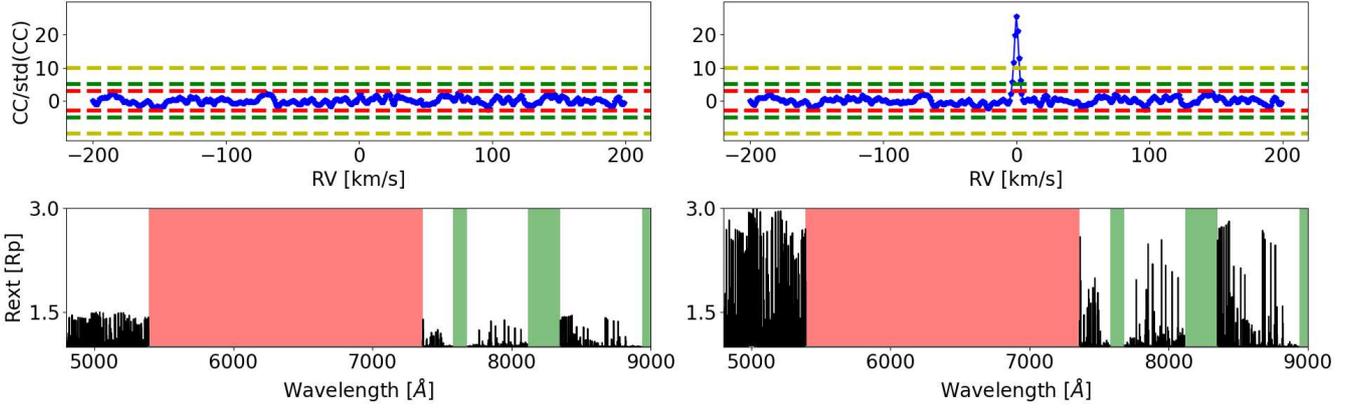}
    \caption{Injection recovery for scaled Ti template spectra (left for  R\textsubscript{ext} = 1.5 R\textsubscript{P} and right for R\textsubscript{ext} = 3.0 R\textsubscript{P}). The bottom panel shows the template spectra in units of R\textsubscript{P}. We excluded from the analysis regions not covered by the observation (red shaded) or strongly contaminated by telluric lines (green shaded). The top panel shows the cross-correlation signal (CC) divided by its standard deviation excluding $\pm$20 km/s around the central region. The dashed lines mark the 3$\sigma$ (red), 5$\sigma$ (green) and 10$\sigma$ (yellow) thresholds.}
    \label{fig:injection}
\end{figure*}

However, there is another approach that enables the derivation of the transmission spectrum, a pixel-by-pixel approach, with the further benefit of removing additionally non-saturated telluric lines, which bases on a similar approach successfully demonstrated in \citet{Strassmeier_2020}. During the observation, the normalized flux value at a certain pixel after wavelength calibration should not change for the entire observation, except at those pixels where the planetary atmosphere introduces additional absorption during the transit and at the positions where the telluric lines arise. Regarding the latter, in our observation, the observer's velocity (i.e. the sum of diurnal, orbital, and Solar System Barycenter velocities of the observer) increased in total linearly by around 350 m/s over the full observational period. Therefore, the telluric lines do not move significantly in our observation compared to the pixel size of around 1250 m/s. Applying a second-order polynomial fit to the flux values at each wavelength resolution element over time and dividing the data by this fit removes therefore efficiently the telluric lines and yields the transmission spectrum. As a possible absorption from the planetary atmosphere would affect only a few data points for a large timely resolved dataset, this is mainly ignored by the second-order polynomial fit that considers the variance from each flux value. Therefore, the pixel-by-pixel approach removes only slight variations overnight, not significantly affecting possible absorption arising from the planetary atmosphere within the transmission spectra. We demonstrate this in Section~\ref{sec:Extent} by recovering artificially injected absorption signature from the data. Figure~\ref{fig:func} shows in the bottom panel the combined transmission spectrum derived using this approach, showing smaller residuals at regions with telluric lines than in the central panel using the formerly mentioned approach. A similar procedure has been applied to create the transmission spectrum as well as to remove telluric line contamination for the planet of interest e.g. by \citet{Esteves_2017} and \citet{Zhang_2021} using the SYSREM tool \citep{Tamuz2005,Mazeh2007} which relies on the method of Principal Component Analysis.

\subsection{Deriving the extension limit of the atmosphere}
\label{sec:Extent}
We estimate a limit on the atmospheric altitude from which a possible absorption of a certain species arises. The absorption strength within the template spectra depends on different factors such as the atmospheric composition, which affects the atmospheric mean-molecular weight and thus the atmospheric scale height. As the composition of the atmosphere might be different than assumed in Section~\ref{sec:Templates}, the different species might be able to reach much higher altitude levels, e.g. due to a increase in the scale height, than calculated within the template spectra. Therefore, the template spectrum of interest is scaled and injected into the data with the aim to recover it. More precisely, after aligning the normalized spectra onto a common wavelength grid in the stellar rest-frame, the spectra from the in-transit phase are multiplied with the template spectrum of interest which we Doppler-shift beforehand with the opposite sign of the line of sight velocity of the planet. The opposite sign velocity is used to avoid smearing with a possible absorption signature \citep{Merritt_2021}. Creating now the combined transmission spectrum and applying the cross-correlation, only injections can be recovered that have introduced a significant absorption into the data. This is due to the circumstance, that the pixel-by-pixel approach makes use of a polynomial fit to derive each of the transmission spectra at the different in-transit phases. The polynomial fit again depends on the data quality. Thus, if the injected absorption is too weak, i.e. the absorption within the planetary atmosphere arises at low altitudes, this will not affect significantly the corresponding pixels where the absorption is introduced and later fitted by the polynomial, thus will not be recovered.

We scale the synthetic template spectra, so that the strongest absorption line in the investigated wavelength region reaches different levels which we denote as the atmospheric extension limits R\textsubscript{ext} and repeat the analysis, i.e. we inject this template again into the observations and repeat the cross-correlation analysis. For larger R\textsubscript{ext} values, the recovery will show larger detection signals. We determine the atmospheric extension limit R\textsubscript{ext} in units of the planetary radius, at which the cross-correlation analysis would yield a 3-$\sigma$ detection for the species of interest. Note, that the investigated wavelength region has only two chunks within the optical wavelength range and that we scale the template only according to the strongest absorption line in the accessible (i.e. not excluded) wavelength range of the observations. The template spectra are shown in Section~\ref{sec:Appendix}, where one can see that most of the species investigated have the strongest absorption line within the accessible wavelength range, and that the differences for the other templates are small. 

Figure~\ref{fig:injection} shows the injection approach, on the left side for \mbox{R\textsubscript{ext} = 1.5 R\textsubscript{P}} and on the right side for R\textsubscript{ext} = 3.0 R\textsubscript{P}. The top panels show the cross-correlation signal divided by its standard deviation and the bottom panel shows the scaled Ti template spectra. The injection recovery yields a detection > 10$\sigma$ for R\textsubscript{ext} = 3.0 R\textsubscript{P} and no detection for R\textsubscript{ext} = 1.5 R\textsubscript{P}. This verification demonstrates that the applied method is suitable to derive possible absorption arising from the planetary atmosphere. 

\section{Results}
\label{sec:Results}
\begin{figure*}
\includegraphics[width=1.0\textwidth]{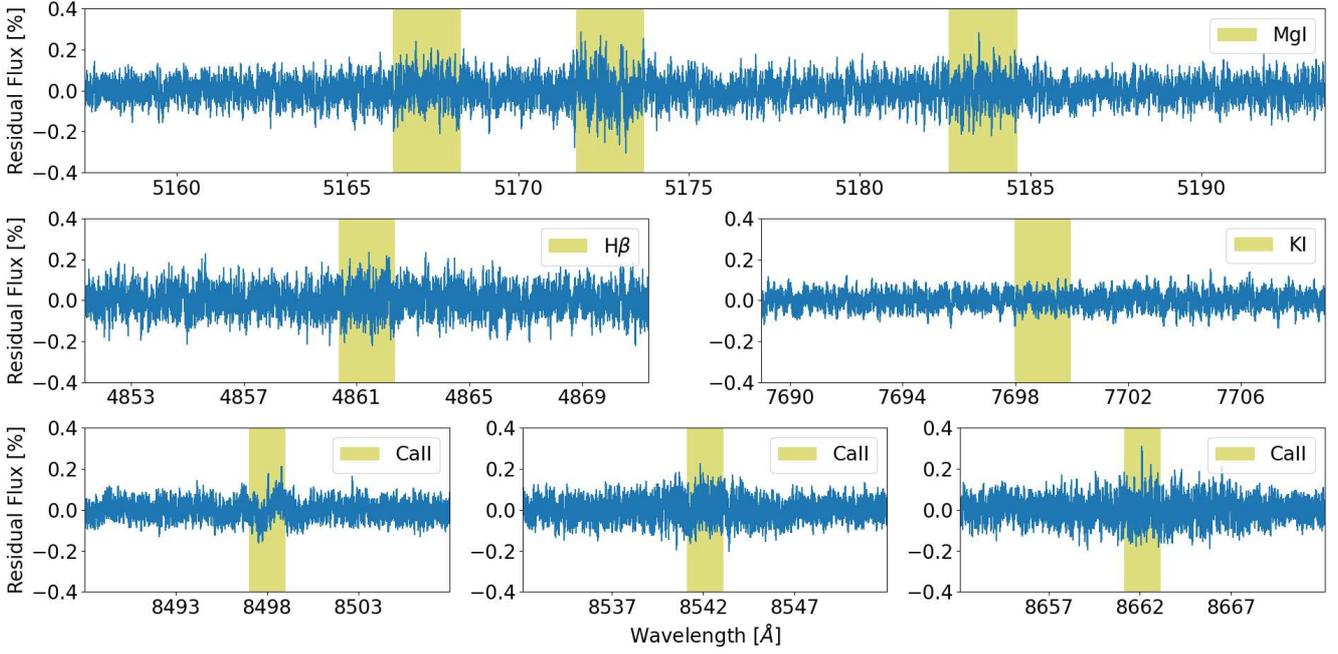}
    \caption{The combined transmission spectra for the \ion{Mg}{I}-triplet, the H$\beta$-line, the \ion{K}{I} line and the \ion{Ca}{II}-IR-triplet lines. The yellow shaded region shows the expected absorption position $\pm$1\AA.}
    \label{fig:transspec}
\end{figure*}

\begin{figure*}
\includegraphics[width=1.0\textwidth]{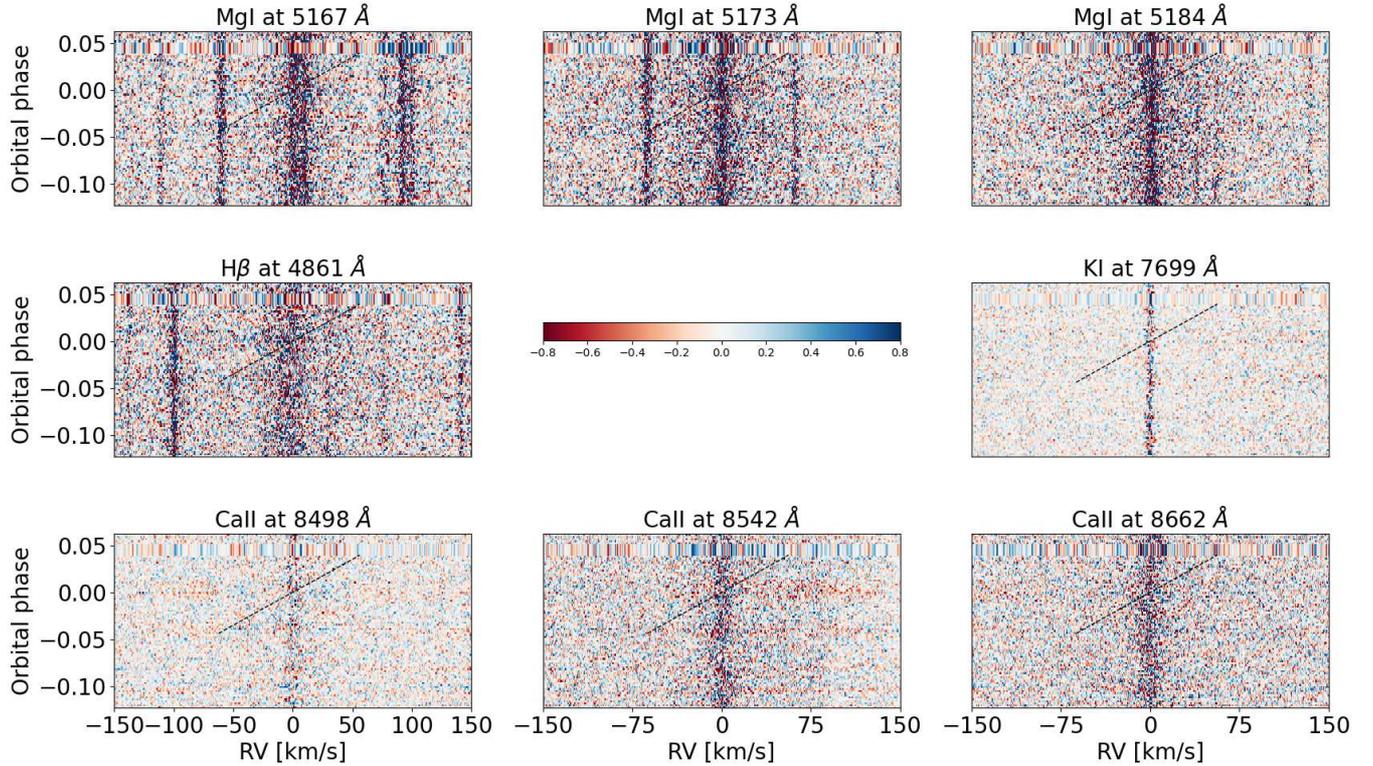}
    \caption{2D maps of the residual spectra at different orbital phases in the velocity range for strong opacity bearing species i.e. \ion{Mg}{I} (top), \ion{H}{$\beta$} and \ion{K}{I} (center) and \ion{Ca}{II} (bottom). The color bar shows the residual Flux in \%. The dashed black line shows the expected absorption trace.}
    \label{fig:2Dmapsingle}
\end{figure*}

Figure~\ref{fig:transspec} shows the combined transmission spectrum around strong opacity bearing species. Here, the standard deviation at $\pm$1\AA\space around the line core is 6.05 $\times$ 10\textsuperscript{-4} at the H$\beta$-line, 2.78 $\times$ 10\textsuperscript{-4} at the \ion{K}{I}-line, 6.49 $\times$ 10\textsuperscript{-4} at the \ion{Mg}{I}-lines and 5.09 $\times$ 10\textsuperscript{-4} at the \ion{Ca}{II}-lines (see Table~\ref{tab:fluxlevel}). Figure~\ref{fig:2Dmapsingle} shows the color-coded 2D map of the residual spectra at different orbital phases for these species. None reveals evidence of absorption.

\begin{table}
              \centering
              \caption{Standard deviation at $\pm$1\AA\space around the line core of certain species in the combined transmission spectrum }
              \label{tab:fluxlevel}
              \begin{tabular}{lr}
              \hline
           Line & Standard deviation \\
        \hline
           H$\beta$-line      &   6.05 $\times$ 10\textsuperscript{-4}  \\
           \ion{K}{I}-line    &   2.78 $\times$ 10\textsuperscript{-4}  \\
           \ion{Mg}{I}-lines  &   6.49 $\times$ 10\textsuperscript{-4}   \\
           \ion{Ca}{II}-lines &   5.09 $\times$ 10\textsuperscript{-4}    \\
           \hline
              \end{tabular}
\end{table}

Figures~\ref{fig:A1}- \ref{fig:A4} show the cross-correlation approach searching for different atmospheric species for the scenario with the atmospheric temperature of T = 2500 K. From left to right in landscape format, the first panels show the template spectrum, the second panels the cross-correlation signal, the third panels the colour coded cross-correlation signal for the transmission spectra at different orbital phases in the velocity range and the fourth panel the colour coded K\textsubscript{P} - RV maps. The K\textsubscript{P} - RV maps are calculated $\pm$100 km/s around the expected velocity-semi amplitude of the planet and are used to verify that a possible absorption signal arises only at the expected K\textsubscript{P} value and is thus of planetary origin. The inspection of the cross-correlation signals shows no evidence of absorption for any of the investigated species. The scatter remains mainly below $\pm$3 standard deviations. Also, no evidence of significant absorption is visible in the third panels or visible at different K\textsubscript{P} positions in the fourth panels.

\begin{figure*}
\includegraphics[width=0.99\textwidth]{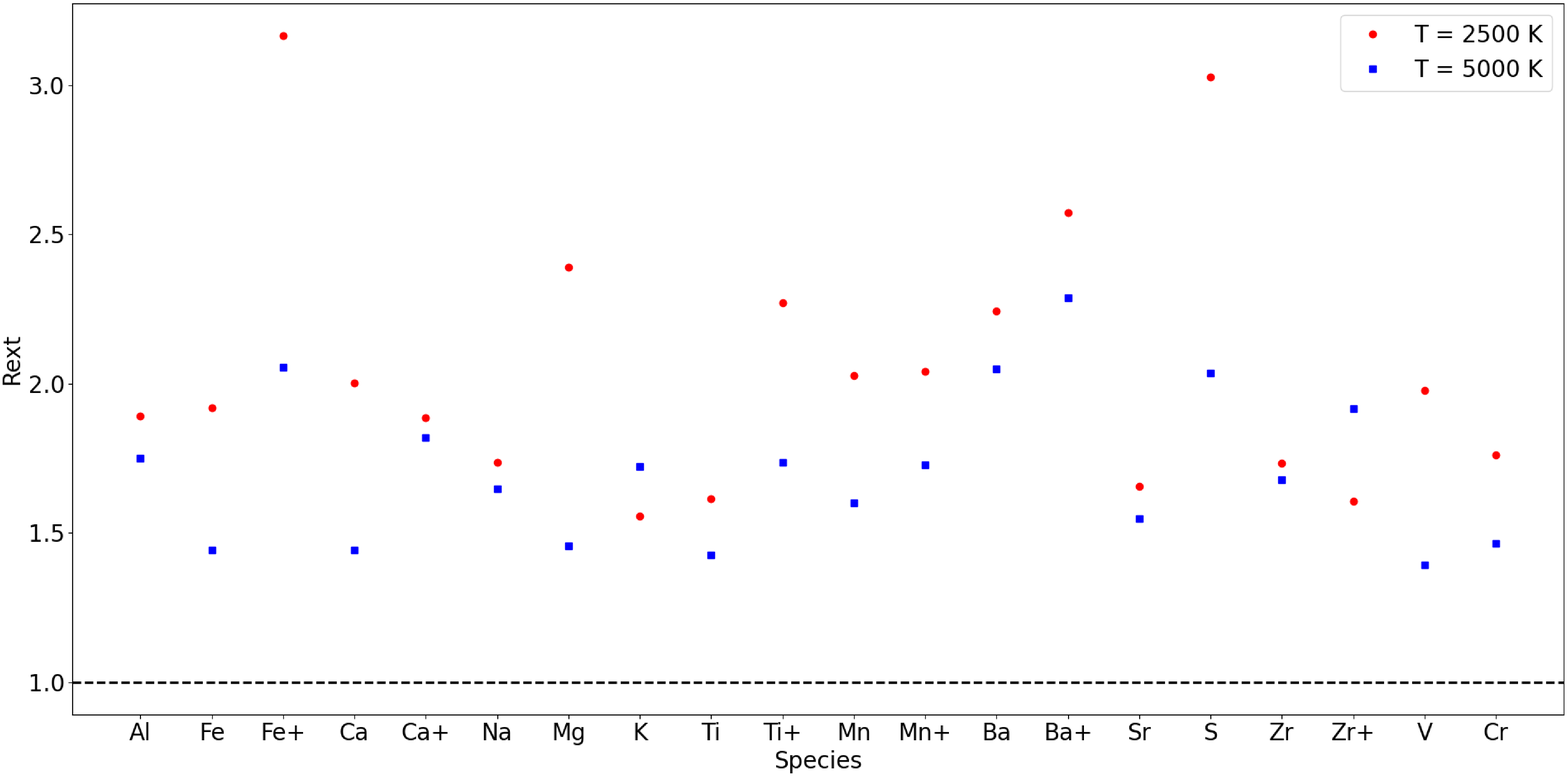}
    \caption{3$\sigma$ limit for atmospheric absorption arising from an extended atmosphere for different species. The red dots correspond to the T = 2500K scenario and the blue squares to the T = 5000K scenario. The dashed black line marks the planetary surface at 1 $\times$ R\textsubscript{P}.}
    \label{fig:extension}
\end{figure*}

The fifth panel shows the detection thresholds for different R\textsubscript{ext} values applied for the injection recovery up to 3 R\textsubscript{P}, with the blue dot showing the significance for the injection of the unscaled template calculated according to the planetary properties as described in Section~\ref{sec:Templates}. Due to the high-mean molecular weight of the atmosphere, the extension of the atmosphere is mainly below 1.4 R\textsubscript{P} in the unscaled templates for all species investigated, not detectable at a 3$\sigma$ confidence level within the observation available. We are able to put an upper limit of the atmospheric extension for the investigated species by injecting scaled templates into the data and recovering them. Figure~\ref{fig:extension} shows the atmospheric extension R\textsubscript{ext} limits which we would able to probe for the investigated species for the T = 2500 K scenario (red dots) and the T = 5000 K scenario (blue squares) at the 3$\sigma$ confidence level. The dashed line marks the planetary surface at 1 $\times$ R\textsubscript{P}. The upper limit of the atmospheric extension shows a decrease for most species going from \mbox{T = 2500 K} scenario to the T = 5000 K scenario, mainly due to the variation in the number of absorption lines, but depends also on the quality of the combined transmission spectrum at the regions compared to the synthetic templates. Table~\ref{tab:Table2} shows the derived limits for the investigated species at a 3$\sigma$ confidence level, with a median value of \mbox{R\textsubscript{ext} = 1.9 $\times$ R\textsubscript{P}} for the T = 2500 K scenario and R\textsubscript{ext} = 1.7 $\times$ R\textsubscript{P} for the T = 5000 K scenario, ruling out absorption from an atmospheric envelope which is widely extended for most of the species.

\begin{table*}
              \centering
              \caption{Derived limits for the atmospheric extension R\textsubscript{ext} at a 3$\sigma$ confidence level for the different species investigated for atmospheric temperature of \mbox{T = 2500 K} and T = 5000 K.}
              \label{tab:Table2}
              \setlength{\tabcolsep}{4.4pt}
              \begin{tabular}{ccccccccccccccccccccc}
              \hline
              Species & Al        & Fe & Fe\textsuperscript{+}    &Ca & Ca\textsuperscript{+} & Na & Mg & K & Ti & Ti\textsuperscript{+} & Mn & Mn\textsuperscript{+} &Ba& Ba\textsuperscript{+} & Sr & S & Zr & Zr\textsuperscript{+} & V & Cr\\
              \hline
              T = 2500K  & 1.88 &  1.91 & 3.16 & 2.00 & 1.88 & 1.73 & 2.39 & 1.55 & 1.61 & 2.27 & 2.02 & 2.04 & 2.24 & 2.57 & 1.65 & 3.02 & 1.73 & 1.60 & 1.97 & 1.75\\
              \vspace{0.5cm}
              T = 5000K & 1.75 &  1.44 & 2.05 & 1.44 & 1.81 & 1.64 & 1.45 & 1.72 & 1.42 & 1.73 & 1.60 & 1.72 &  2.04 & 2.28 & 1.54  & 2.03 & 1.67 & 1.91 &  1.39 & 1.46\\
              \end{tabular}
\end{table*}
\section{Discussion and Conclusion}
\label{sec:Discussion}
The search for absorption by silicate species in the transmission spectrum of \mbox{55 Cnc e}, for instance released by a magma ocean, did not show an absorption for any of the investigated species. To account for different conditions in the upper atmosphere of \mbox{55 Cnc e}, we repeated the cross-correlation analysis with template spectra calculated for T = 5000 K i.e. $\sim$2 $\times$ T\textsubscript{eq}, yielding the same non-detection of any absorption signature in all cases. 

Our non-detection of H$\beta$ agrees with previous investigations not finding evidence of hydrogen absorption for this planet \citep{Ehrenreich_2012,Tabernero_2020}. Other absorption signatures from species with strong opacities previously resolved on gaseous exoplanets such as \ion{Mg}{I} \citep{Cauley2019}, \ion{K}{I} \citep{Keles2019} or \ion{Ca}{II} \citep{Yan_2020} are not detected either for 55 Cnc e. 

For short period exoplanets, high escape rates of \ion{Na}{I}, \ion{Ca}{II} and \ion{Mg}{II} can be expected to build up large planetary tails on hot rocky exoplanets \citep{Mura_2011}. In this study, we derived limits for the atmospheric extension related to the absorption of major species (see Table~\ref{tab:Table2}), including the major abundant species in the Earth's crust such as Al, Fe, Ca, Na, Mg and K (see Table~\ref{tab:Table1}), which are far below the Roche lobe radius of R\textsubscript{Roche} = 5.35 $\times$ R\textsubscript{$\oplus$} \citep{Ridden_Harper_2016}, i.e. R\textsubscript{Roche} $\approx$ 2.85 $\times$ R\textsubscript{P}, showing that this might be not the case for 55 Cnc e. 

We compare our results to the findings by \citet{Ridden_Harper_2016} who showed a possible 3-$\sigma$ sodium signature at the \ion{Na}{D} lines and a variable 4.1-$\sigma$ Ca\textsuperscript{+} signature at the \ion{Ca}{H\&K} lines on \mbox{55 Cnc e}. As \citet{Ridden_Harper_2016} uses a larger planetary radius which bases on \citet{Gillon_2012}, we compare the inferred probed altitudes in terms of R\textsubscript{$\oplus$}. The \ion{Na}{D} signal from \citet{Ridden_Harper_2016} corresponds to R\textsubscript{P} = 5 $\times$ R\textsubscript{$\oplus$}. To be able to compare our findings,we scale the Na- template so that the \ion{Na}{D} lines in the templates reach this altitude and repeat the analysis. Note, that we have no access to the resonance lines in our observations, but the scaling affects all lines within the wavelength range accessible. For \ion{Na}{D} lines reaching 5 $\times$ R\textsubscript{$\oplus$} in the template, the injection recovery would yield a 5.8-$\sigma$ detection. Therefore, we can rule out the presence of sodium at such extended altitudes. In our work, the inferred 3-$\sigma$ limit for the Na investigation shows an upper limit of \mbox{R\textsubscript{ext} = 1.73 $\times$ R\textsubscript{P}}, which would correspond to \ion{Na}{D} lines reaching R\textsubscript{ext} = 1.99 $\times$ R\textsubscript{P} and thus R\textsubscript{P} = 3.75 $\times$ R\textsubscript{$\oplus$} compared to the \mbox{R\textsubscript{P} = 5 $\times$ R\textsubscript{$\oplus$}} presented by \citet{Ridden_Harper_2016}. We apply the same comparison for the Ca\textsuperscript{+} lines. The \ion{Ca}{H\&K} lines reach around  R\textsubscript{P} = 25 $\times$ R\textsubscript{$\oplus$} in \citet{Ridden_Harper_2016}, corresponding to R\textsubscript{ext} $\approx$ 13.30 $\times$ R\textsubscript{P}. We scale the ionized calcium- template so that the \ion{Ca}{H\&K} lines in the templates reach this altitude and repeat the analysis. Note, that we have also no access to these resonance lines in our observations. For such an atmospheric extension, we would yield a > 64-$\sigma$ detection, which is not the case. A 4.1-$\sigma$ detection of ionized calcium would correspond to an upper limit of around R\textsubscript{ext} = 2.1 $\times$ R\textsubscript{P} in this work, which would correspond to \ion{Ca}{H\&K} lines reaching R\textsubscript{ext} = 2.39 $\times$ R\textsubscript{P} and thus R\textsubscript{P} = 4.5 $\times$ R\textsubscript{$\oplus$}, compared to the R\textsubscript{P} = 25 $\times$ R\textsubscript{$\oplus$} presented by \citet{Ridden_Harper_2016}. Thus, within the direct comparison to \citet{Ridden_Harper_2016} regarding the Na- and Ca\textsuperscript{+}- absorption, we can only put a slightly lower absorption limit regarding the presence of sodium, but more than five times lower limit regarding the presence of the ionized calcium in the exosphere of 55 Cnc e, probably due to the larger S/N values achieved by the observation and the cross-correlation analysis, although \citet{Ridden_Harper_2016} have access to stronger resonance absorption lines.

Note, that \citet{Ridden_Harper_2016} did not claim detection and reported the variability of the \ion{Ca}{H\&K} absorption, which might be due to stellar activity. Probably also induced by stellar variability, a broad downward showing feature is visible in the cross-correlation analysis for the ionized \ion{Ca}{II} lines (see Figure~\ref{fig:A1}) at a significance of \mbox{$\sim$4 $\sigma$}. As the template spectrum and the transmission spectrum have the same alignment for the cross-correlation analysis, a downward peaking feature suggests emission rather than absorption. Inspecting the \ion{Ca}{II}- infrared triplet lines that have the major contribution within the cross-correlation analysis, in the combined transmission spectrum (Figure~\ref{fig:transspec}), only the \ion{Ca}{II} line around 8498\AA\space shows a hint for an emission. We repeat the cross-correlation analysis for the ionized \ion{Ca}{II} excluding this line, which results in the decrease of the significance of the feature down to $\sim$3 $\sigma$. Therefore, this strengthens the possibility that the emission feature does not arise from the planet or from an ionized calcium-rich envelope, but is rather due to variability in the stellar calcium lines, also strengthened by the fact that the emission is not following the atmospheric trace (see the third panel in Figure~\ref{fig:A1}).

The atmospheric conditions on 55 Cnc e are still puzzling: is there a detectable atmosphere on this planet? Possible scenarios include a purely rocky surface, a rocky surface with a magma ocean on top releasing species and building up a thin silicate-vapor envelope, or a atmosphere e.g. with a water envelope or a N\textsubscript{2} or CO\textsubscript{2} rich atmosphere \citep{Angelo_2017} with a surface pressure similar to Earth. The non-detection of water \citep{Esteves_2017,Jindal_2020} with the non-detection of light elements \citep{Ehrenreich_2020,Zhang_2021} seems to make the water envelope scenario unlikely. A detectable thin silicate-vapor atmosphere scenario can not be strengthened either by \citet{Tabernero_2020} who did not find evidence for \ion{Na}{I} absorption or by this work not finding evidence of absorption for a variety of species. Overall, our work provide indications that favor the heavyweight atmospheric scenario. This would be in agreement with the shift detection of the hot spot east of the substellar point on 55 Cnc e detected by \citet{Demory_2016}. Such a shift is usually driven by atmospheric circulation processes, including a significant amount of atmospheric mass \citep{Showman_and_Polvani,Kempton_and_Rauscher_2012,Keles2021}. Although we cannot rule out a heavy silicate-atmosphere existing below the presented altitude limits inferred in this work, with a median value of \mbox{ R\textsubscript{ext} = 1.9 $\times$ R\textsubscript{P}}, or a deck of mineral clouds \citep{Mahapatra_2017} that hide absorption signatures, this study showed that very high S/N data can deliver informative upper limits on the extension of the possibly existing silicate atmospheres probing deeply the atmospheres of terrestrial exoplanets.

\section*{Acknowledgements}
This survey is a collaborative effort from AIP, MPIA, INAF, University of Arizona, Ohio State University and the LBT. 
\newline
EK is thankful for the support and granting the LBT time for this observation by the Rat Deutscher Sternwarten (RDS). 
\newline
XA is grateful for the financial support from the Potsdam Graduate School (PoGS) in the form of a doctoral scholarship.
\newline
KM: This research was supported by the Excellence Cluster ORIGINS which is funded by the Deutsche Forschungsgemeinschaft (DFG, German Research Foundation) under Germany's Excellence Strategy - EXC-2094 - 390783311.
\newline
KP and LK acknowledge support from the \textit{Leibniz-Gemeinschaft} under project number P67/2018.

\section*{Data Availability}
The data underlying this article will be shared on reasonable request
to the corresponding author.



\bibliographystyle{mnras}
\bibliography{kelesbib} 



\appendix
\label{sec:Appendix}
\section{}


\begin{figure*}%
  \centering
  \subfloat{\includegraphics[scale = 0.12, angle =90 ]{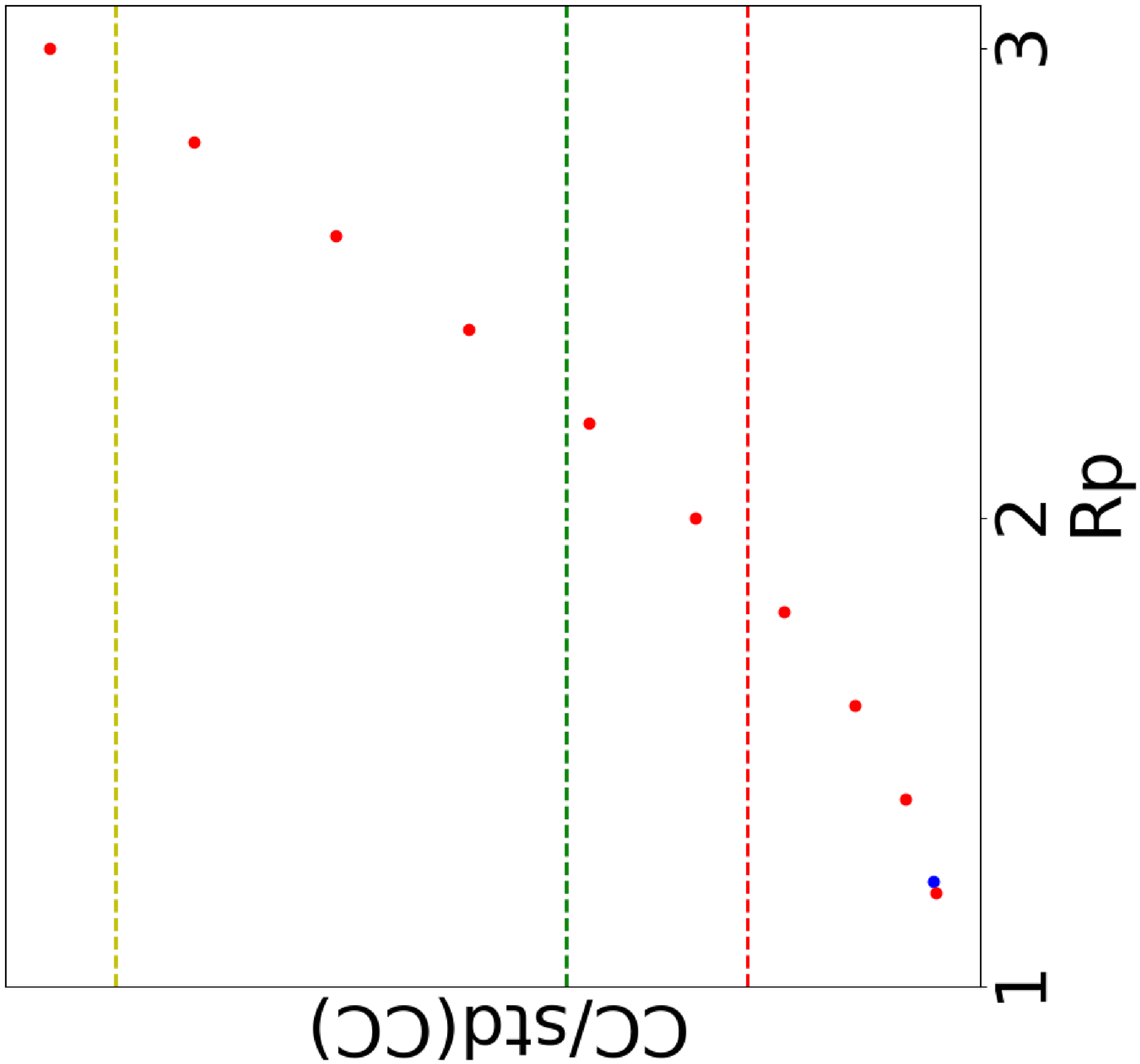}}%
  \qquad
  \subfloat{\includegraphics[scale = 0.12, angle =90 ]{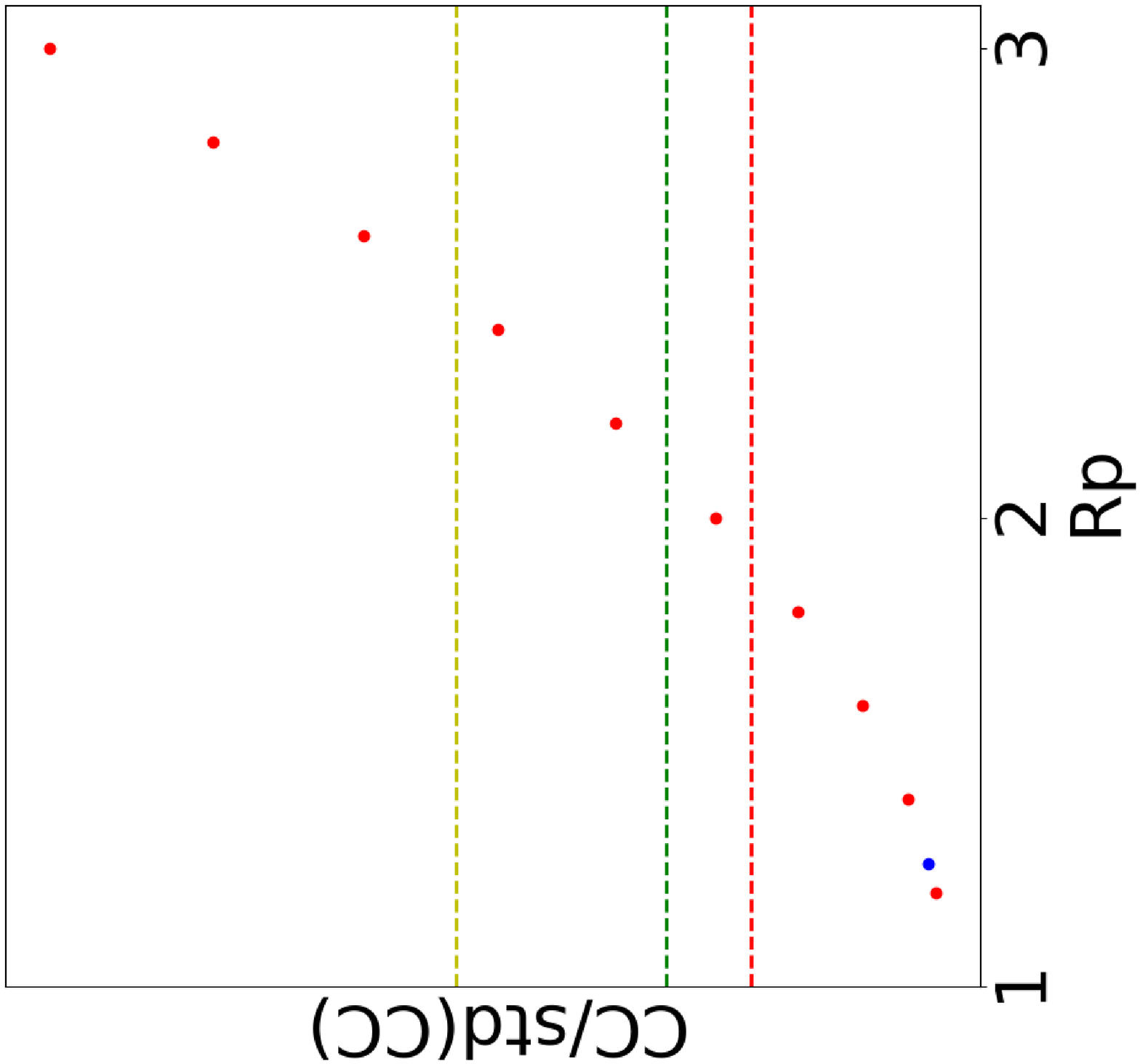}}%
  \qquad
  \subfloat{\includegraphics[scale = 0.12, angle =90 ]{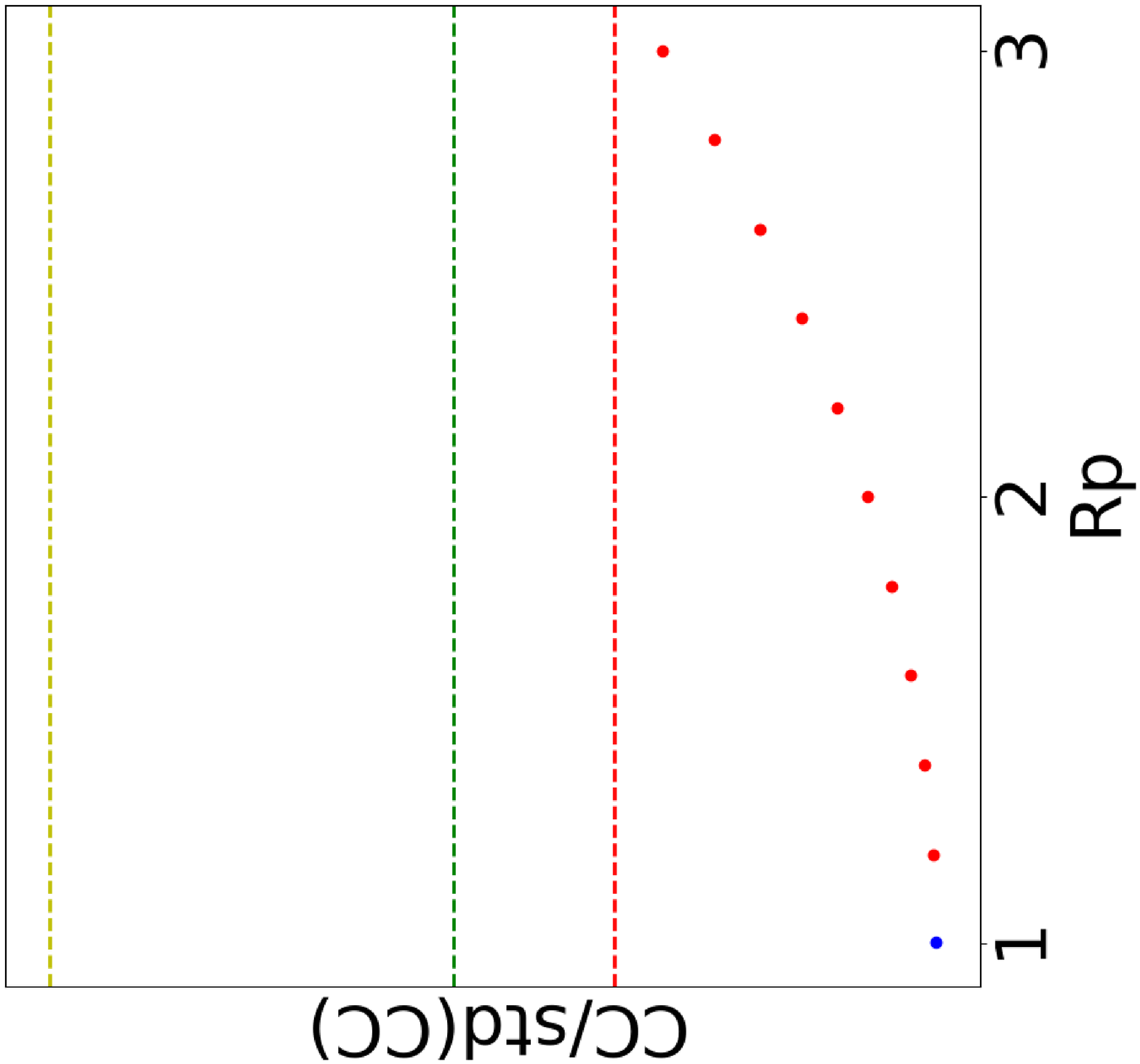}}%
  \qquad
  \subfloat{\includegraphics[scale = 0.12, angle =90 ]{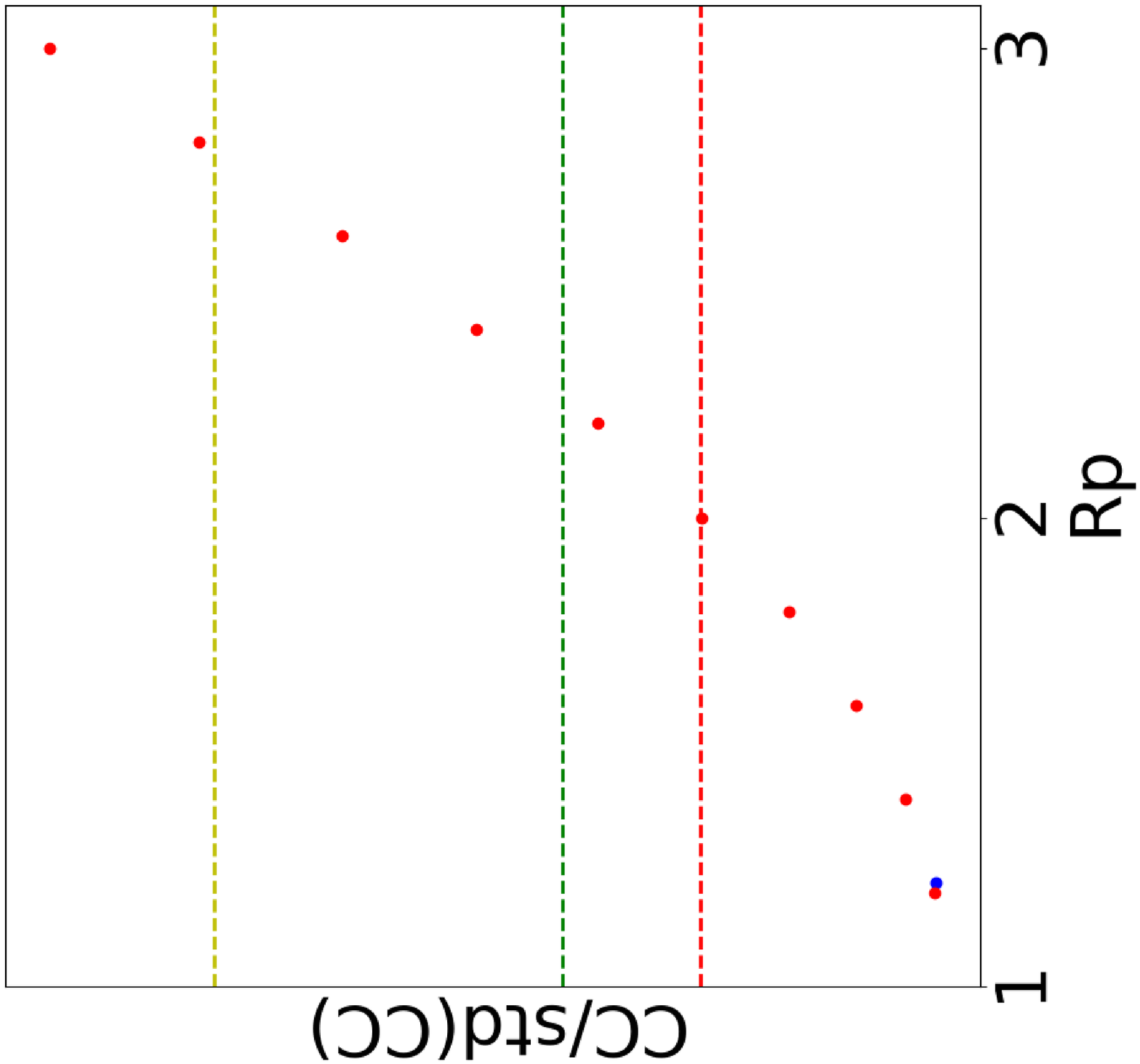}}%
  \qquad
  \subfloat{\includegraphics[scale = 0.12, angle =90 ]{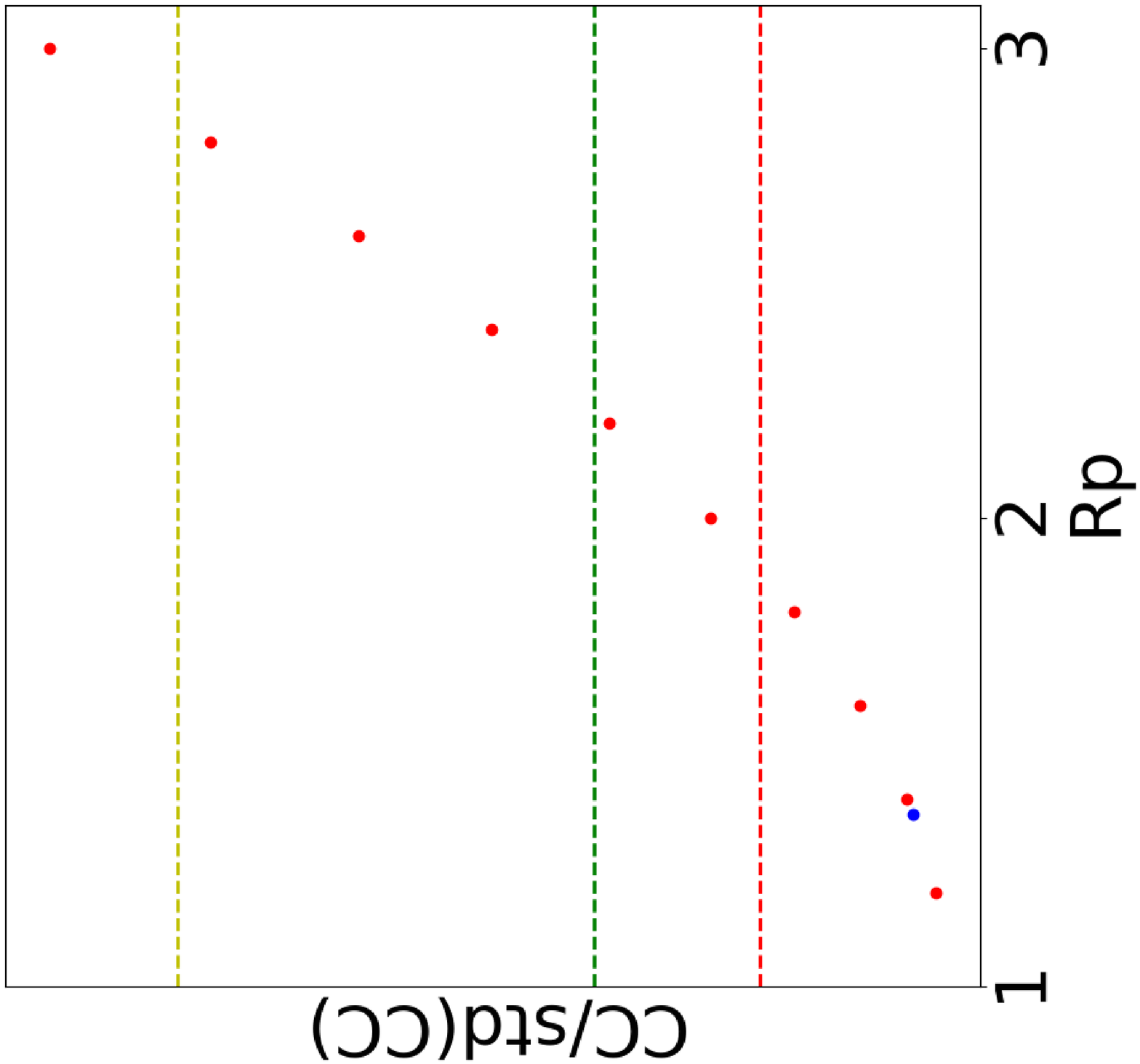}}%
  \\
  \vspace{0.5cm}
  \centering
  \subfloat{\includegraphics[scale = 0.12, angle =90 ]{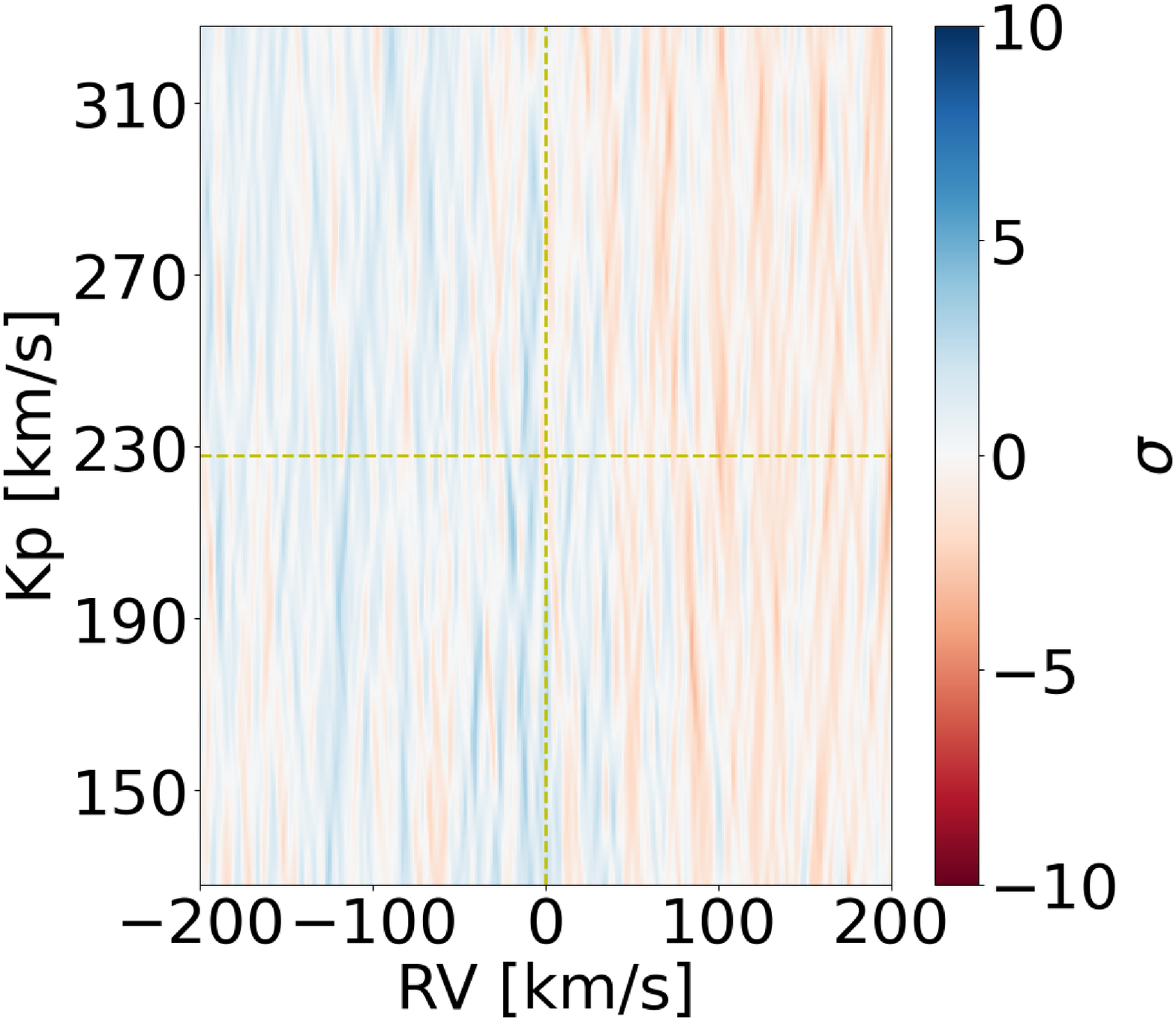}}%
  \qquad
  \subfloat{\includegraphics[scale = 0.12, angle =90 ]{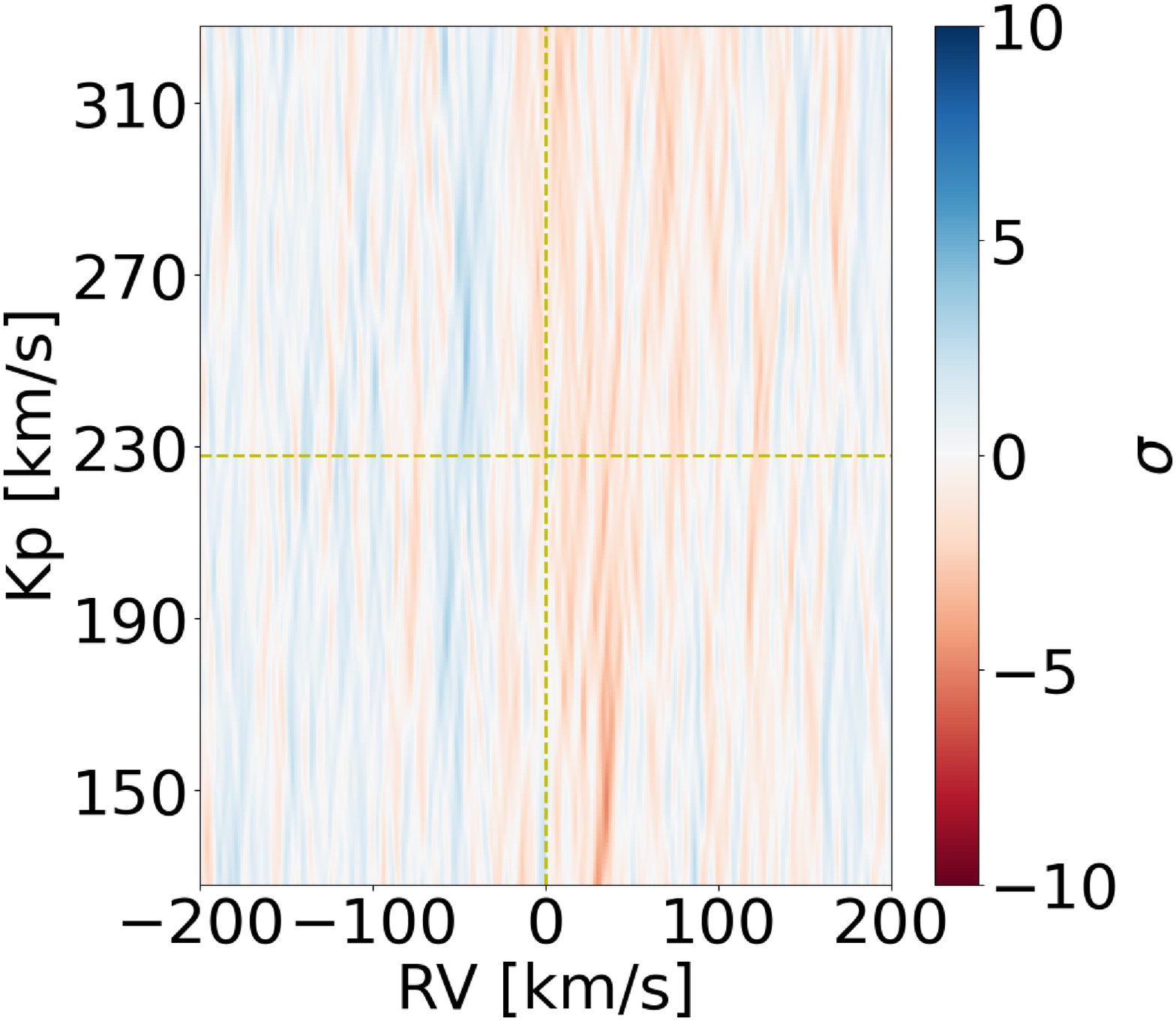}}%
  \qquad
  \subfloat{\includegraphics[scale = 0.12, angle =90 ]{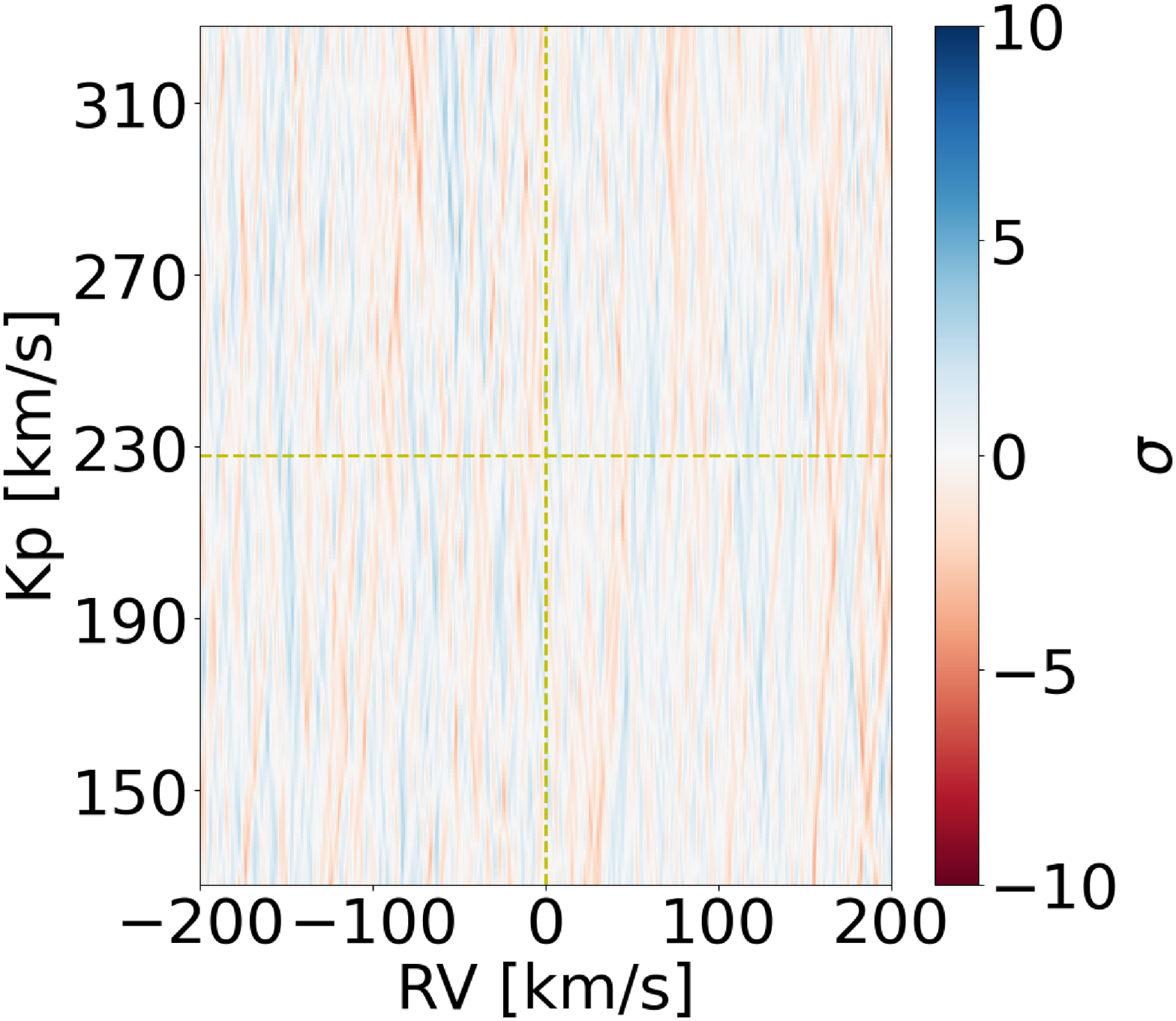}}%
  \qquad
  \subfloat{\includegraphics[scale = 0.12, angle =90 ]{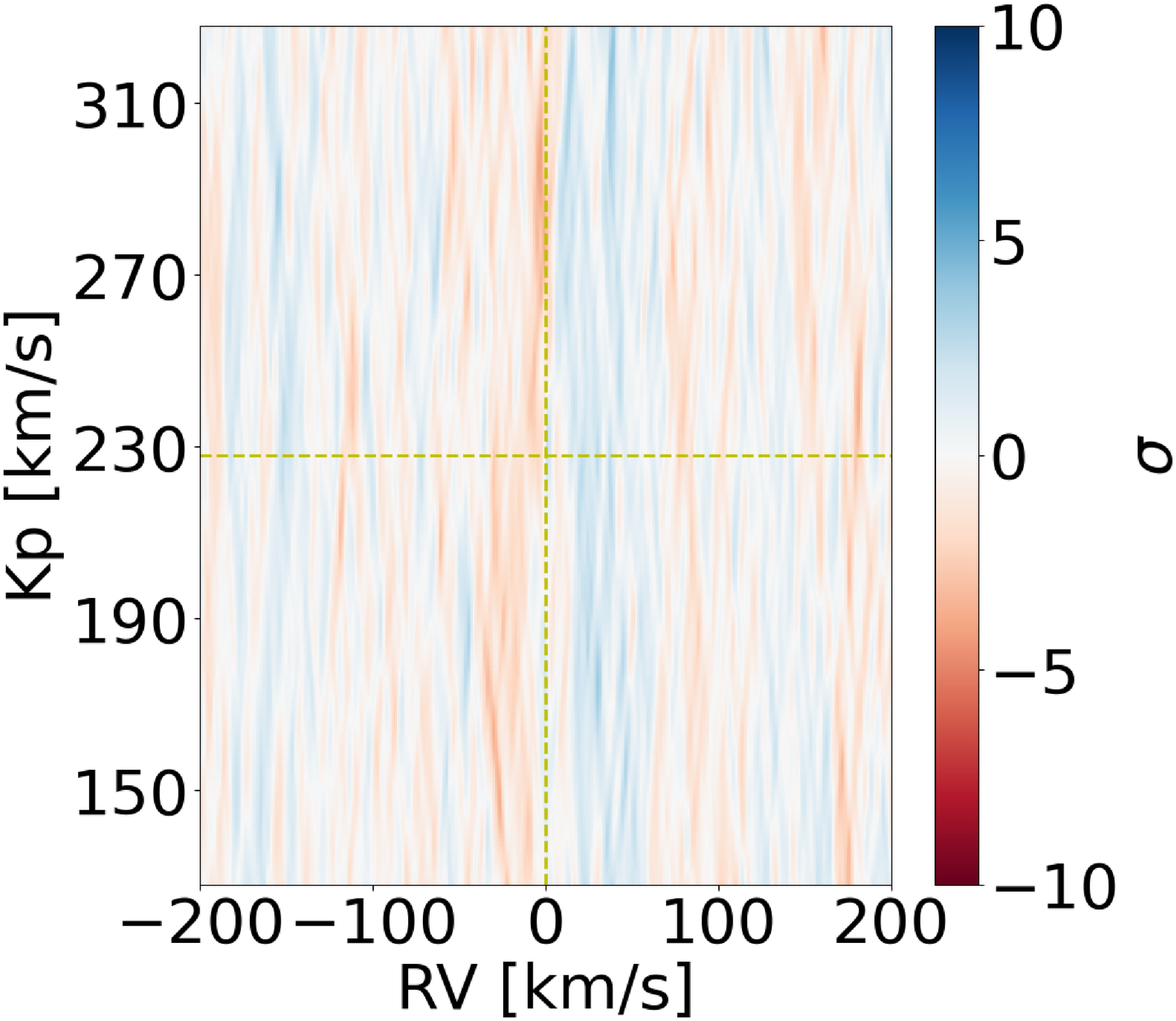}}%
  \qquad
  \subfloat{\includegraphics[scale = 0.12, angle =90 ]{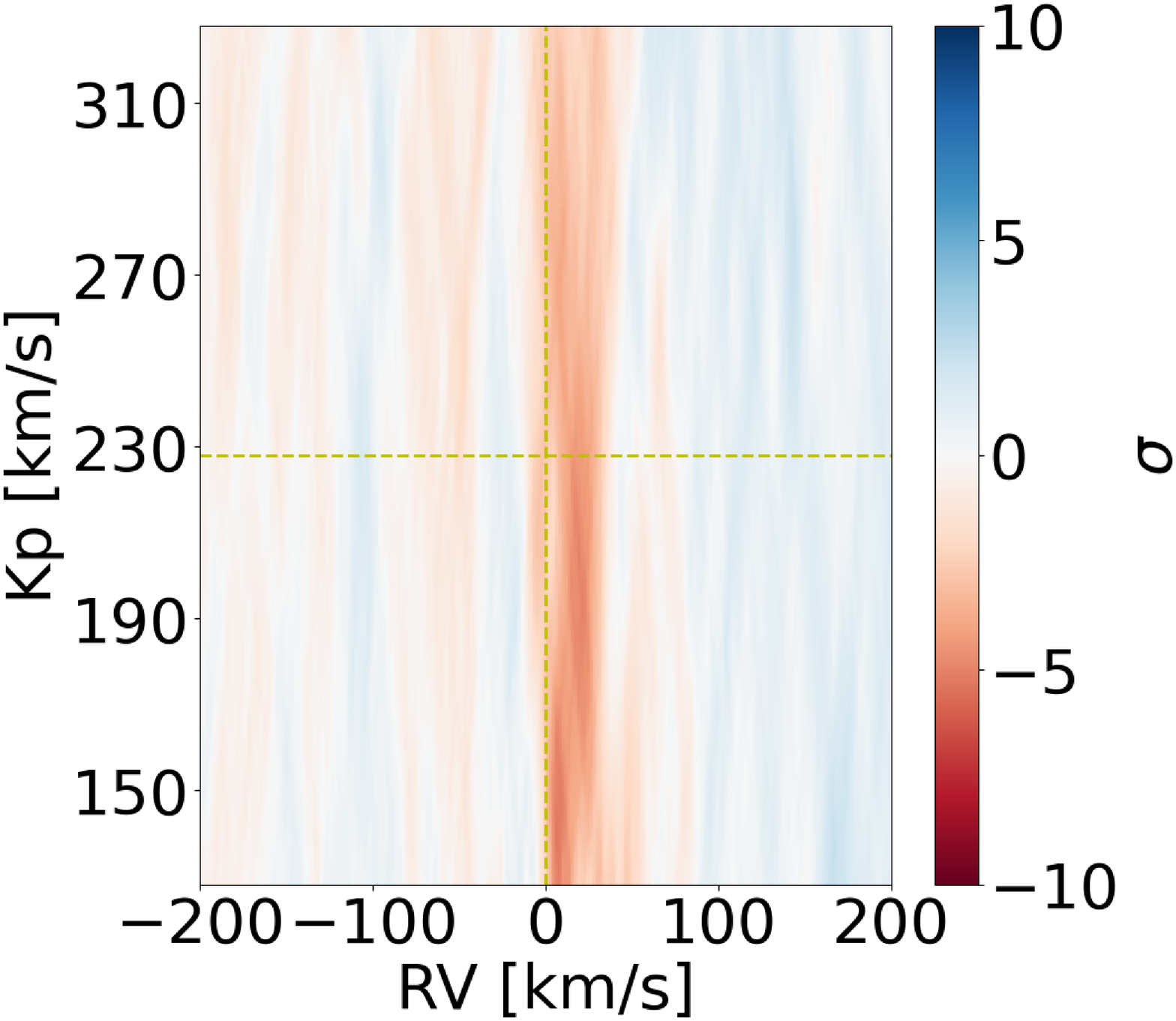}}%
  \\
  \vspace{0.5cm}
  \centering
  \subfloat{\includegraphics[scale = 0.12, angle =90 ]{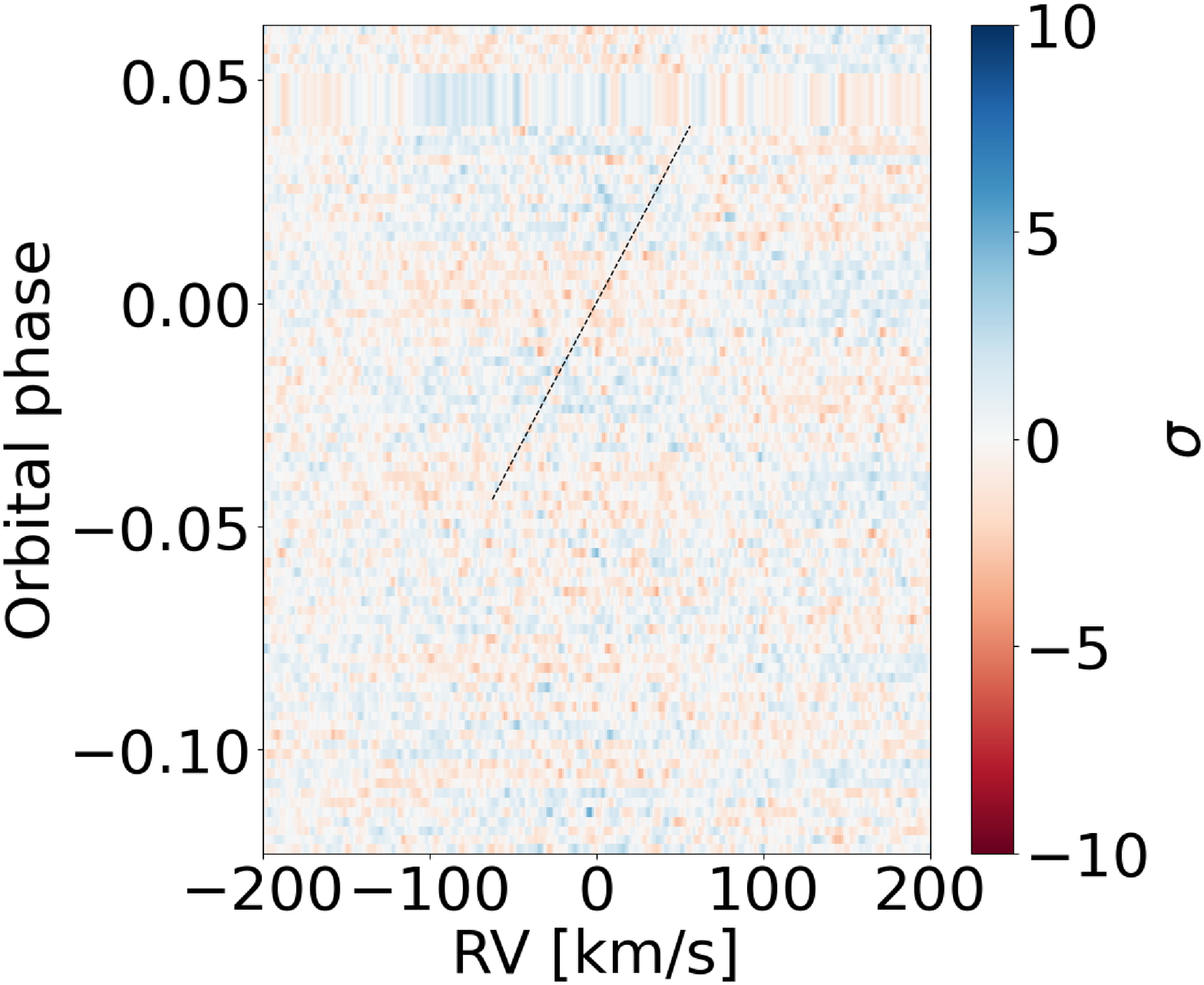}}%
  \qquad
  \subfloat{\includegraphics[scale = 0.12, angle =90 ]{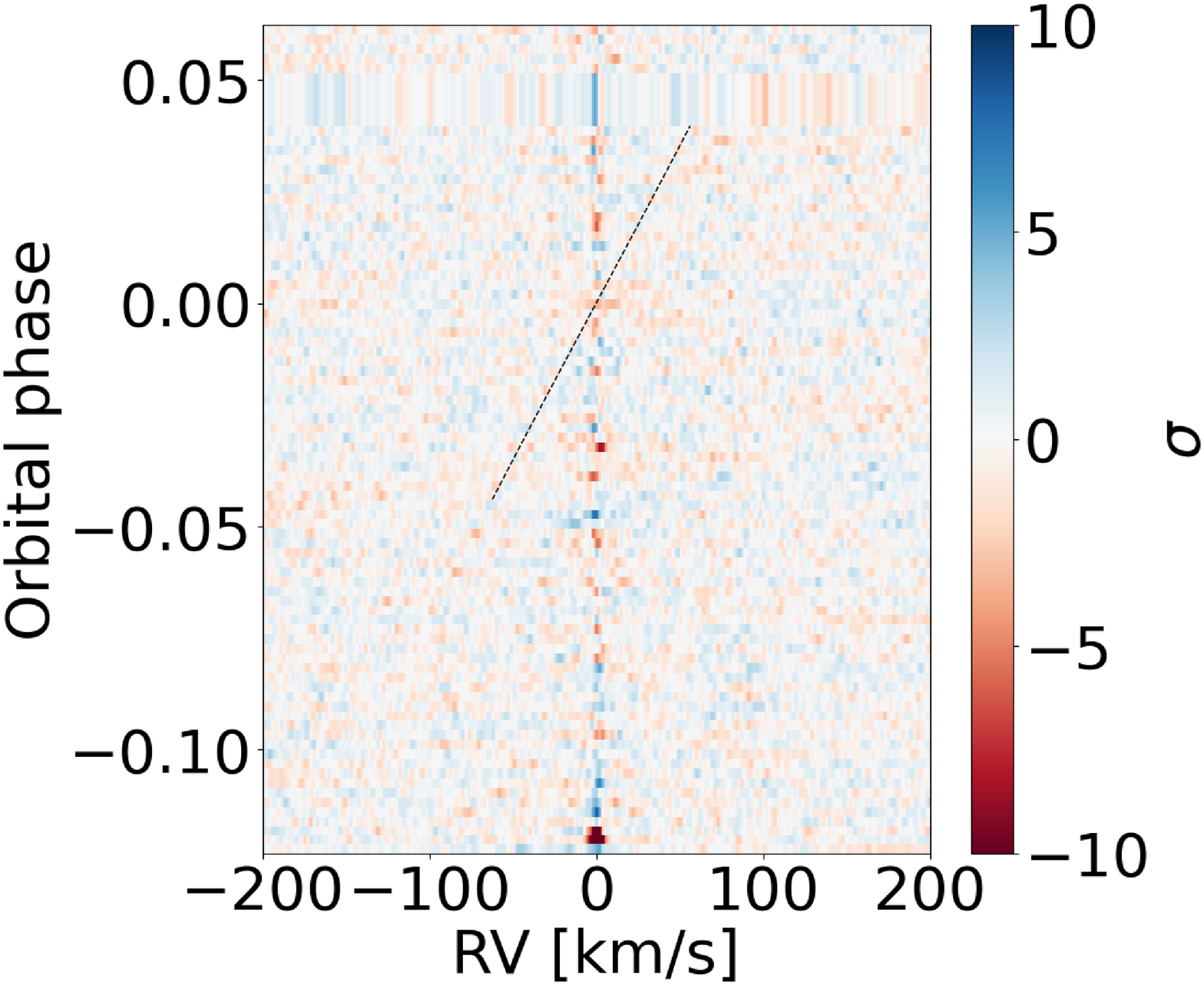}}%
  \qquad
  \subfloat{\includegraphics[scale = 0.12, angle =90 ]{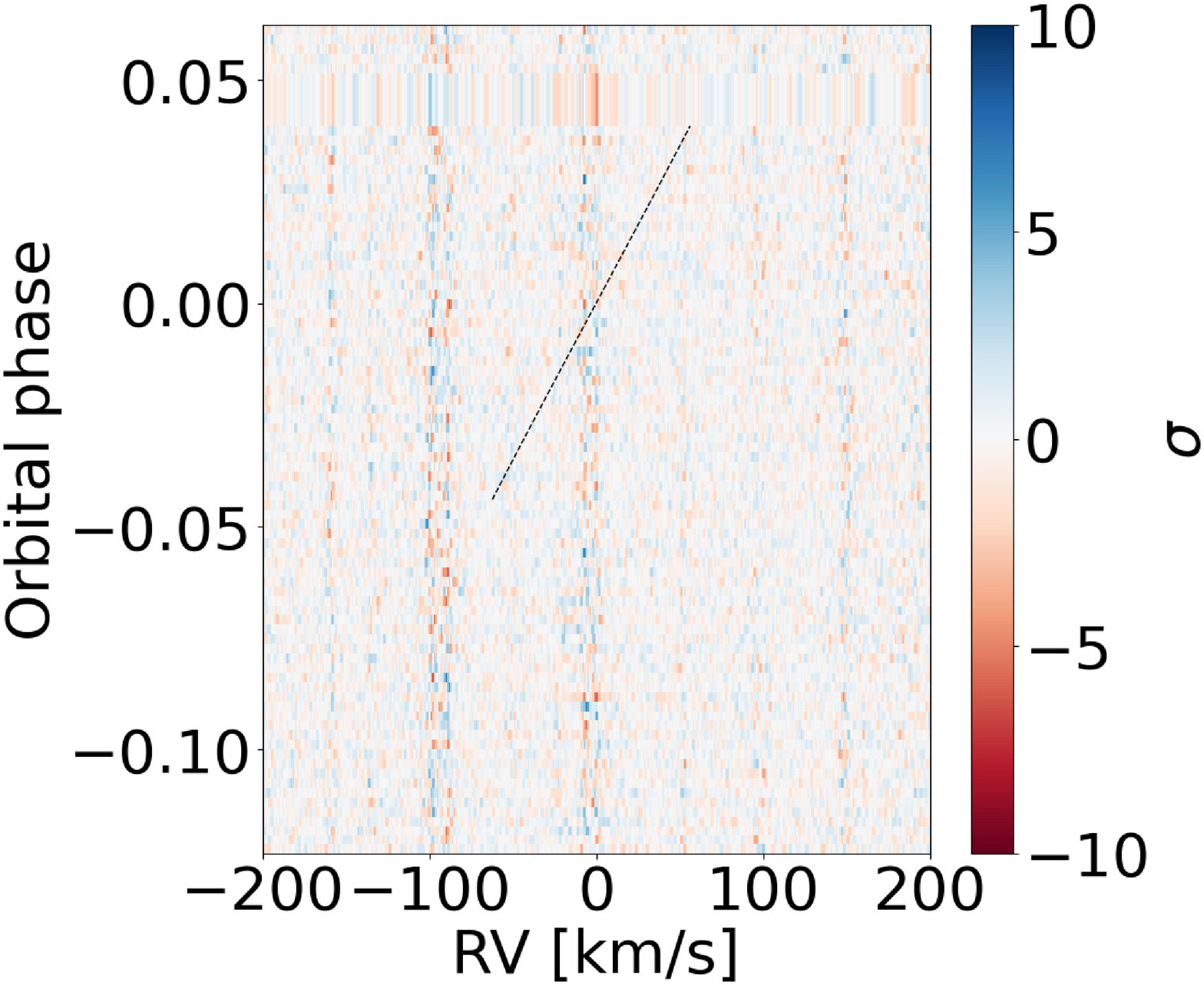}}%
  \qquad
  \subfloat{\includegraphics[scale = 0.12, angle =90 ]{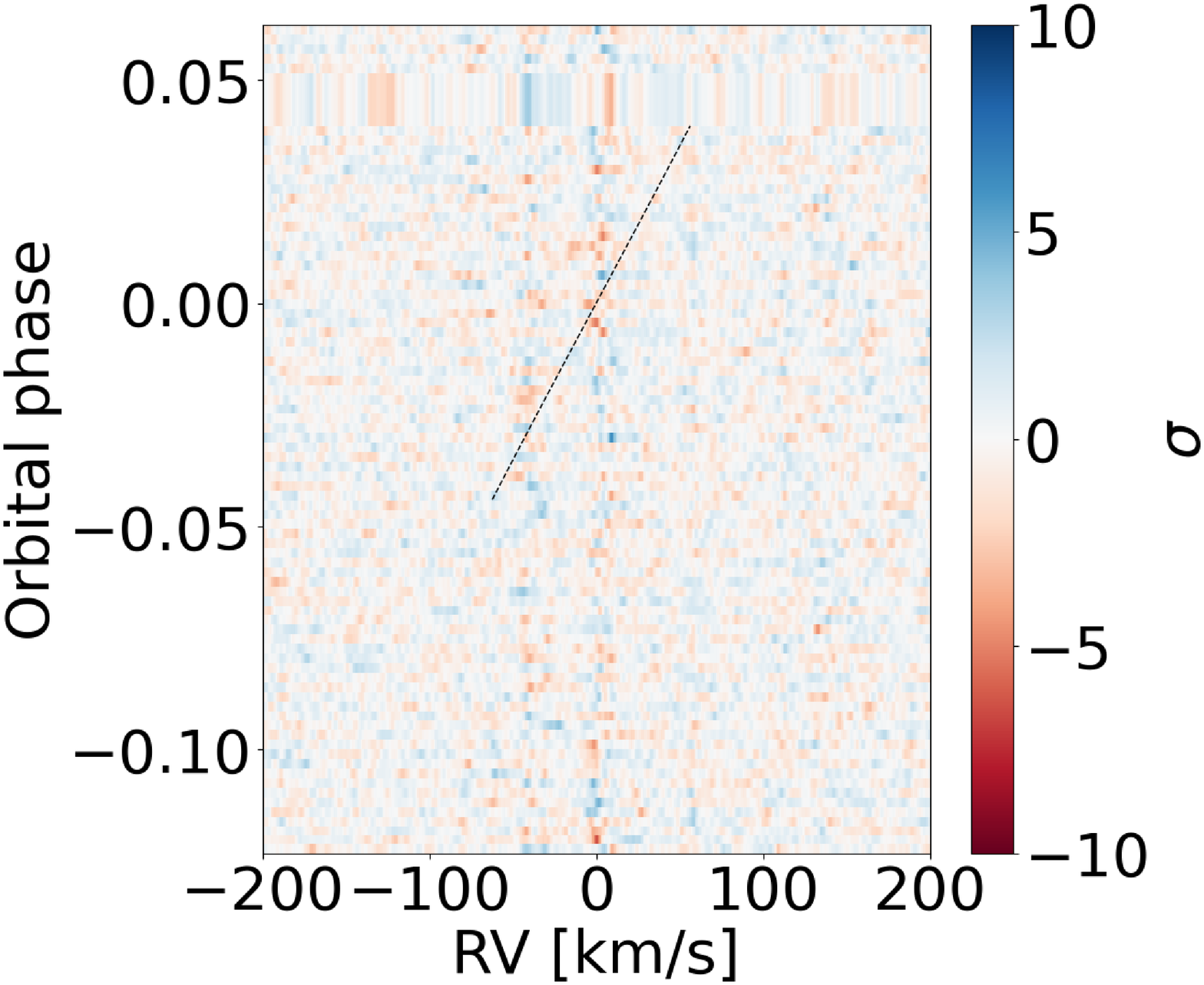}}%
  \qquad
  \subfloat{\includegraphics[scale = 0.12, angle =90 ]{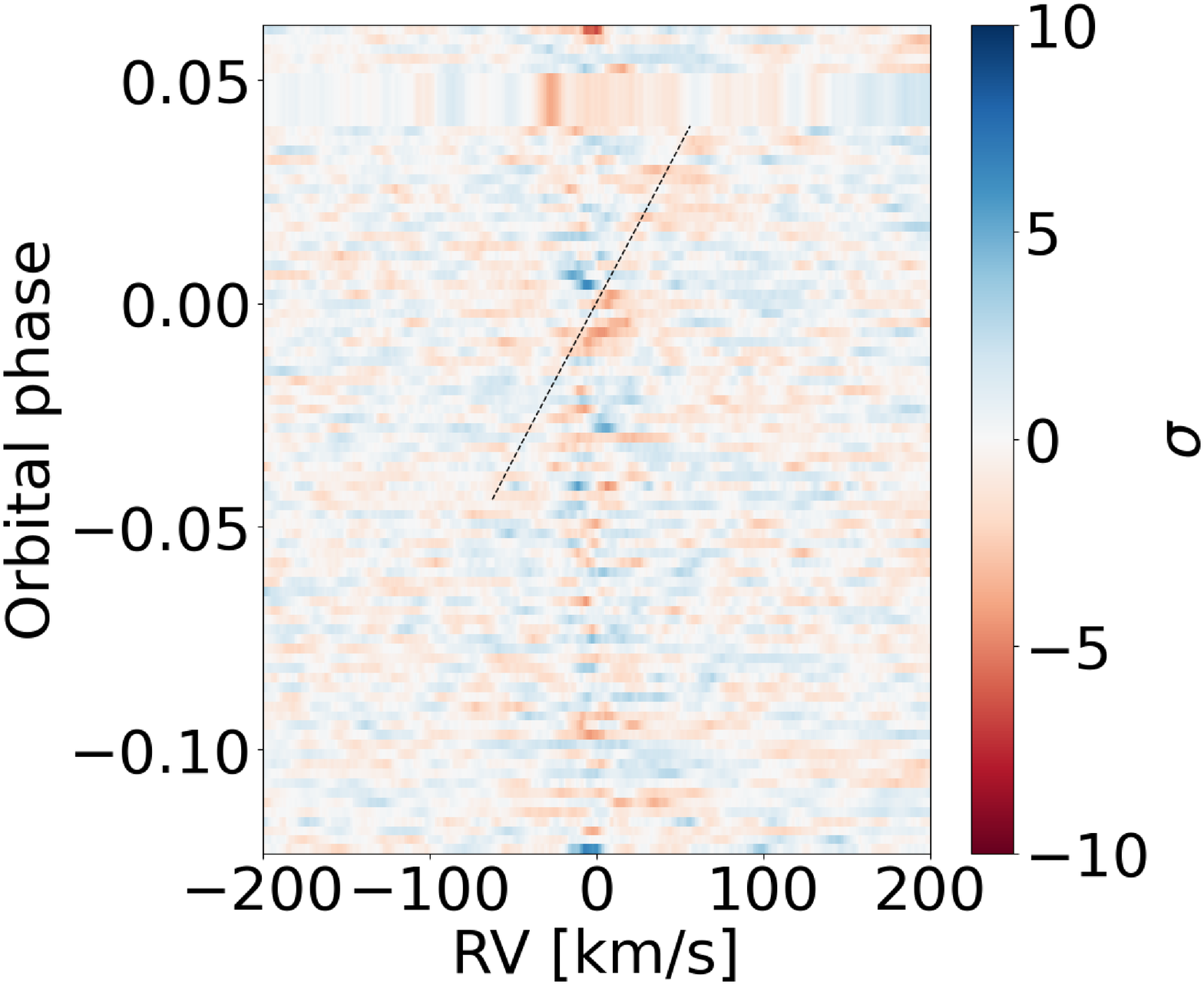}}%
  \\
  \vspace{0.5cm}
  \centering
  \subfloat{\includegraphics[scale = 0.12, angle =90 ]{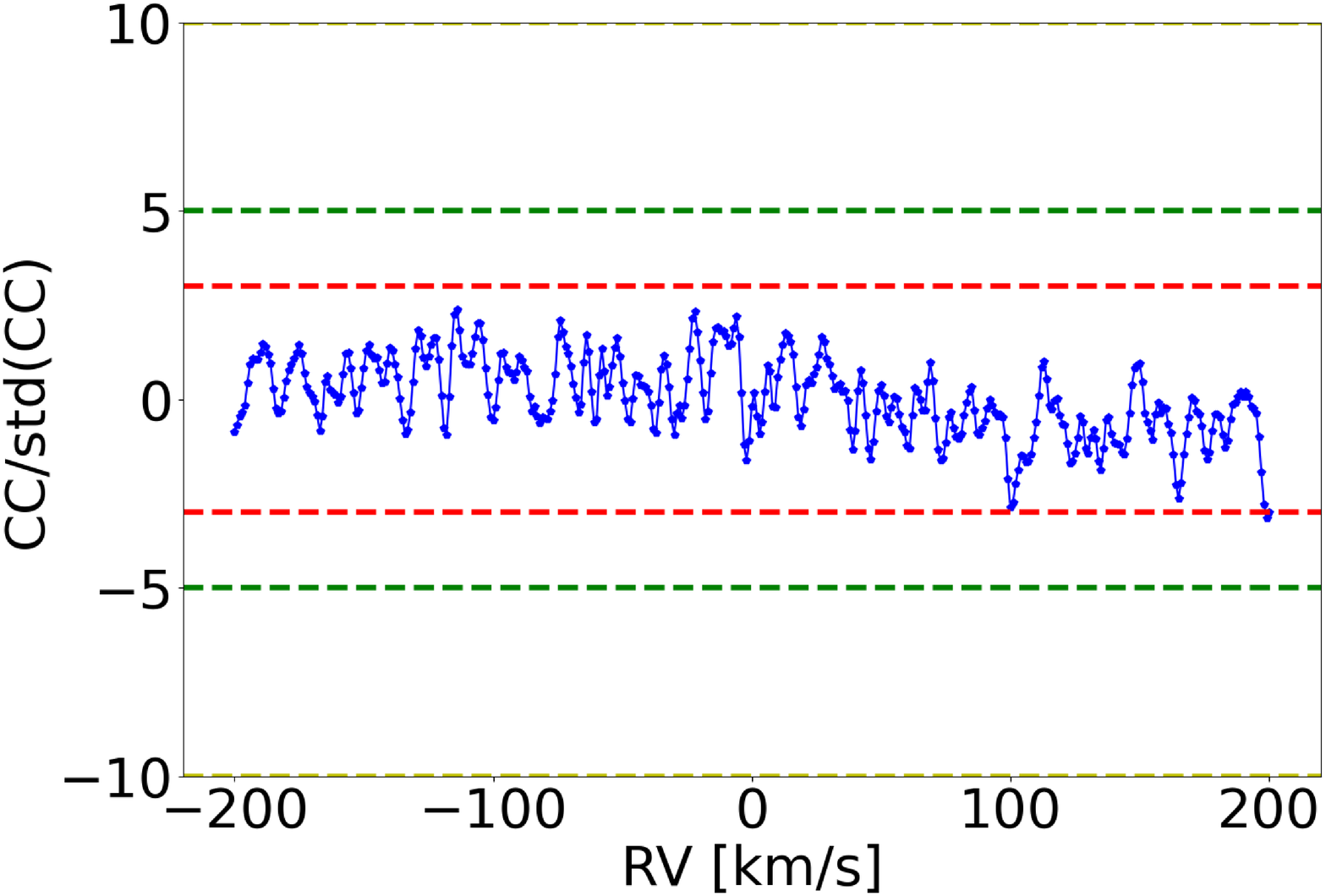}}%
  \qquad
  \subfloat{\includegraphics[scale = 0.12, angle =90 ]{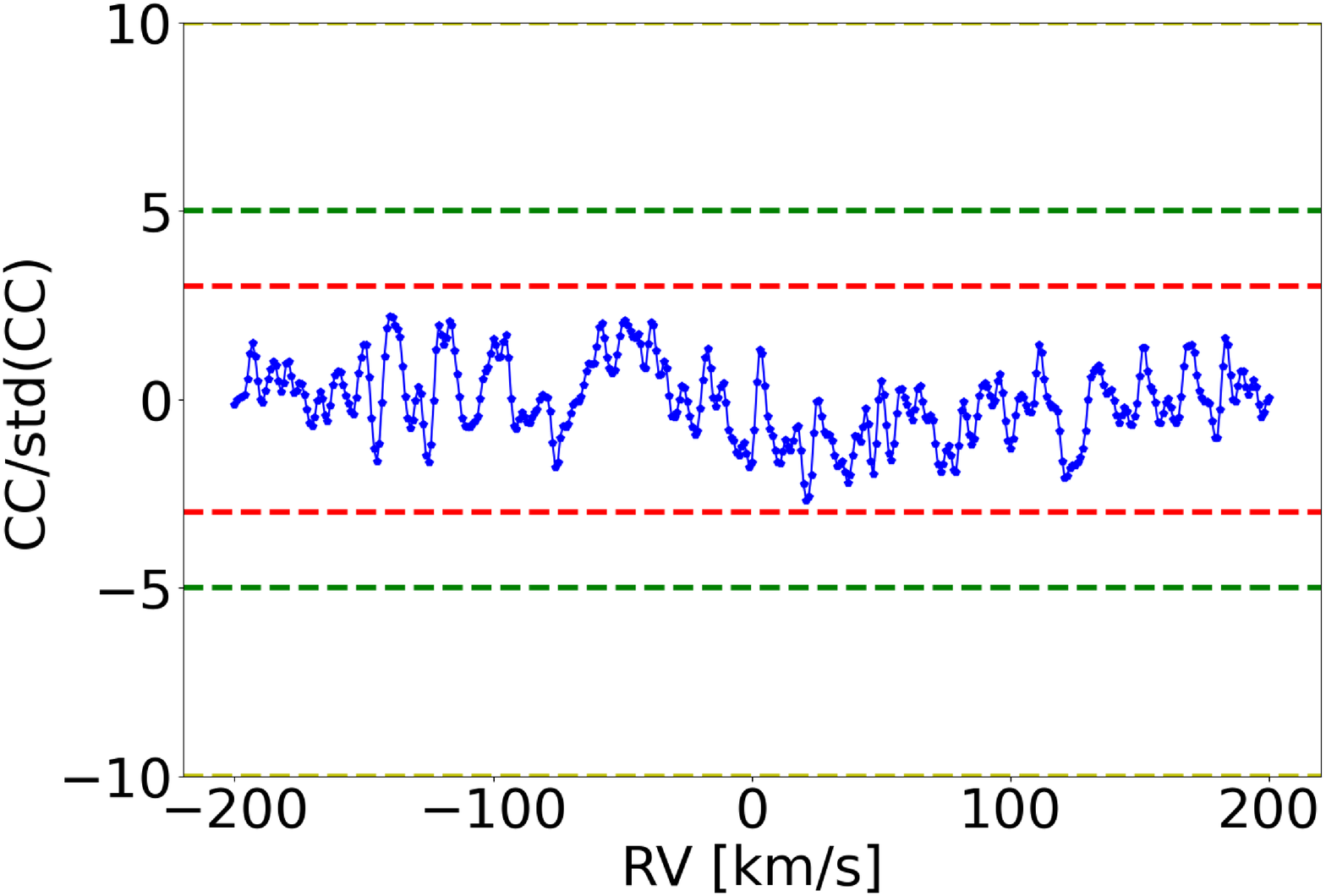}}%
  \qquad
  \subfloat{\includegraphics[scale = 0.12, angle =90 ]{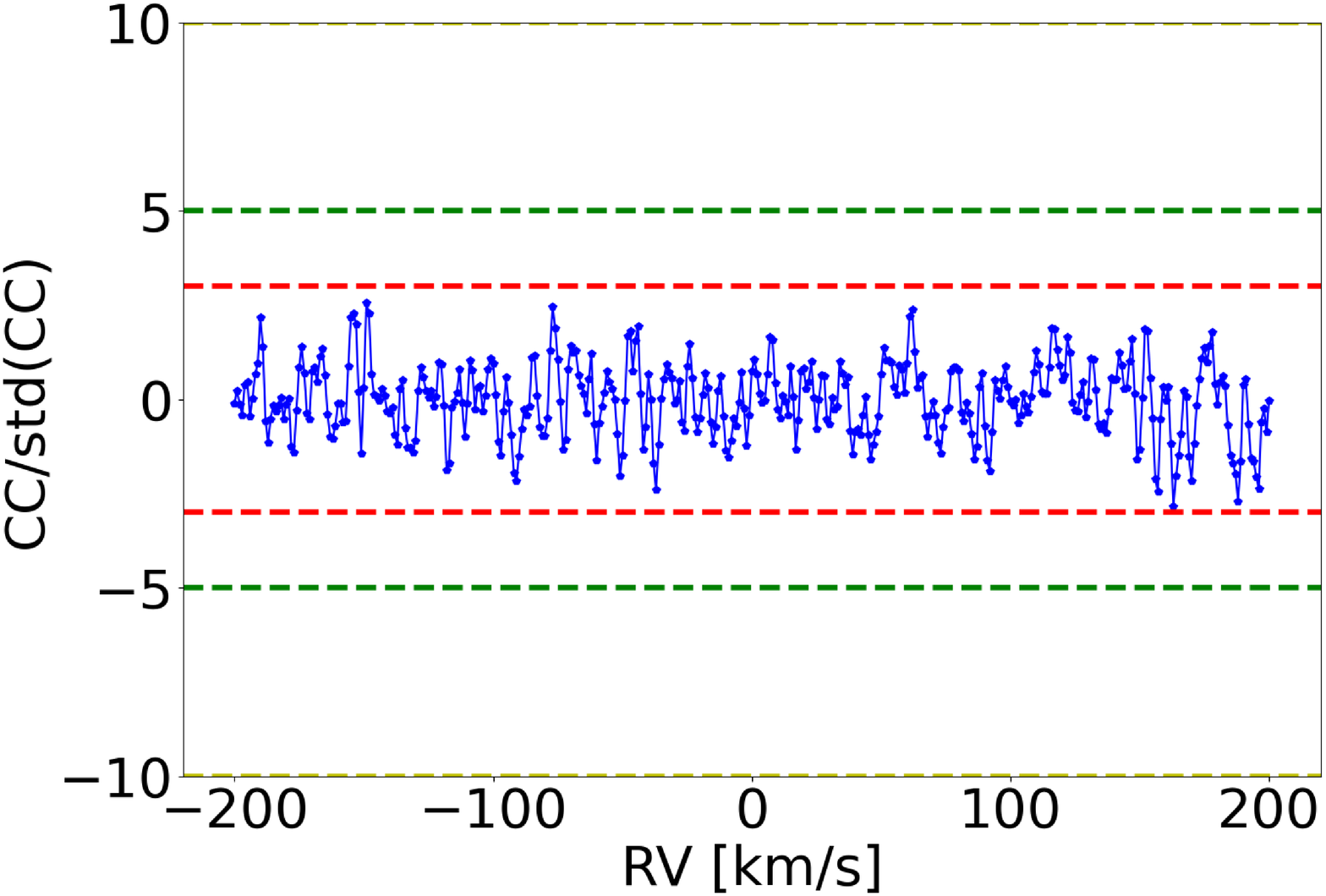}}%
  \qquad
  \subfloat{\includegraphics[scale = 0.12, angle =90 ]{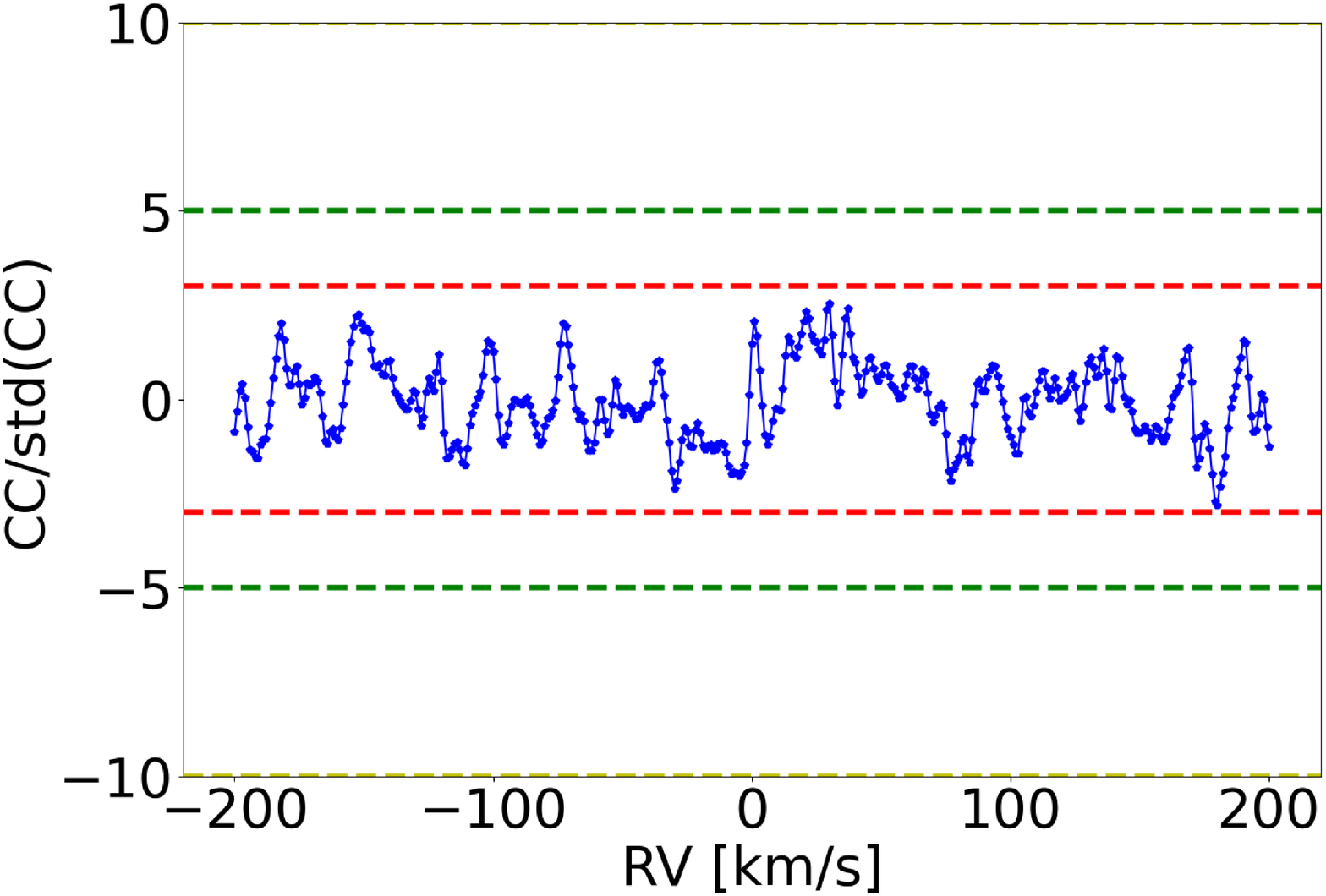}}%
  \qquad
  \subfloat{\includegraphics[scale = 0.12, angle =90 ]{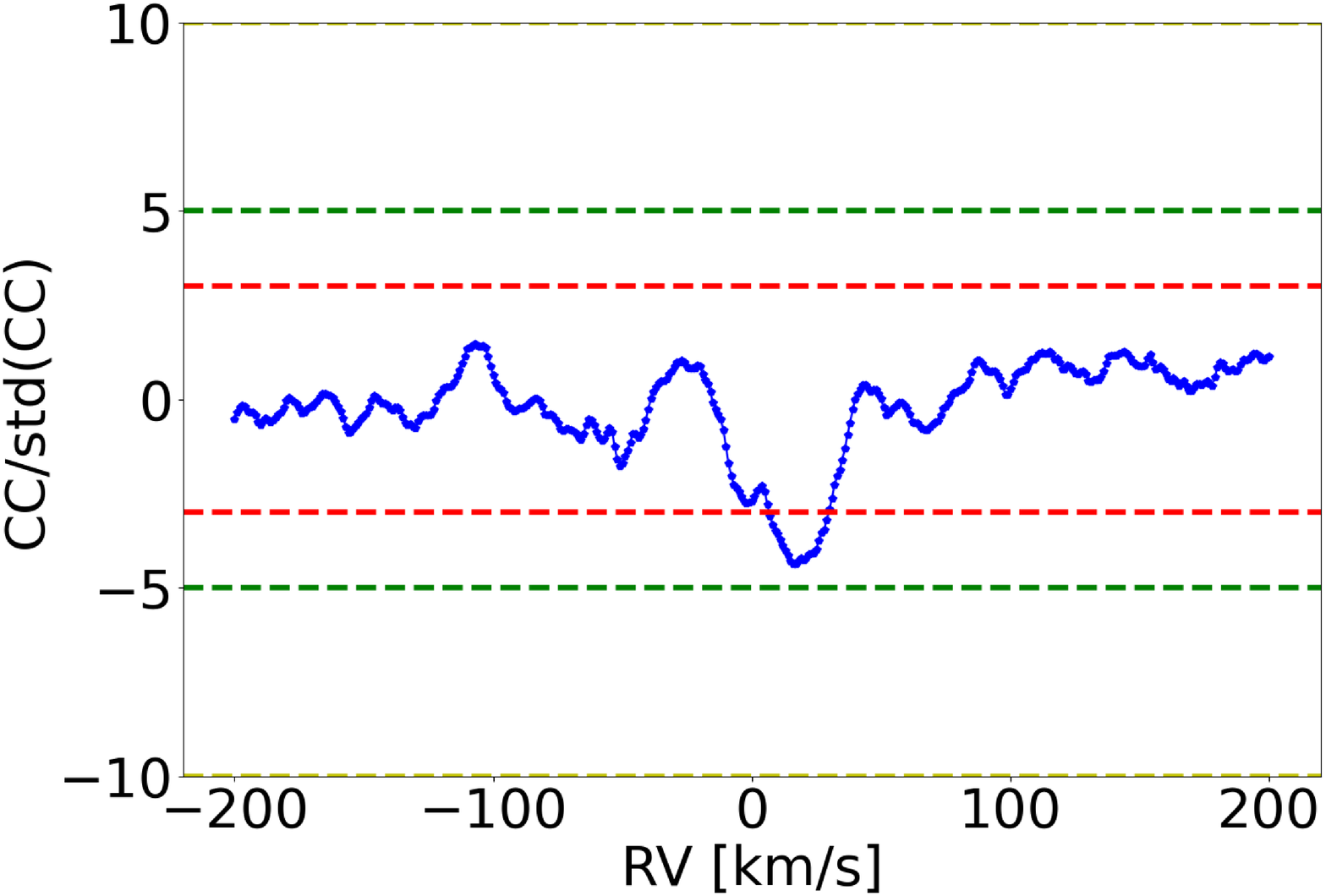}}%
  \\
  \vspace{0.5cm}
  \centering
  \subfloat{\includegraphics[scale = 0.12, angle =90 ]{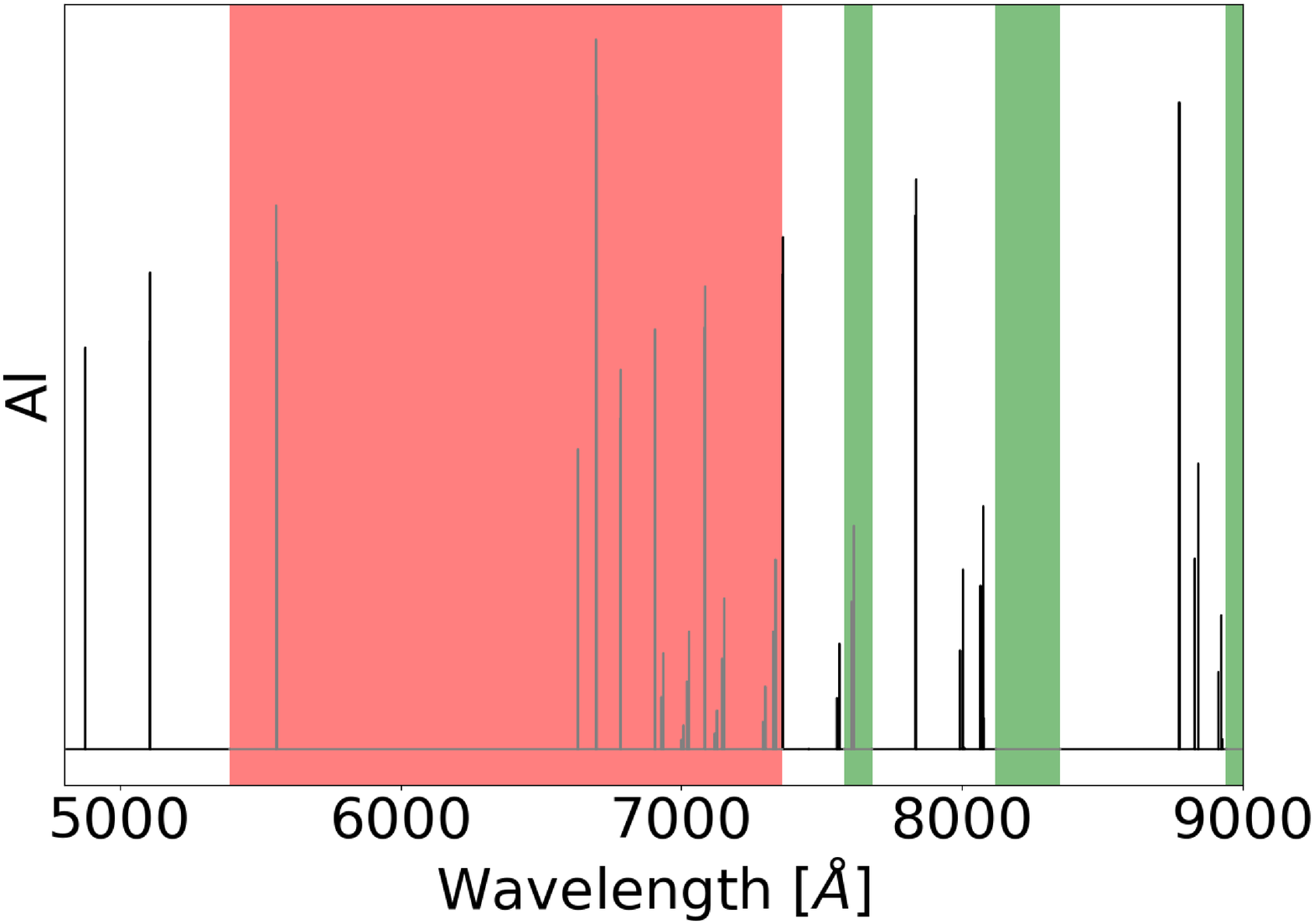}}%
  \qquad
  \subfloat{\includegraphics[scale = 0.12, angle =90 ]{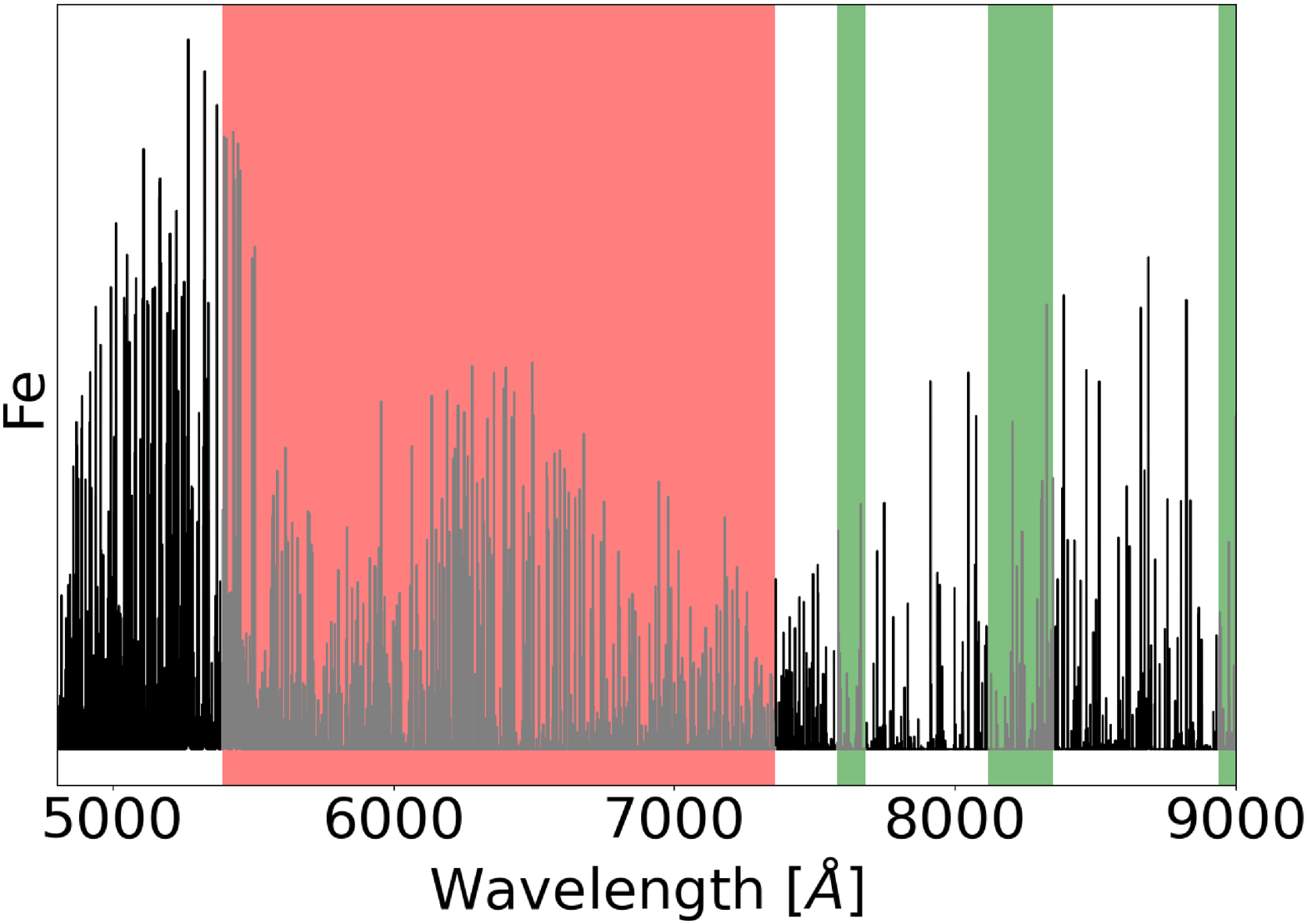}}%
  \qquad
  \subfloat{\includegraphics[scale = 0.12, angle =90 ]{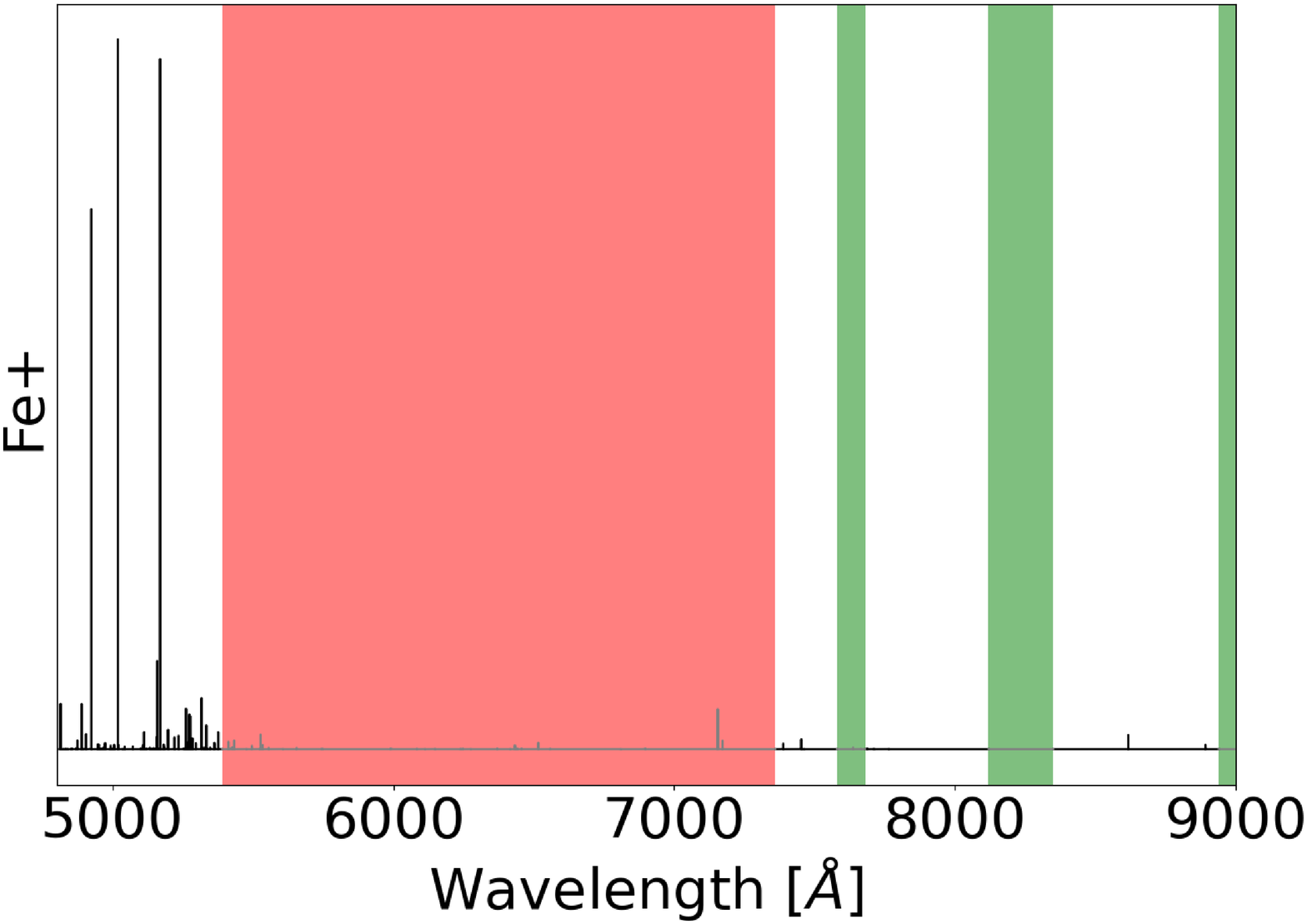}}%
  \qquad
  \subfloat{\includegraphics[scale = 0.12, angle =90 ]{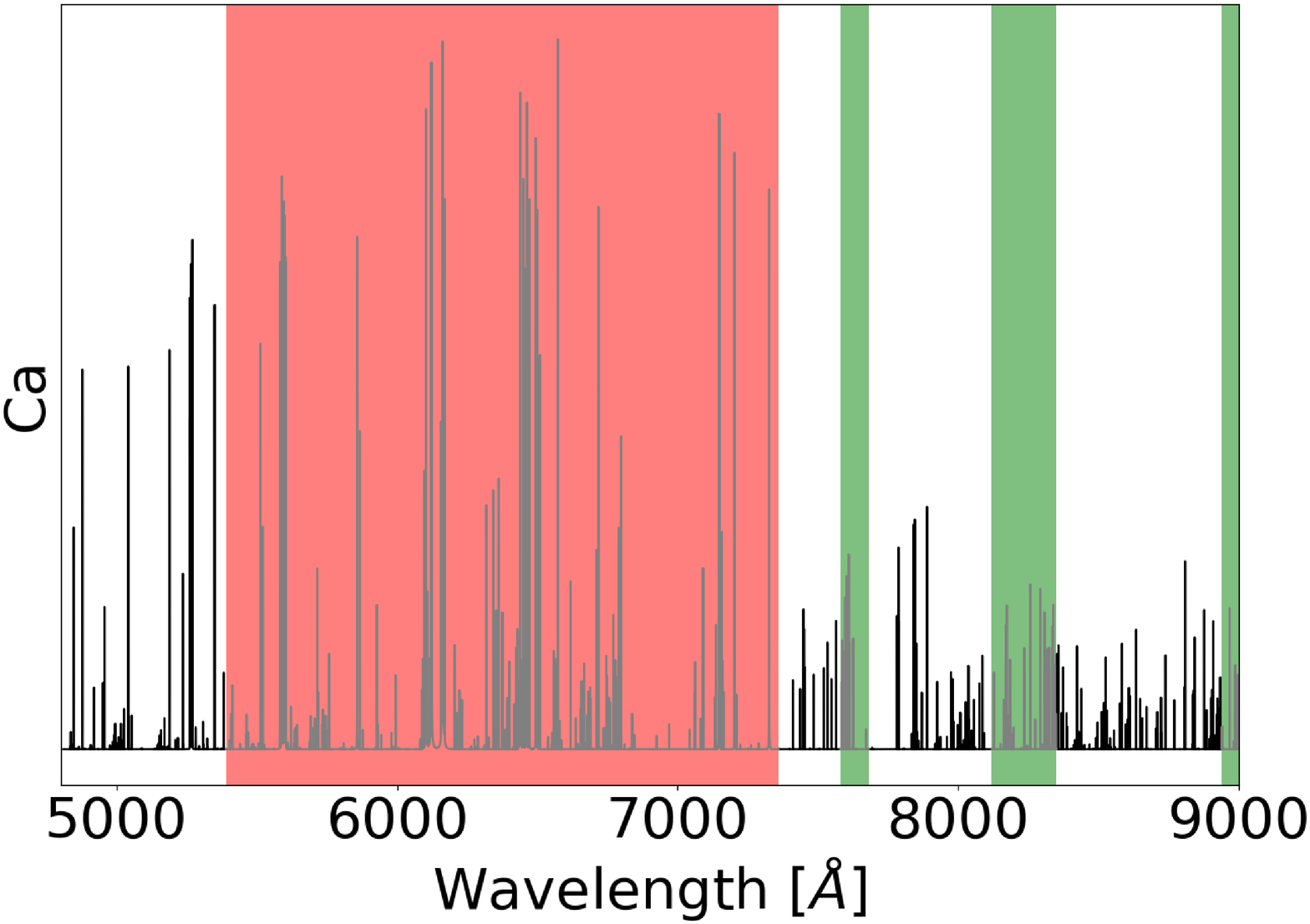}}%
  \qquad
  \subfloat{\includegraphics[scale = 0.12, angle =90 ]{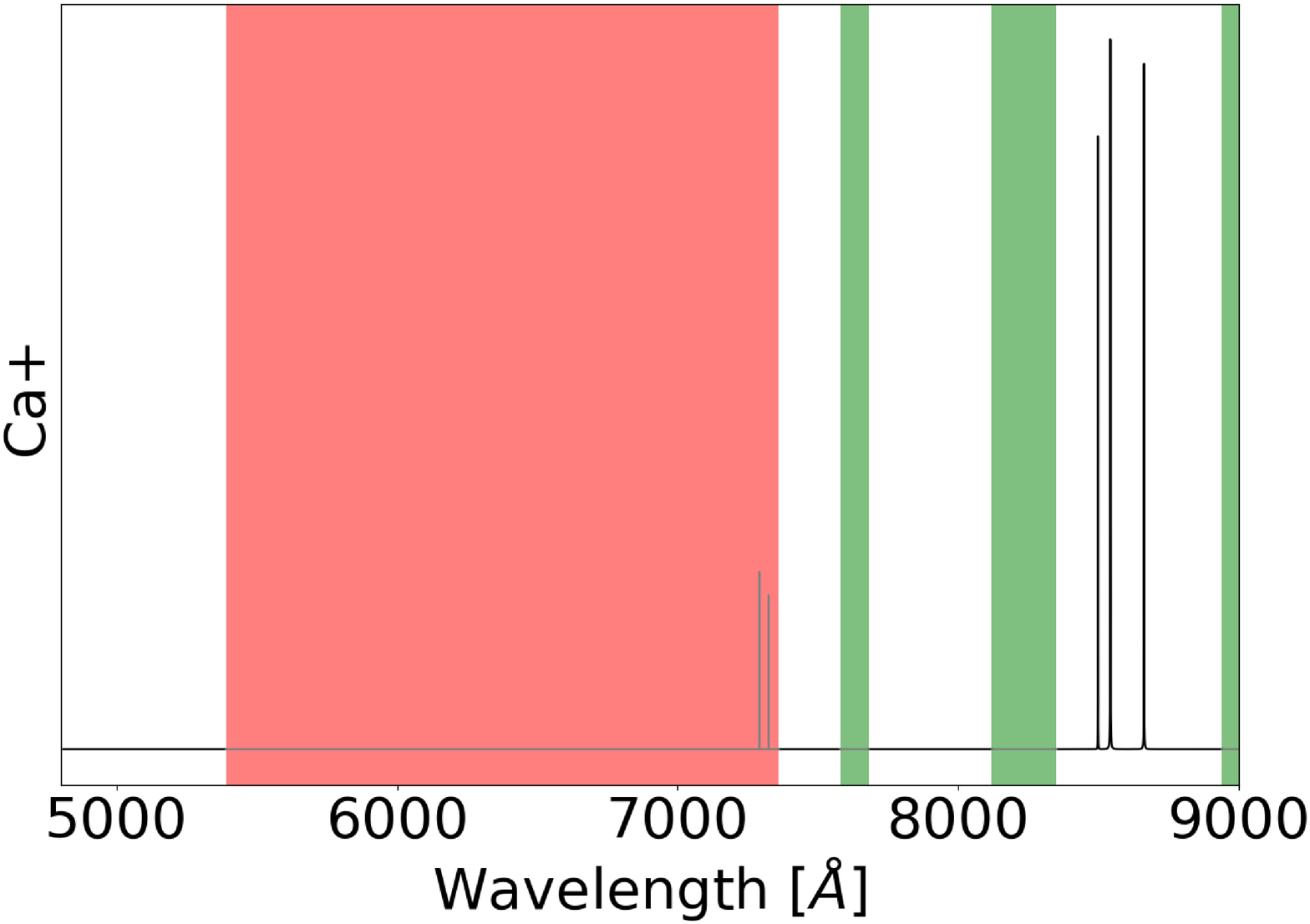}}%
  \caption{Cross-correlation analysis for individual species. In landscape format, from left to right, the first panel shows the template spectrum for the scenario with T = 2500K (the shaded regions excluded from the analysis), the second panel shows the cross-correlation signal (CC) divided by its standard deviation excluding $\pm$20 km/s around the center (dashed lines mark the 3$\sigma$ (red), 5$\sigma$ (green) and 10$\sigma$ (yellow) thresholds), the third panel shows the colour coded CC for the transmission spectra at different orbital phases in the velocity range (the color bar shows the significance and the dashed black line shows the expected planetary atmospheric absorption trace), the fourth panel shows the colour coded K\textsubscript{P} - RV map (the color bar shows the significance and the yellow dashed lines mark the expected K\textsubscript{P} value) and the fifth panel shows the detection thresholds for different R\textsubscript{ext} values (in steps of 0.2 $\times$ R\textsubscript{P}) applied for for the injection recovery up to 3 R\textsubscript{P} (dashed lines mark the 3$\sigma$ (red), 5$\sigma$ (green) and 10$\sigma$ (yellow) thresholds and the blue dot shows the value of the unscaled template).}%
  \label{fig:A1}
\end{figure*}%

\begin{figure*}%
  \centering
  \subfloat{\includegraphics[scale = 0.12, angle =90 ]{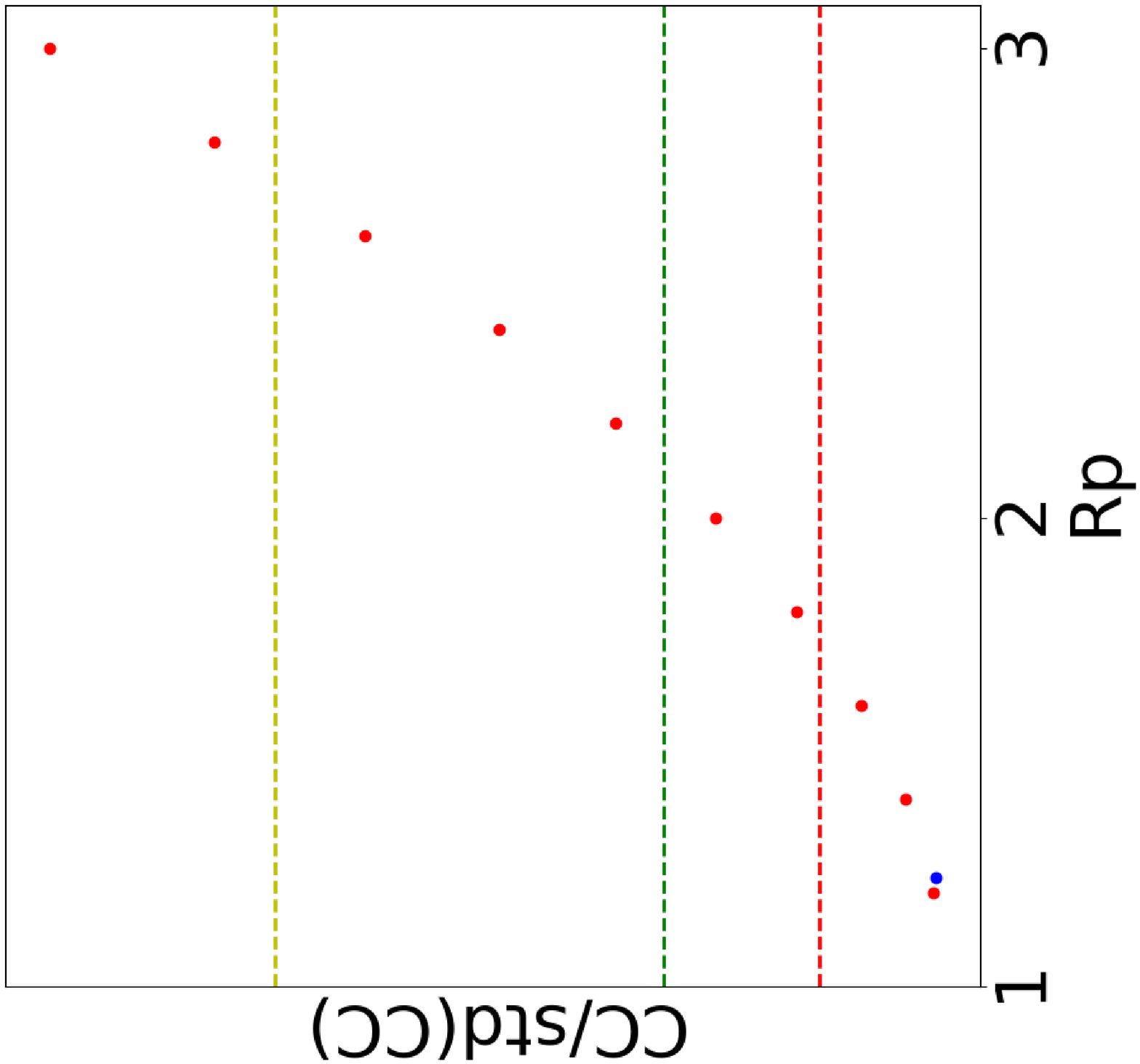}}%
  \qquad
  \subfloat{\includegraphics[scale = 0.12, angle =90 ]{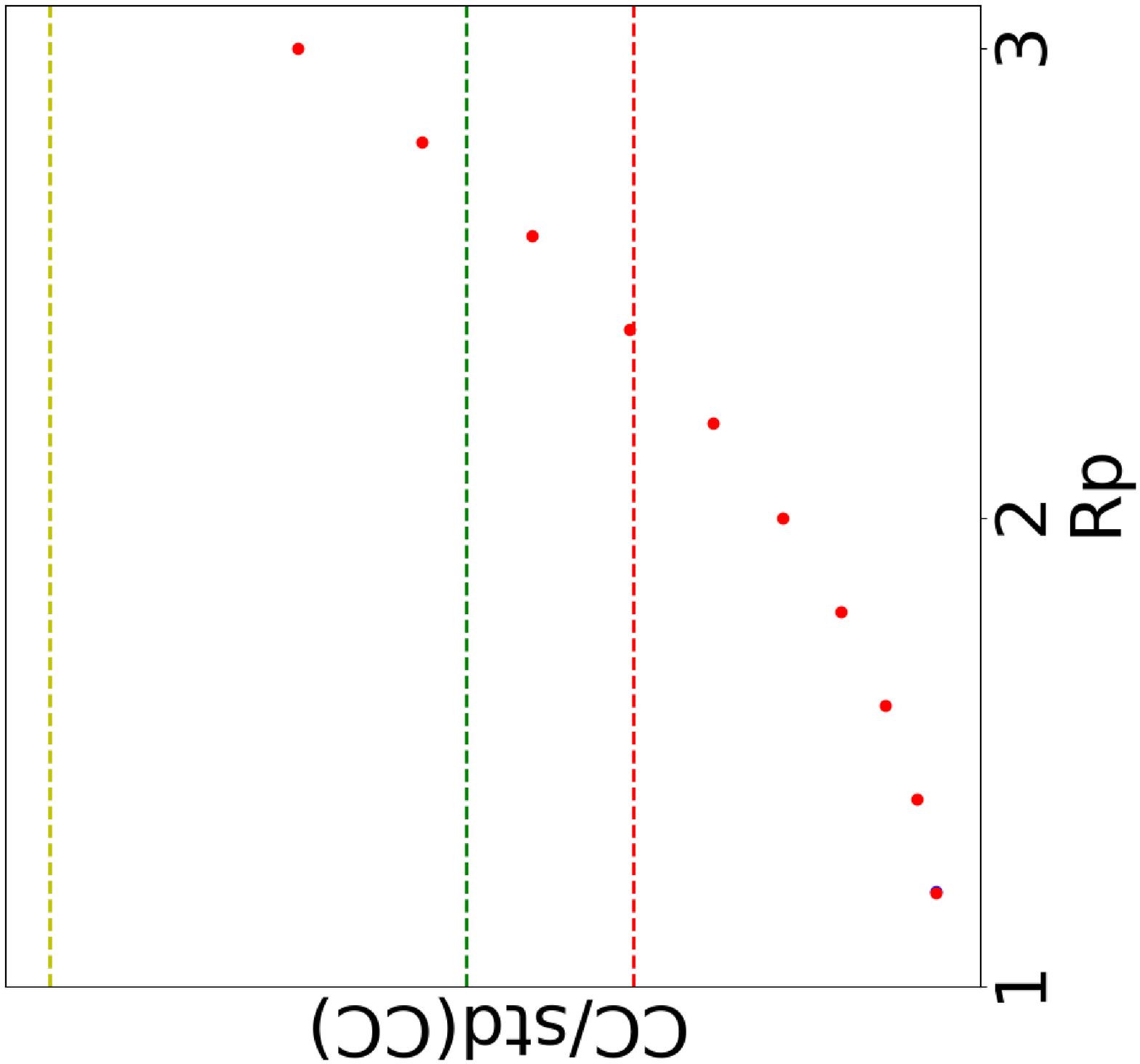}}%
  \qquad
  \subfloat{\includegraphics[scale = 0.12, angle =90 ]{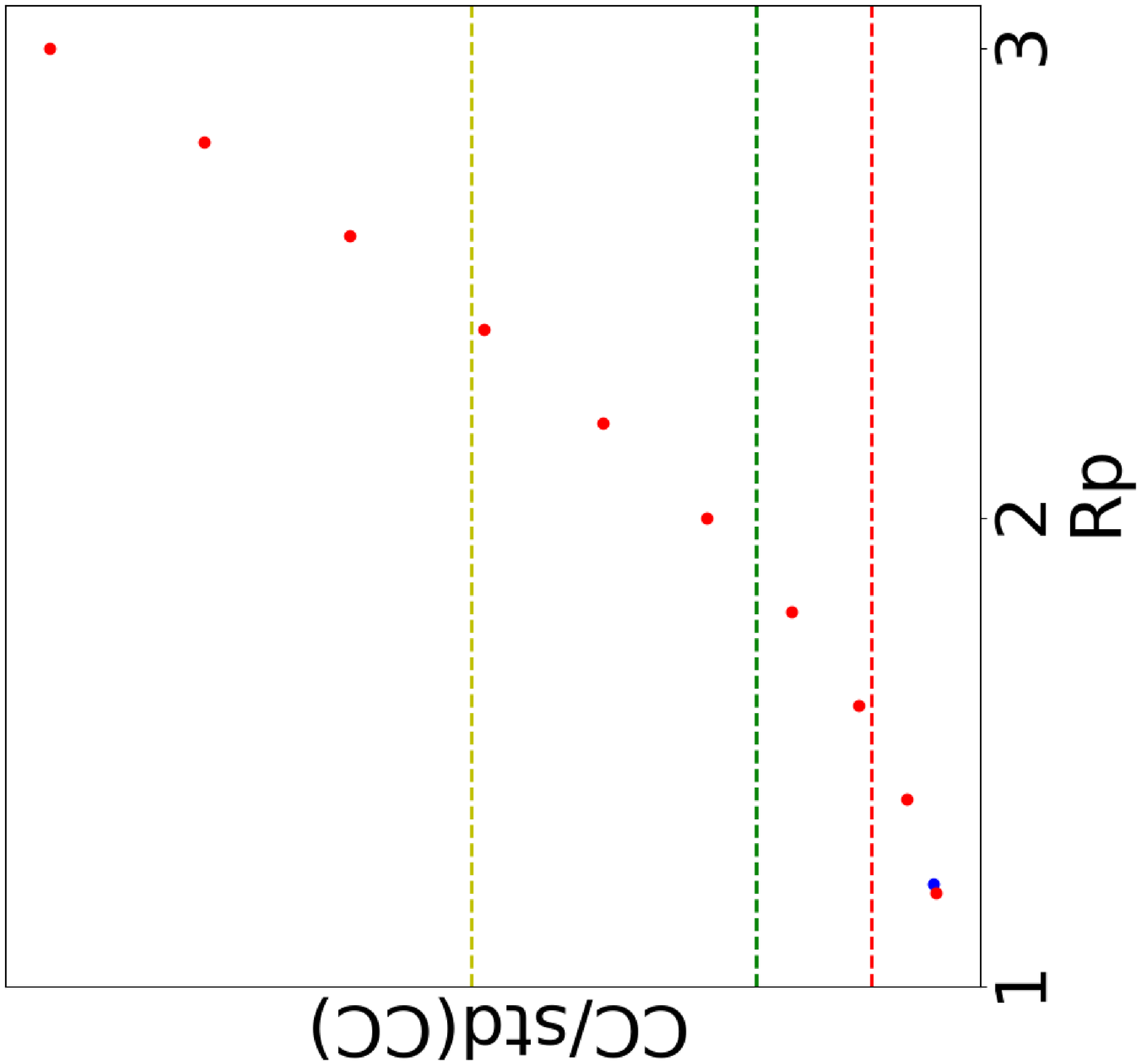}}%
  \qquad
  \subfloat{\includegraphics[scale = 0.12, angle =90 ]{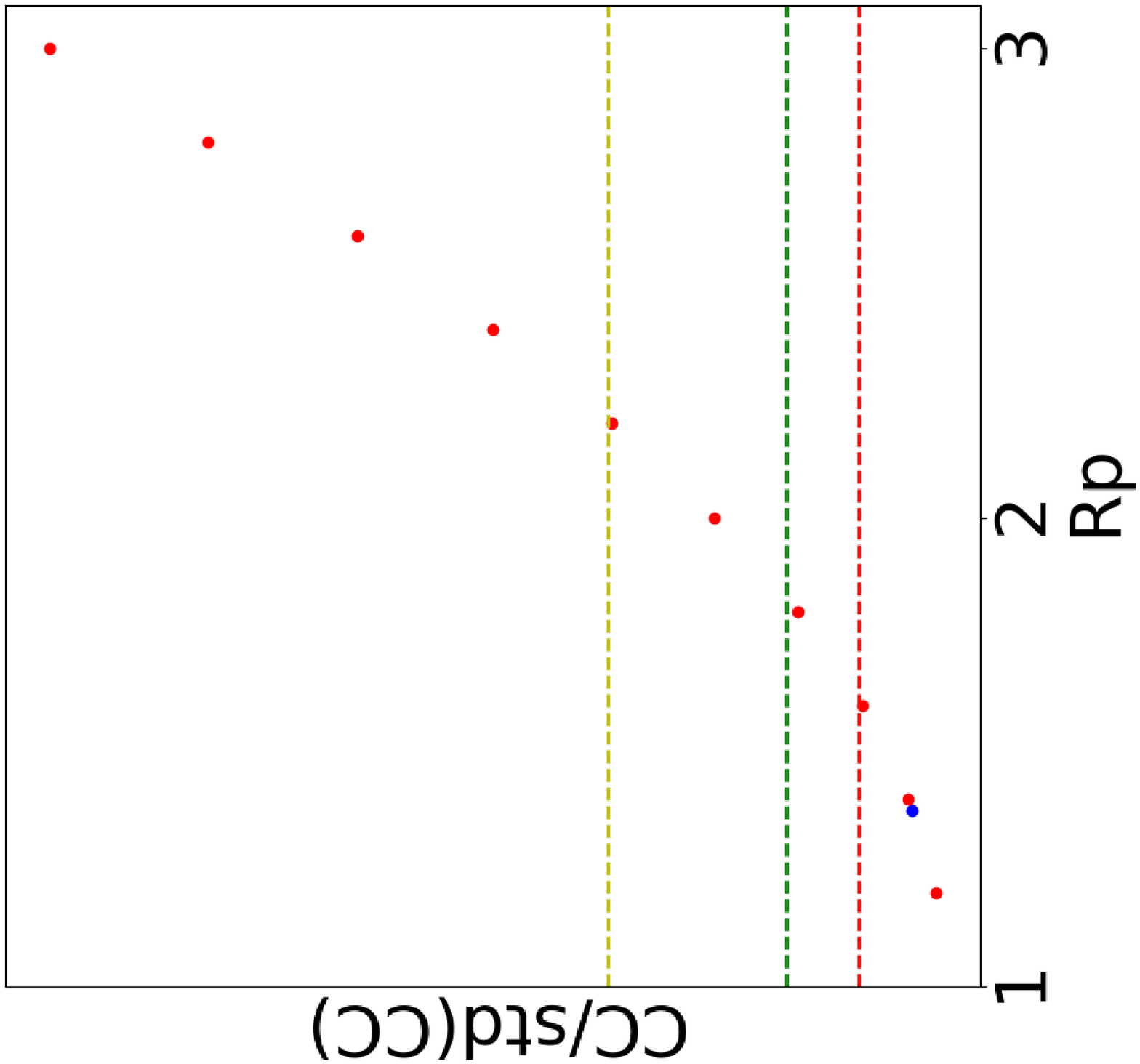}}%
  \qquad
  \subfloat{\includegraphics[scale = 0.12, angle =90 ]{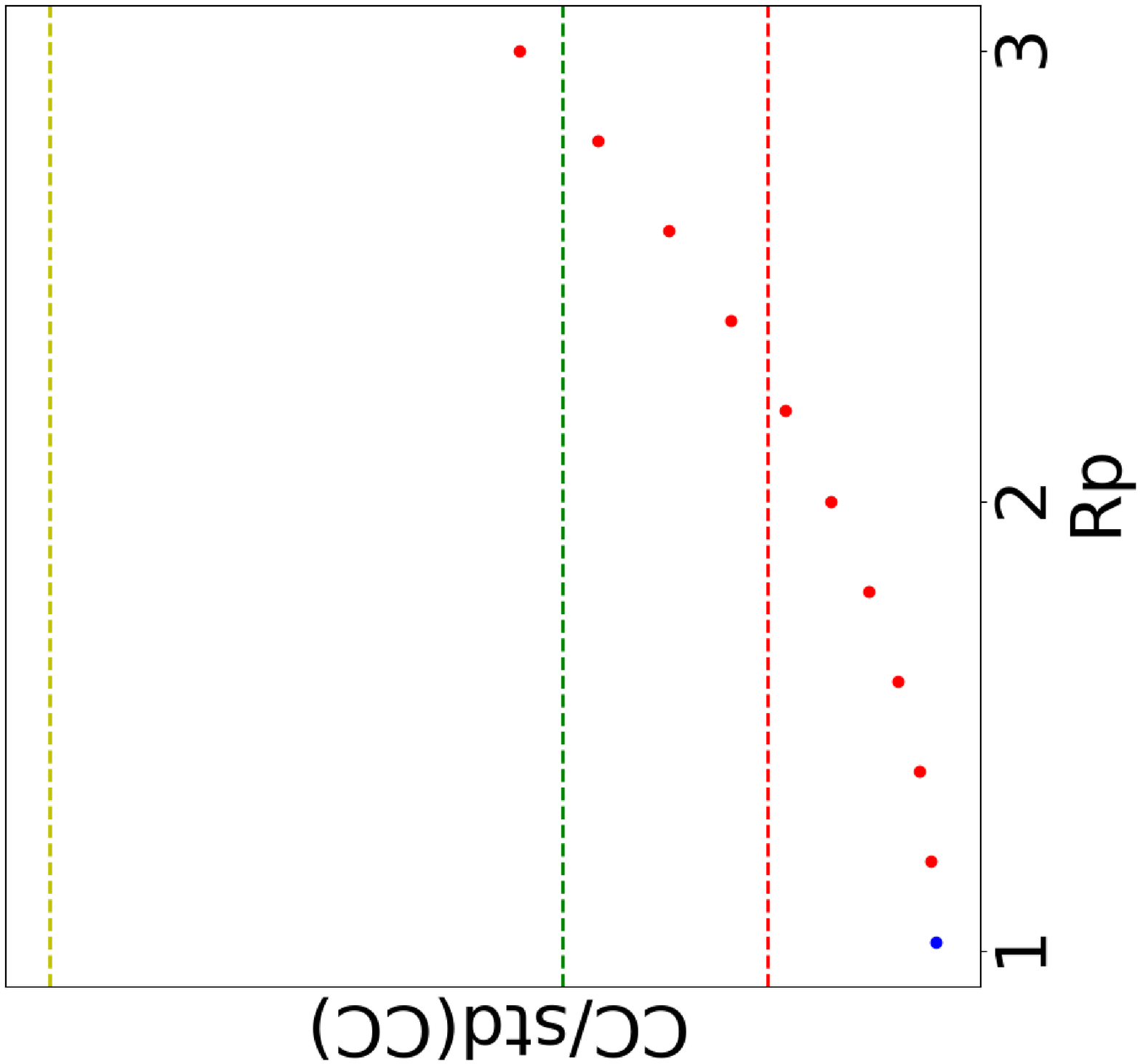}}%
  \\
  \vspace{0.5cm}
  \centering
  \subfloat{\includegraphics[scale = 0.12, angle =90 ]{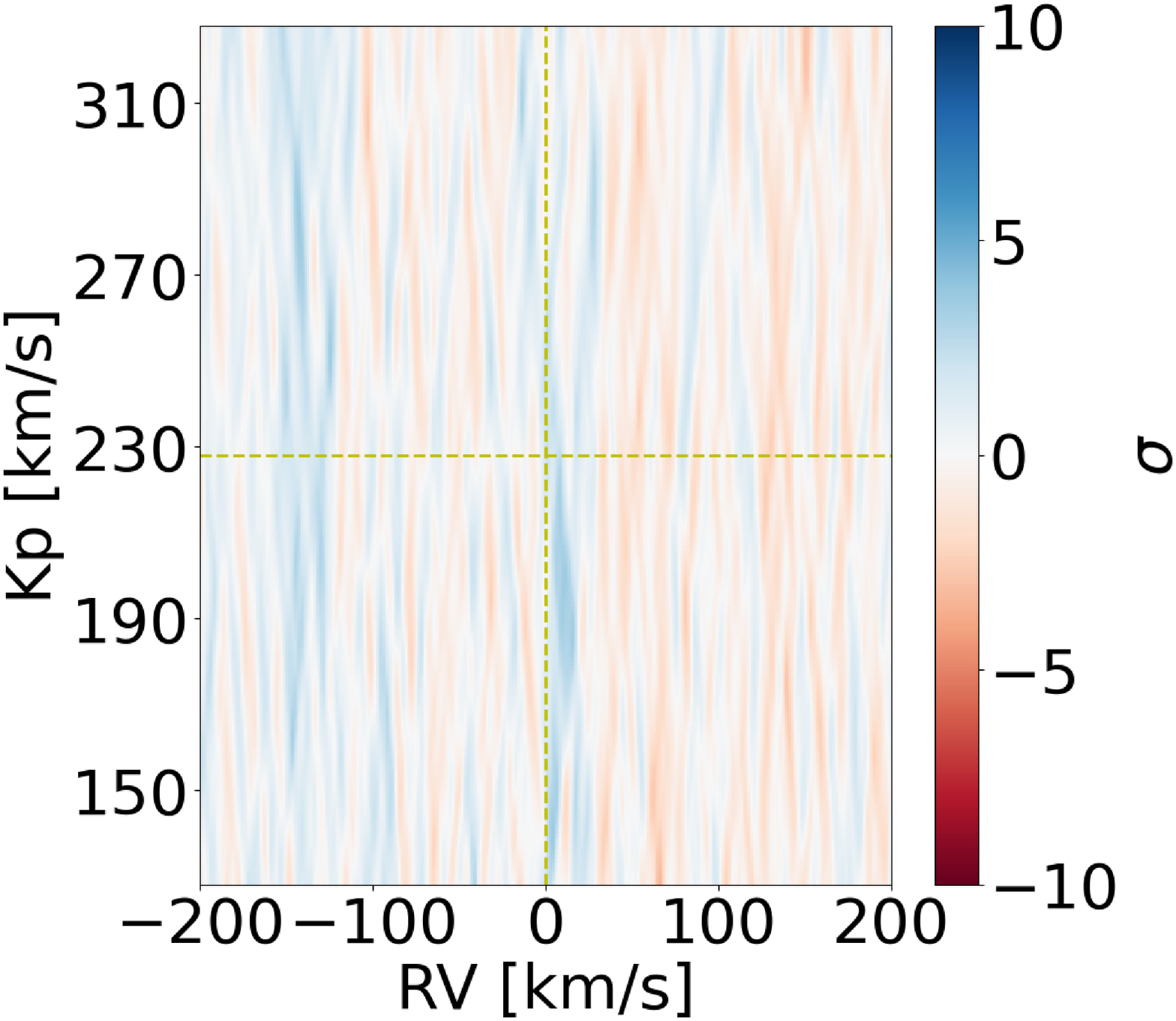}}%
  \qquad
  \subfloat{\includegraphics[scale = 0.12, angle =90 ]{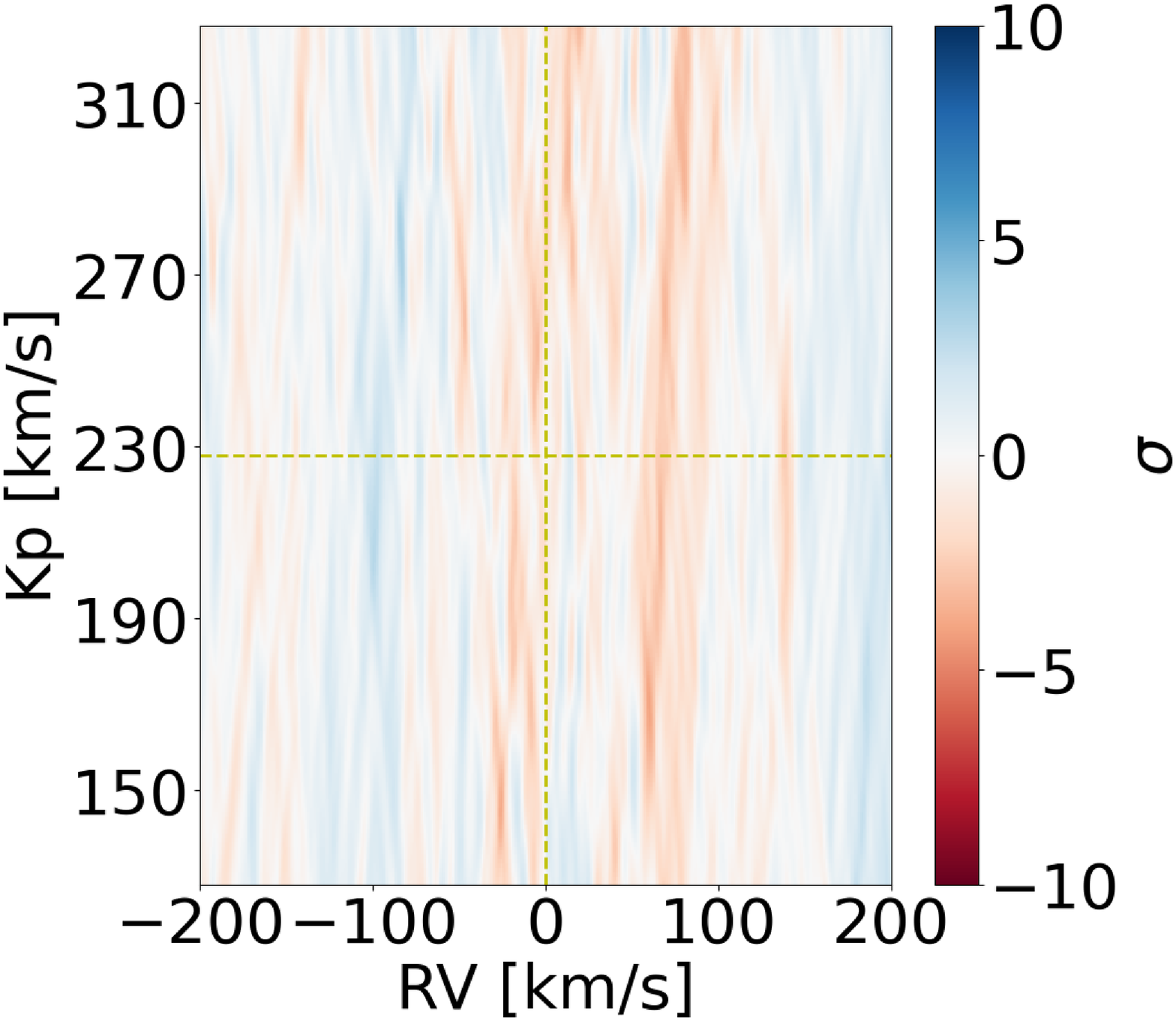}}%
  \qquad
  \subfloat{\includegraphics[scale = 0.12, angle =90 ]{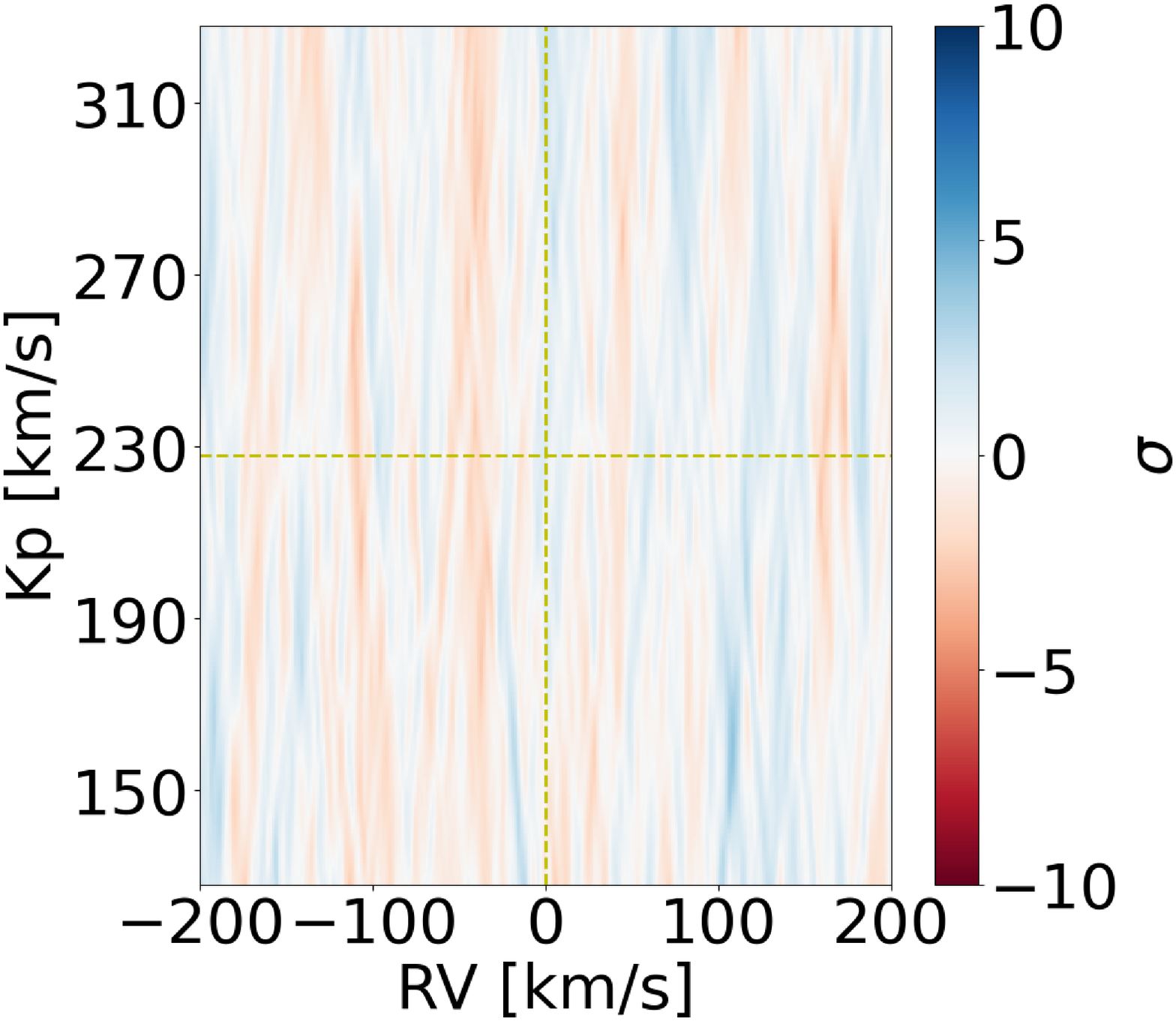}}%
  \qquad
  \subfloat{\includegraphics[scale = 0.12, angle =90 ]{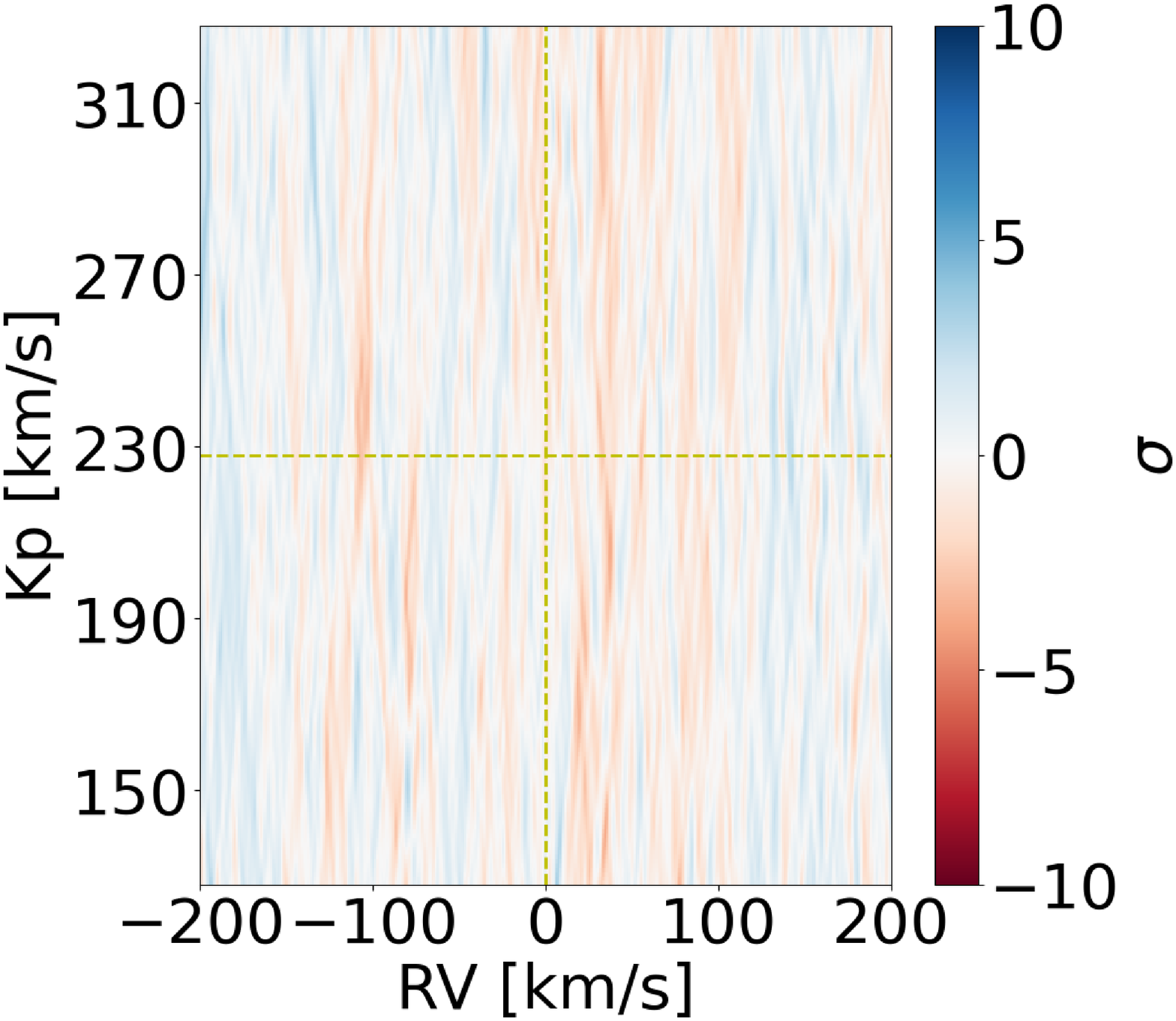}}%
  \qquad
  \subfloat{\includegraphics[scale = 0.12, angle =90 ]{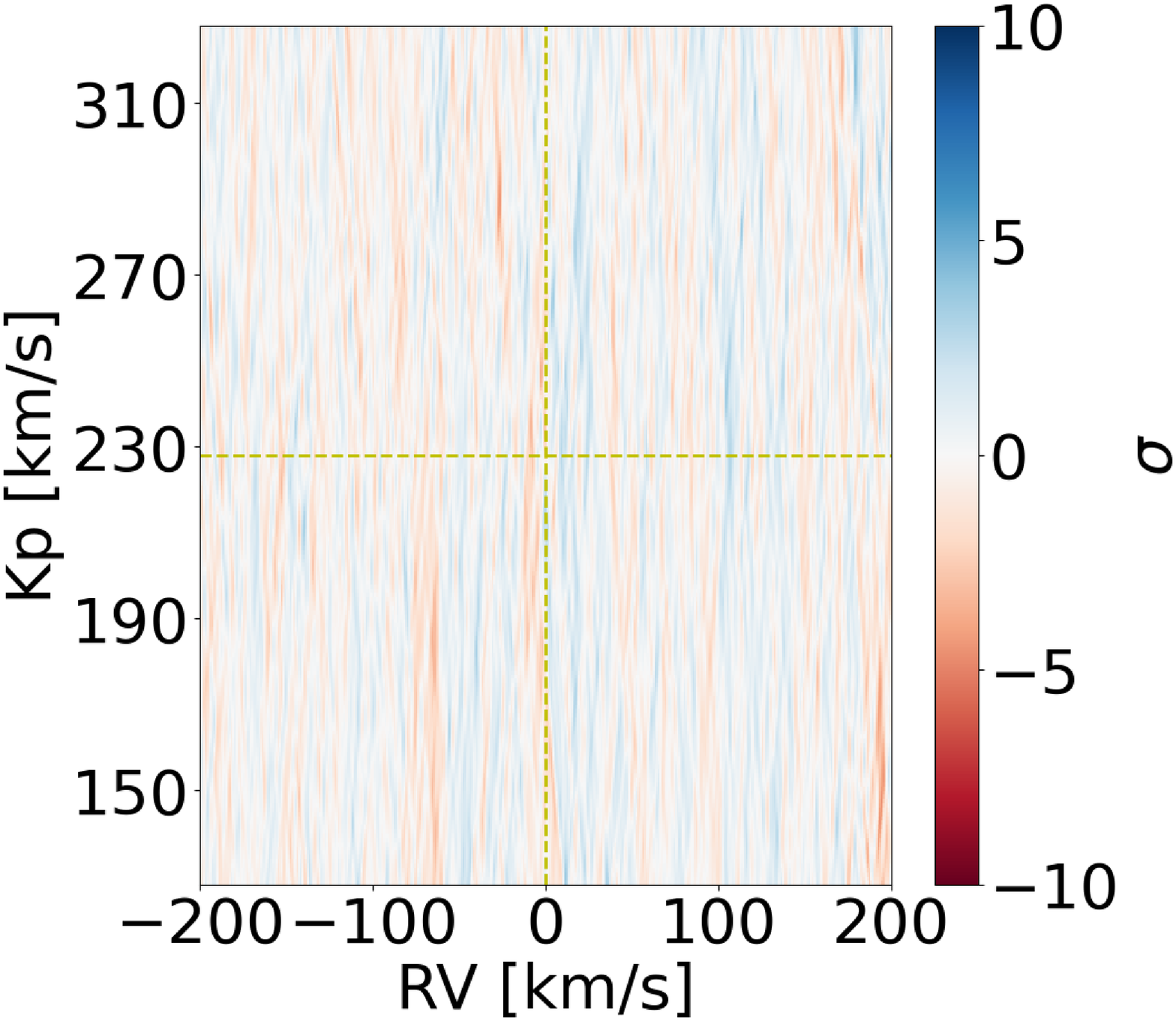}}%
  \\
  \vspace{0.5cm}
  \centering
  \subfloat{\includegraphics[scale = 0.12, angle =90 ]{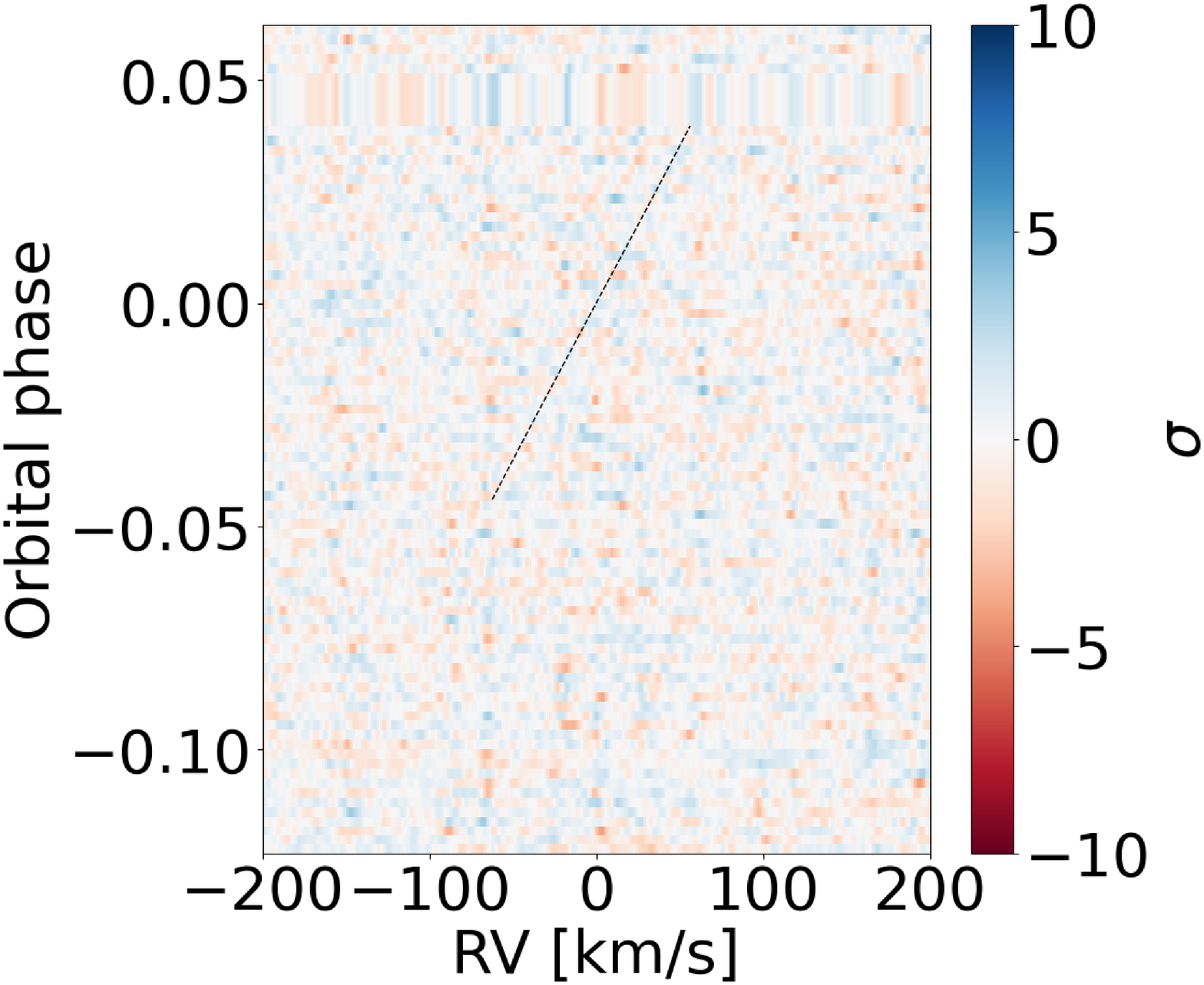}}%
  \qquad
  \subfloat{\includegraphics[scale = 0.12, angle =90 ]{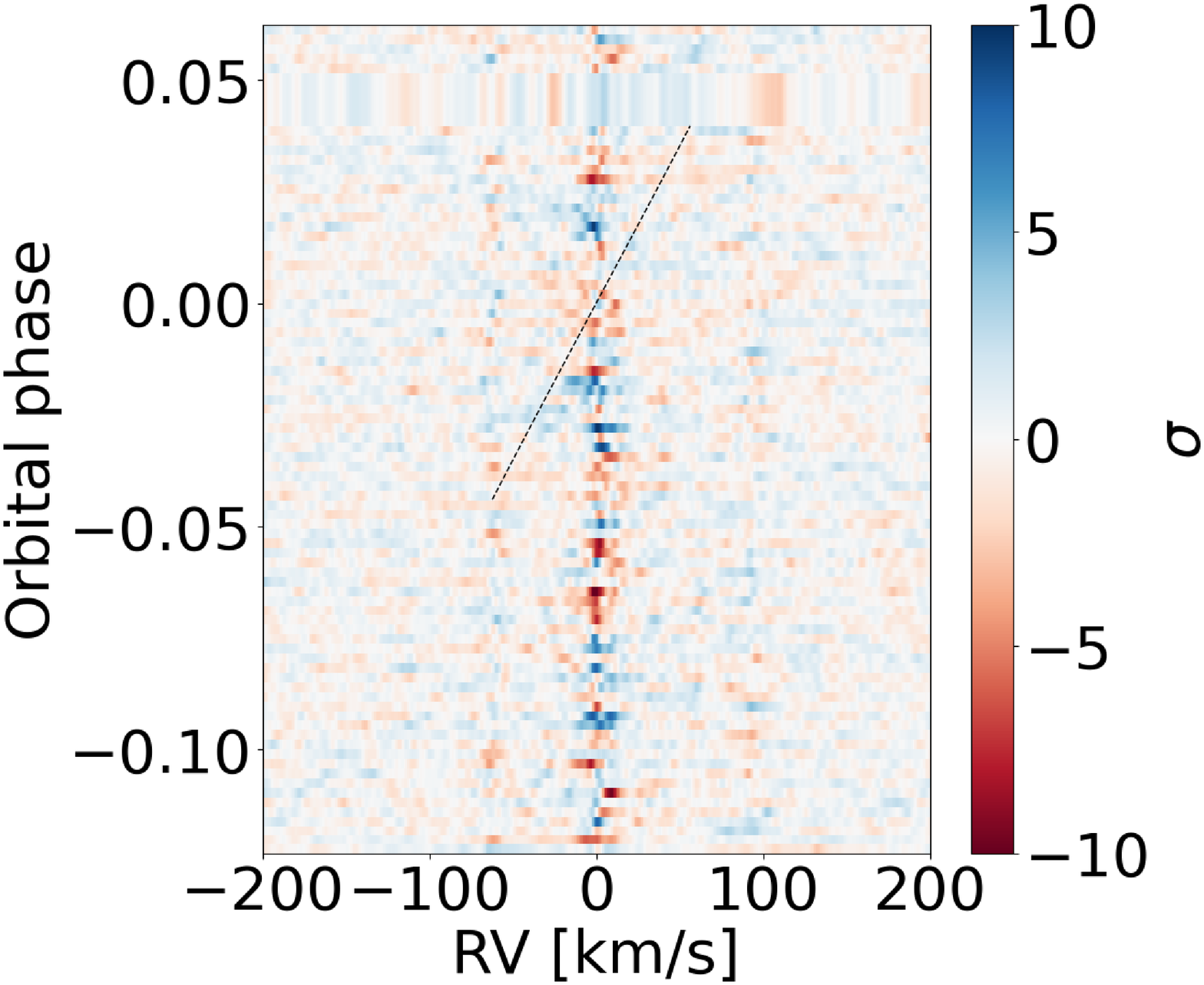}}%
  \qquad
  \subfloat{\includegraphics[scale = 0.12, angle =90 ]{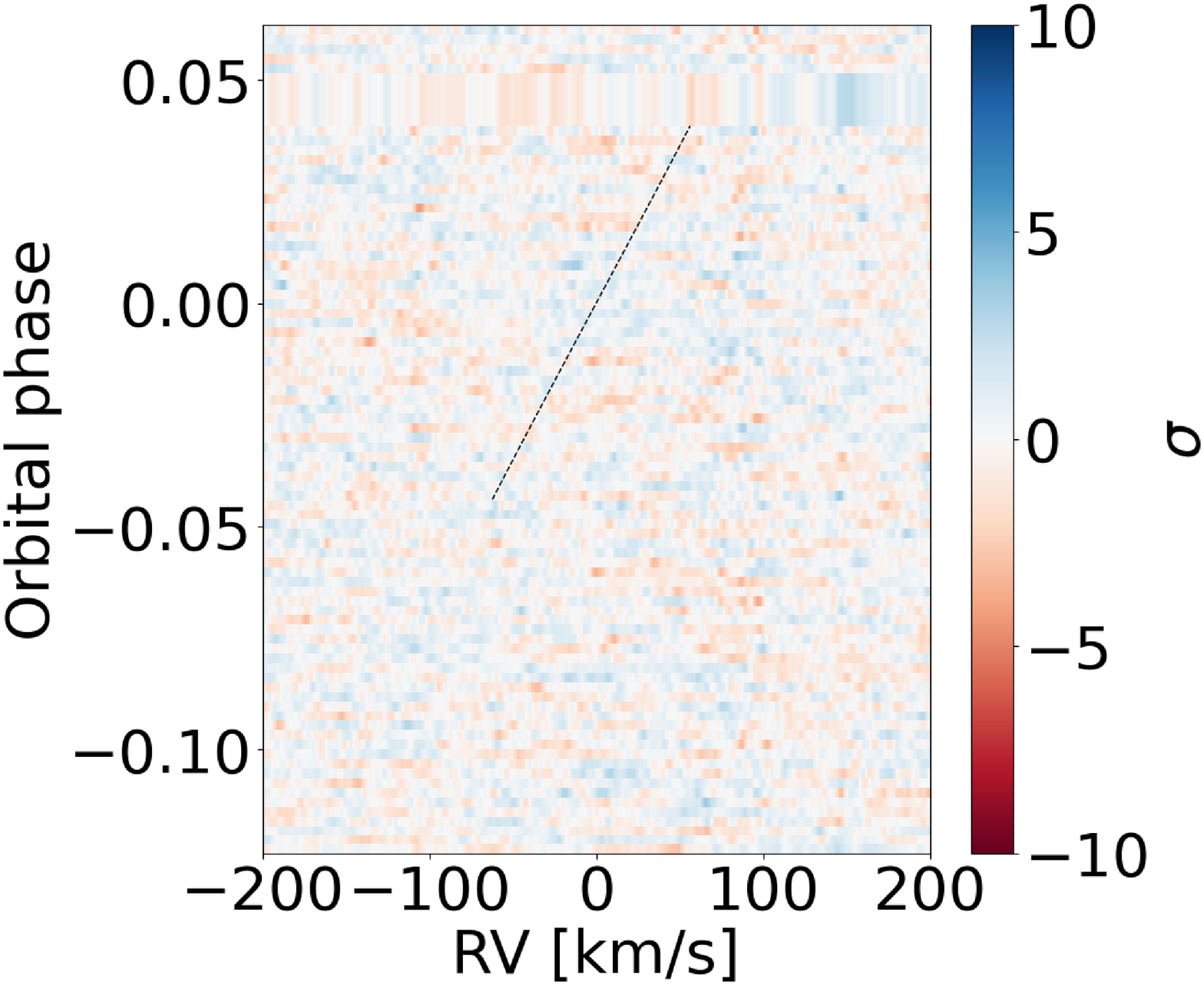}}%
  \qquad
  \subfloat{\includegraphics[scale = 0.12, angle =90 ]{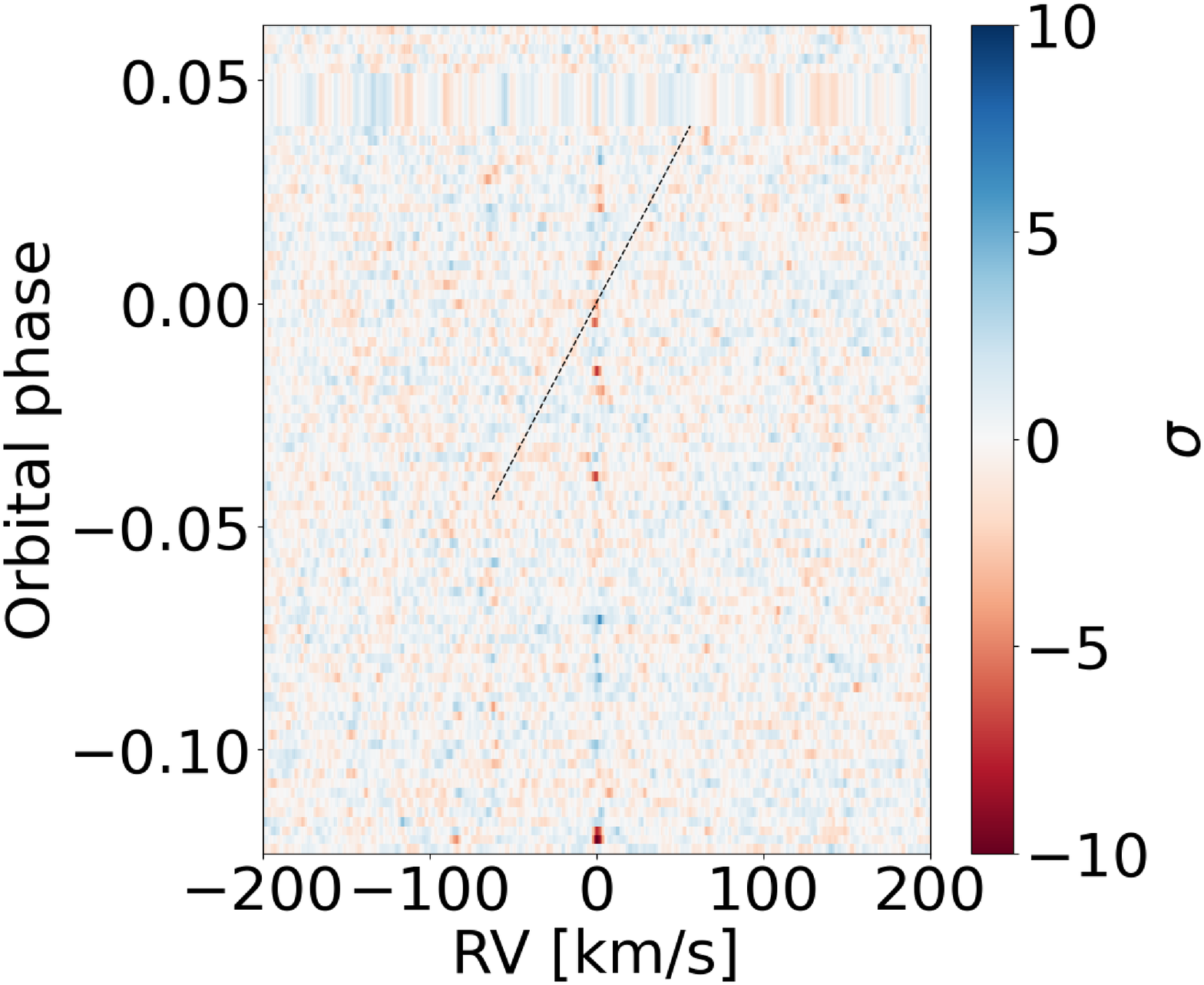}}%
  \qquad
  \subfloat{\includegraphics[scale = 0.12, angle =90 ]{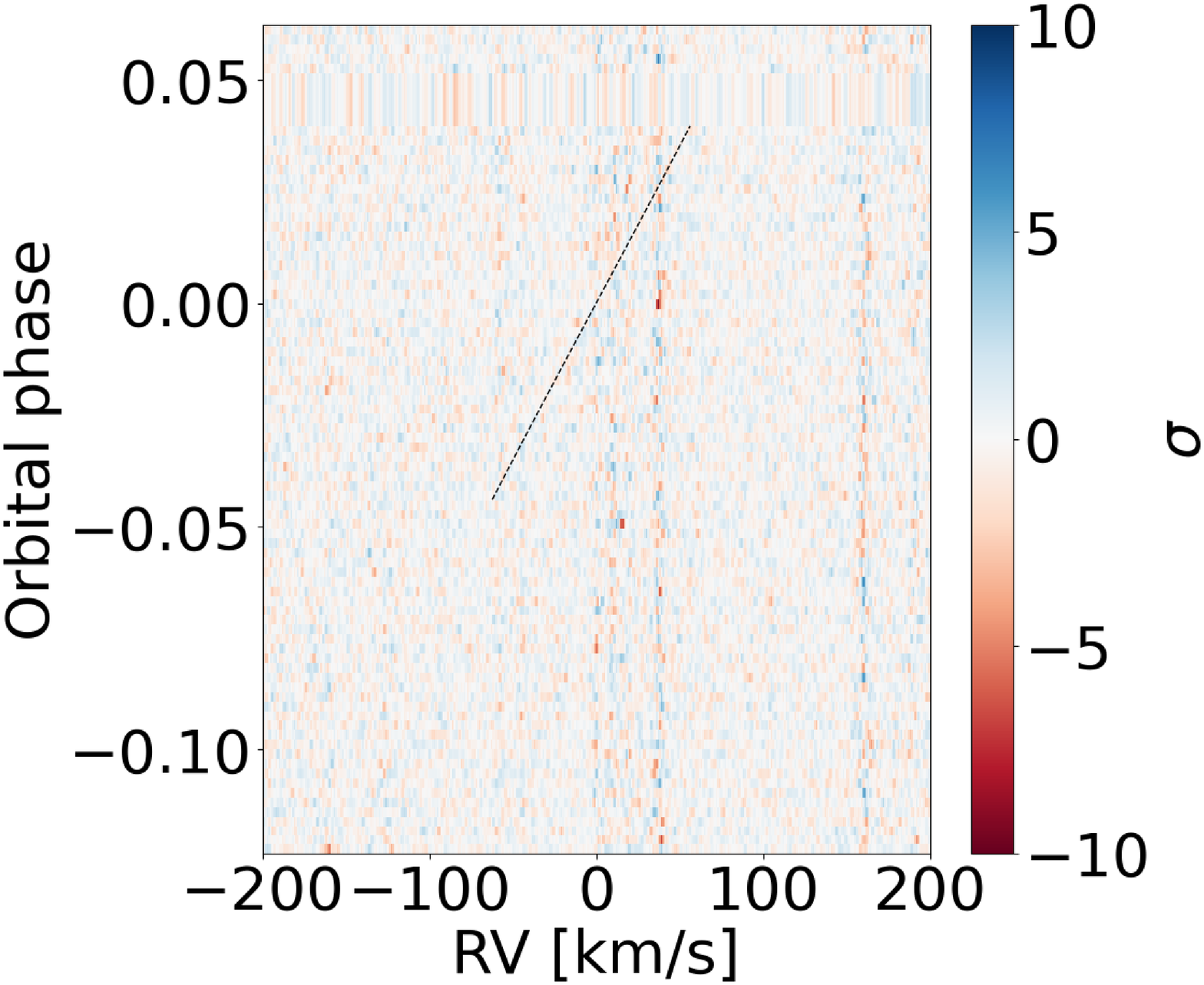}}%
  \\
  \vspace{0.5cm}
  \centering
  \subfloat{\includegraphics[scale = 0.12, angle =90 ]{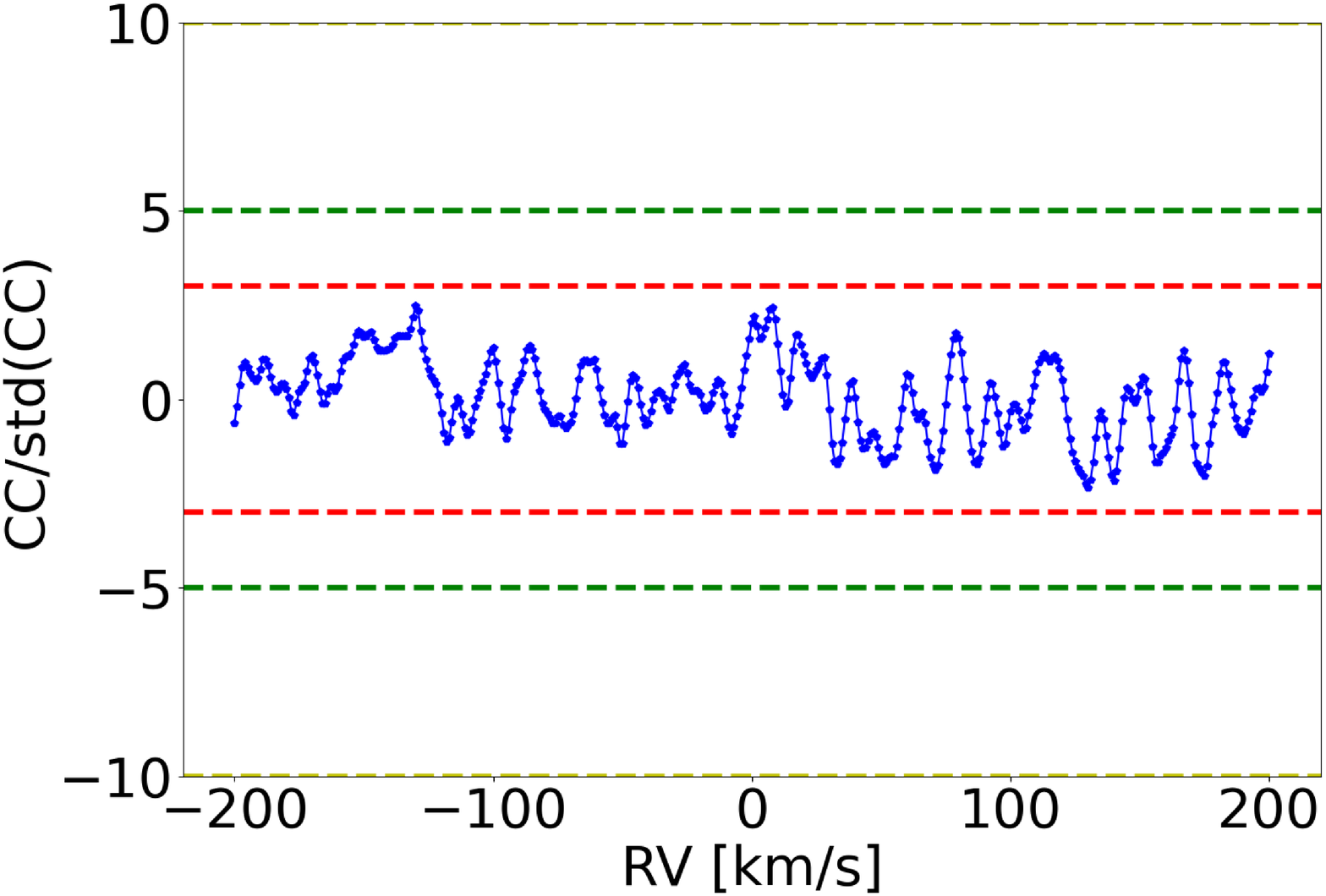}}%
  \qquad
  \subfloat{\includegraphics[scale = 0.12, angle =90 ]{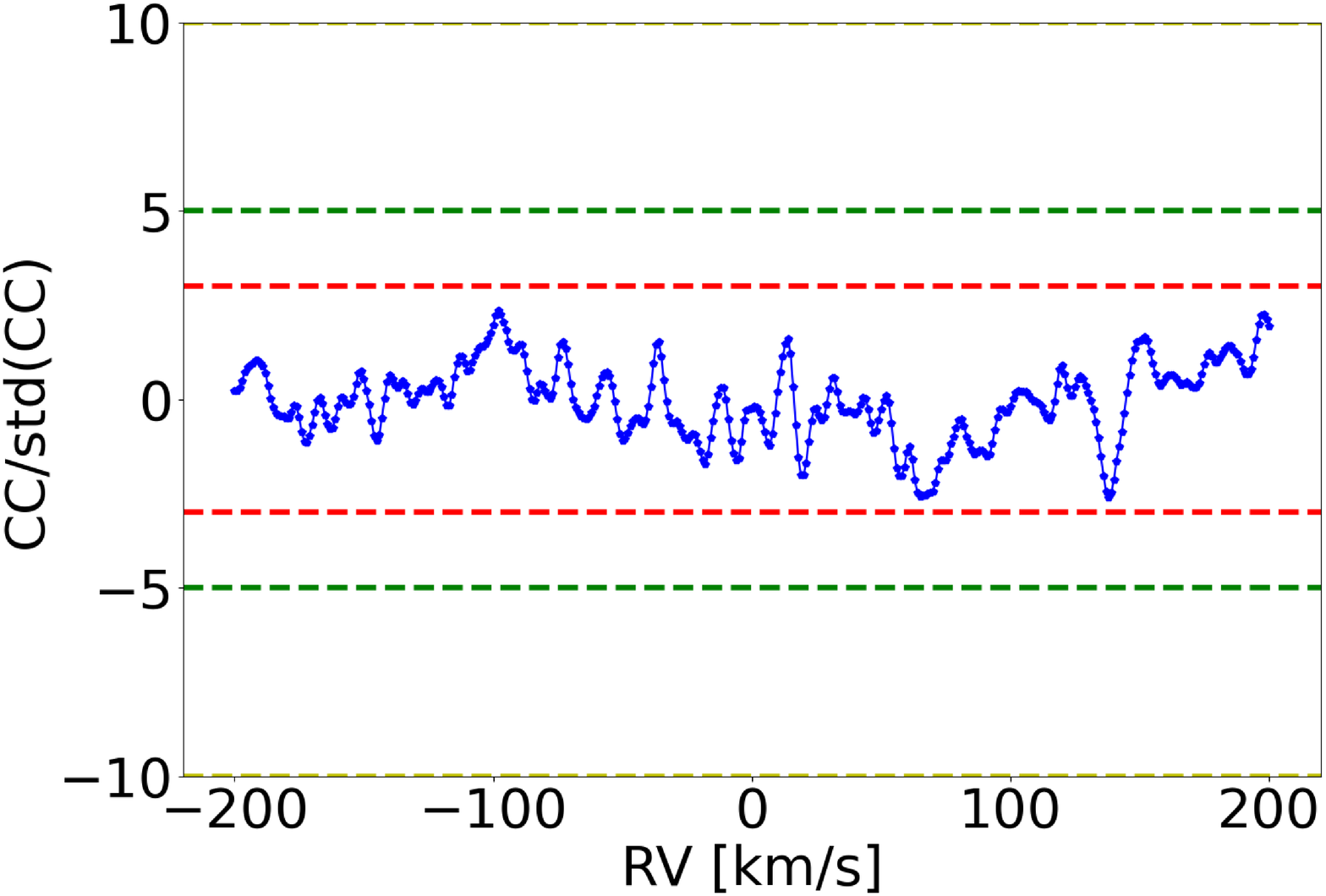}}%
  \qquad
  \subfloat{\includegraphics[scale = 0.12, angle =90 ]{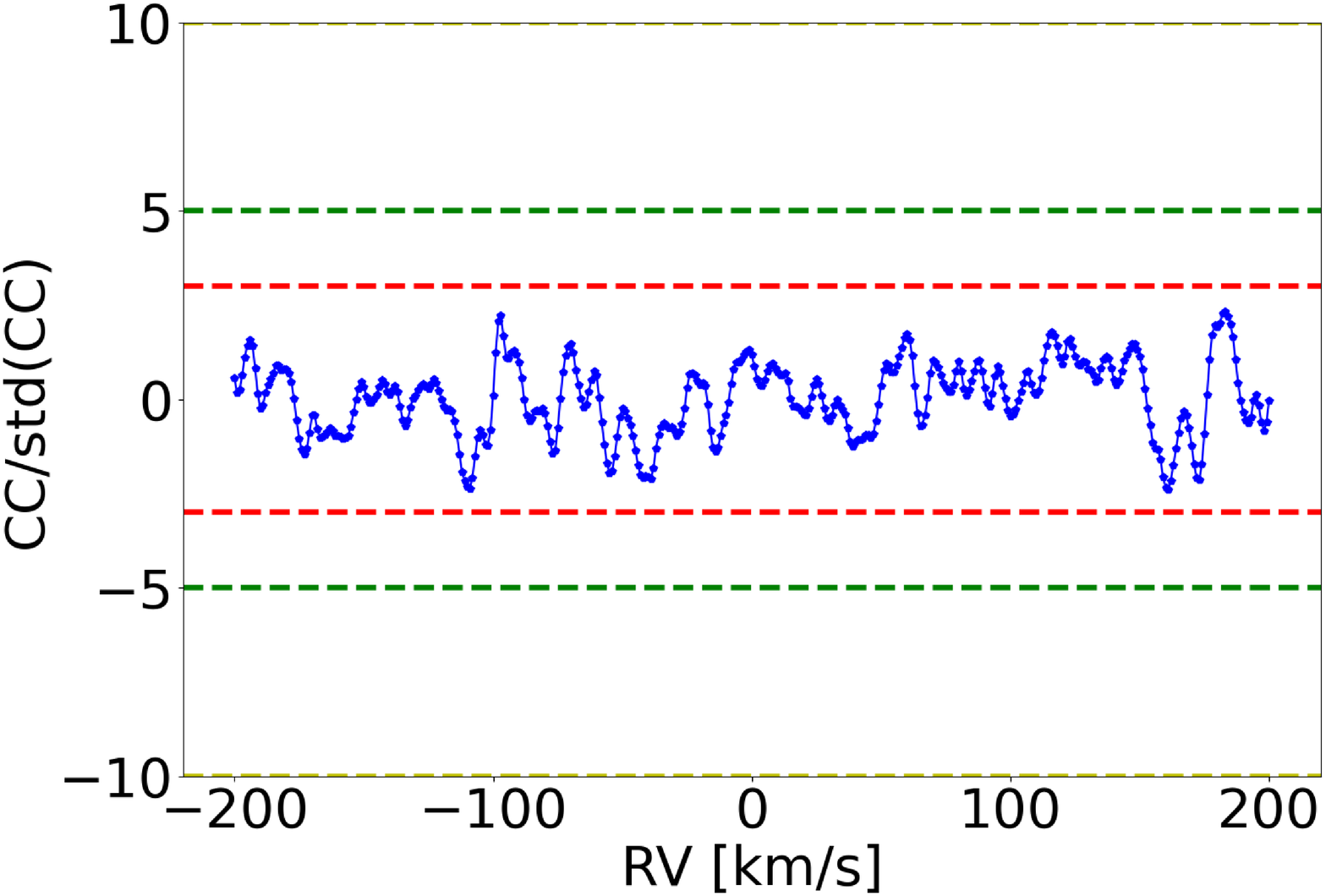}}%
  \qquad
  \subfloat{\includegraphics[scale = 0.12, angle =90 ]{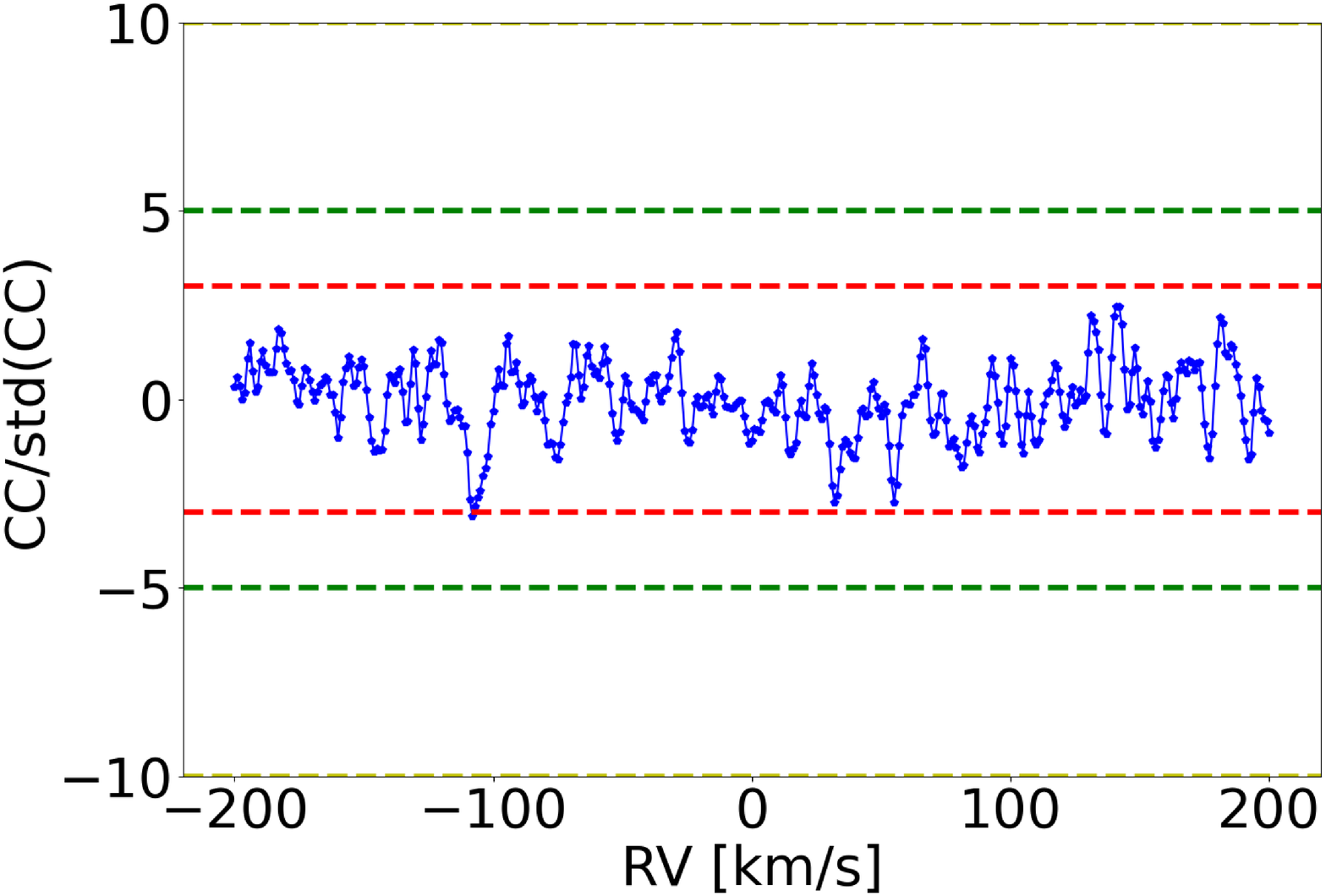}}%
  \qquad
  \subfloat{\includegraphics[scale = 0.12, angle =90 ]{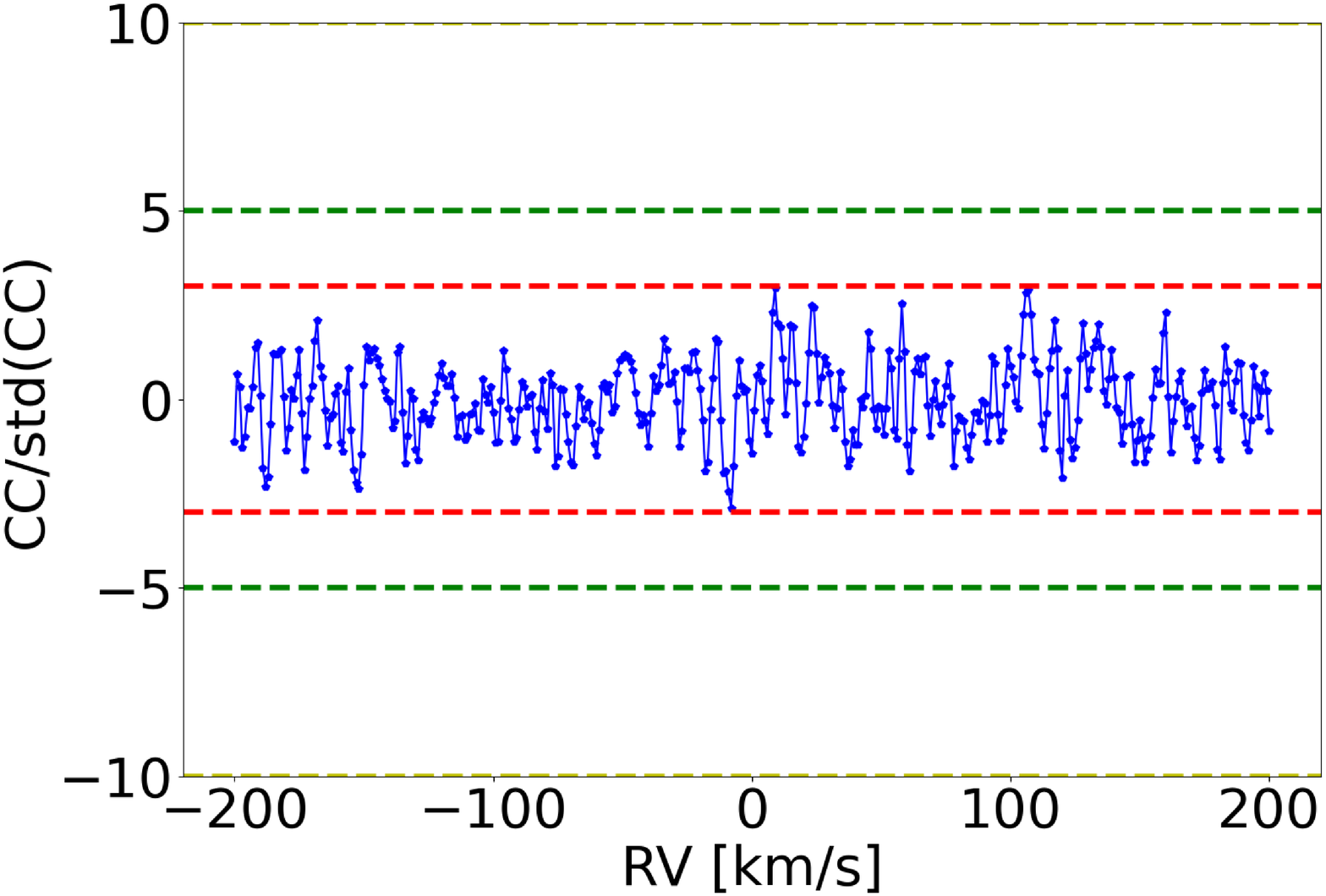}}%
  \\
  \vspace{0.5cm}
  \centering
  \subfloat{\includegraphics[scale = 0.12, angle =90 ]{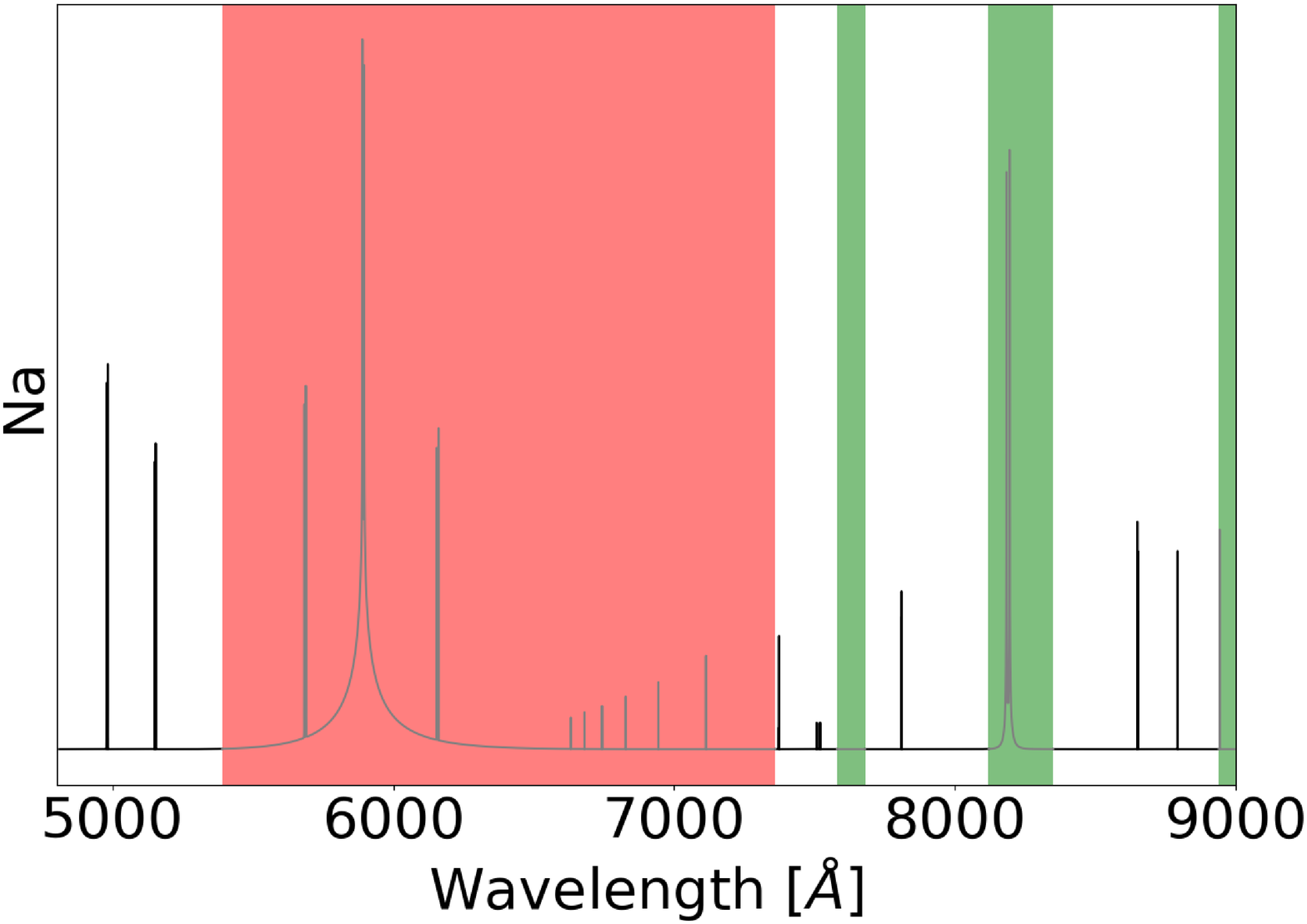}}%
  \qquad
  \subfloat{\includegraphics[scale = 0.12, angle =90 ]{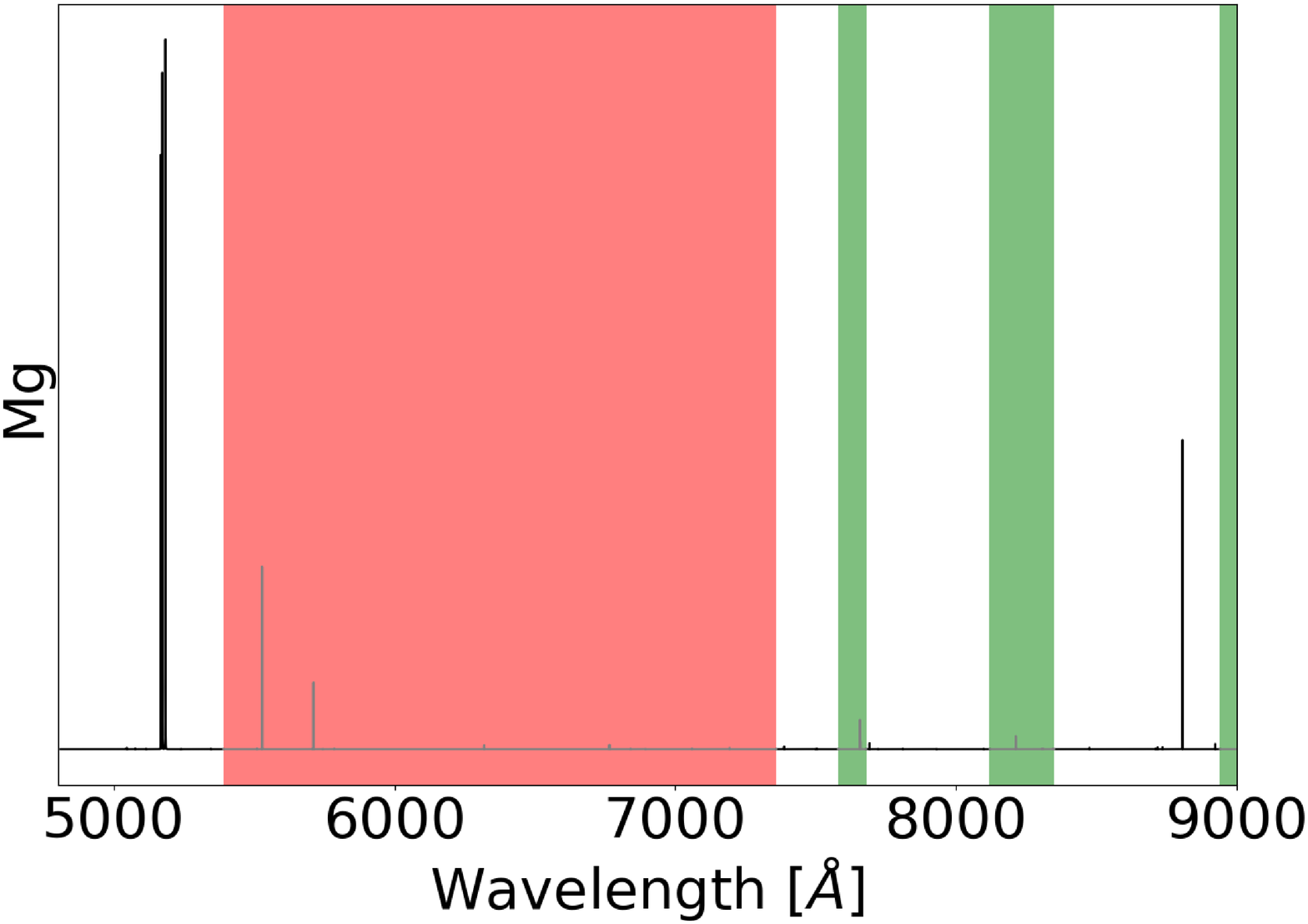}}%
  \qquad
  \subfloat{\includegraphics[scale = 0.12, angle =90 ]{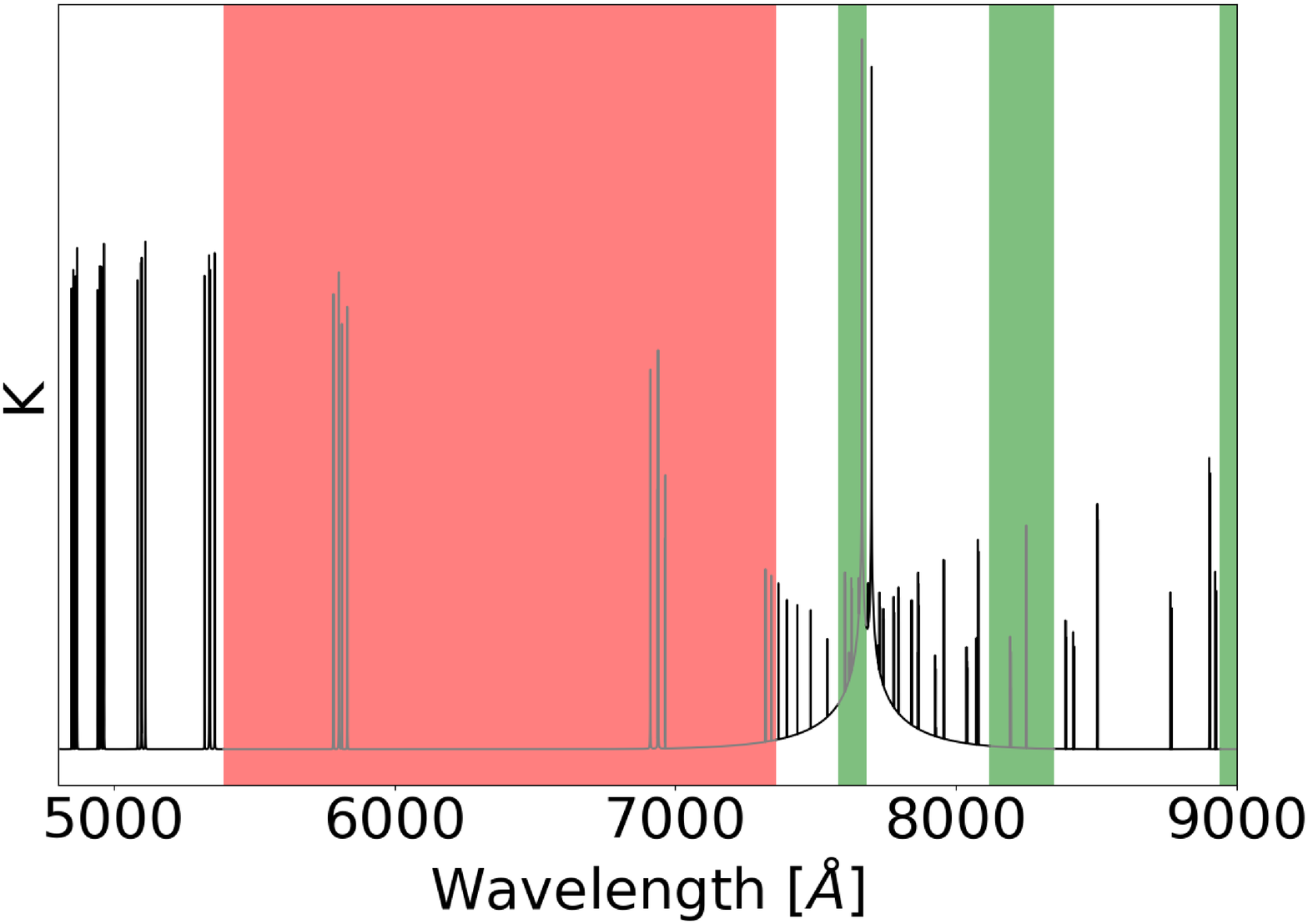}}%
  \qquad
  \subfloat{\includegraphics[scale = 0.12, angle =90 ]{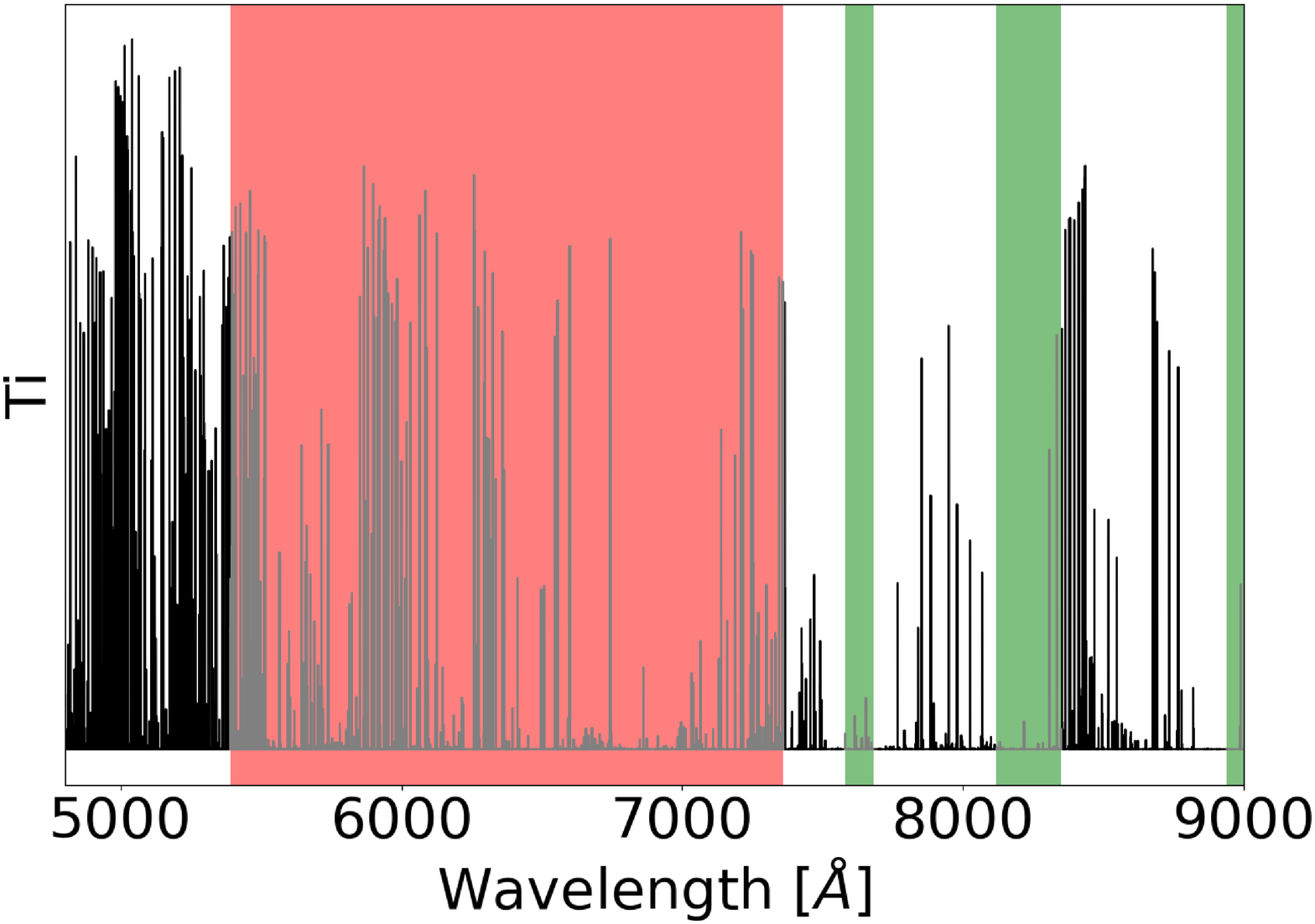}}%
  \qquad
  \subfloat{\includegraphics[scale = 0.12, angle =90 ]{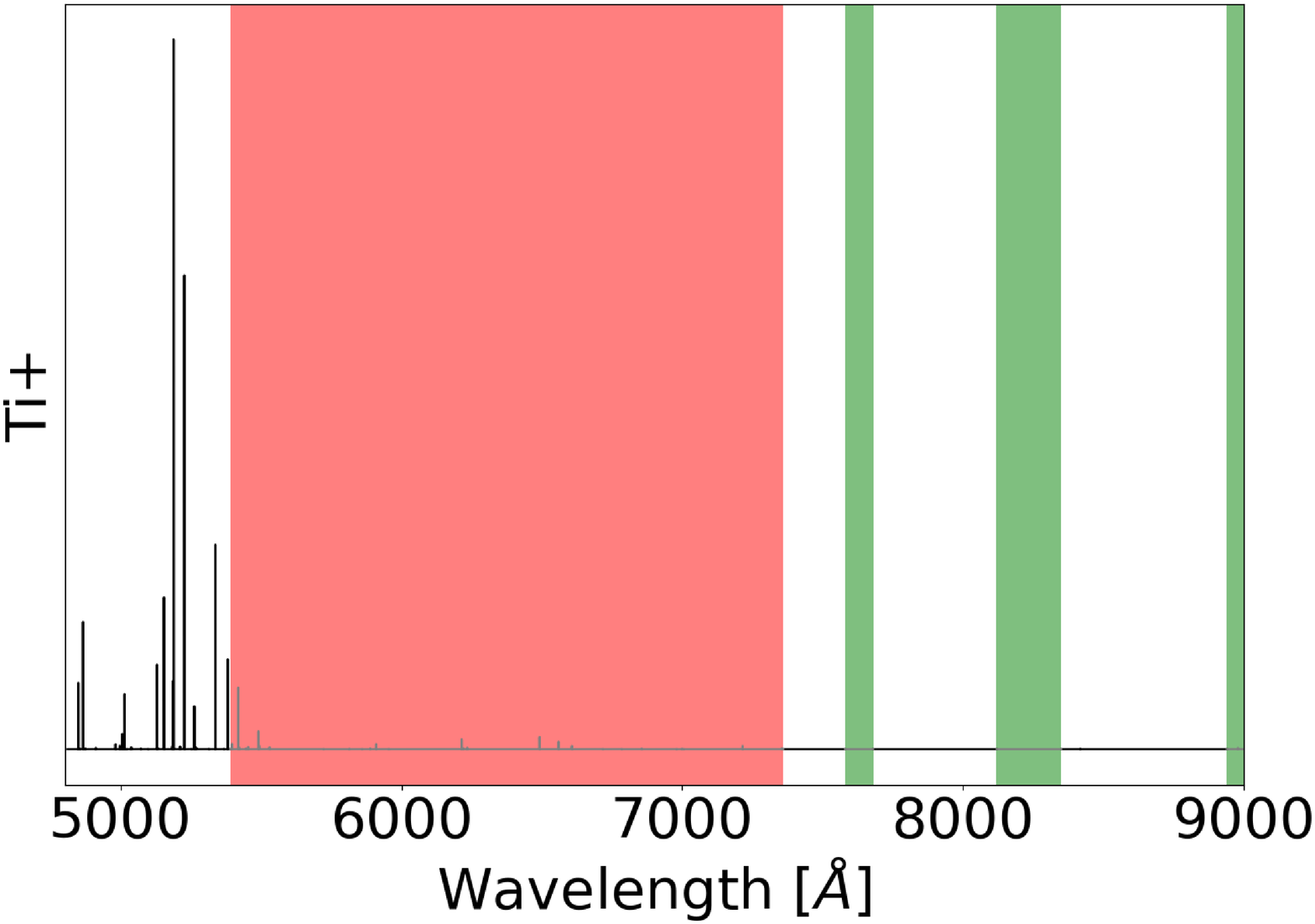}}%
  \caption{Same as Figure~\ref{fig:A1}}%
  \label{fig:A2}
\end{figure*}%

\begin{figure*}%
  \centering
  \subfloat{\includegraphics[scale = 0.12, angle =90 ]{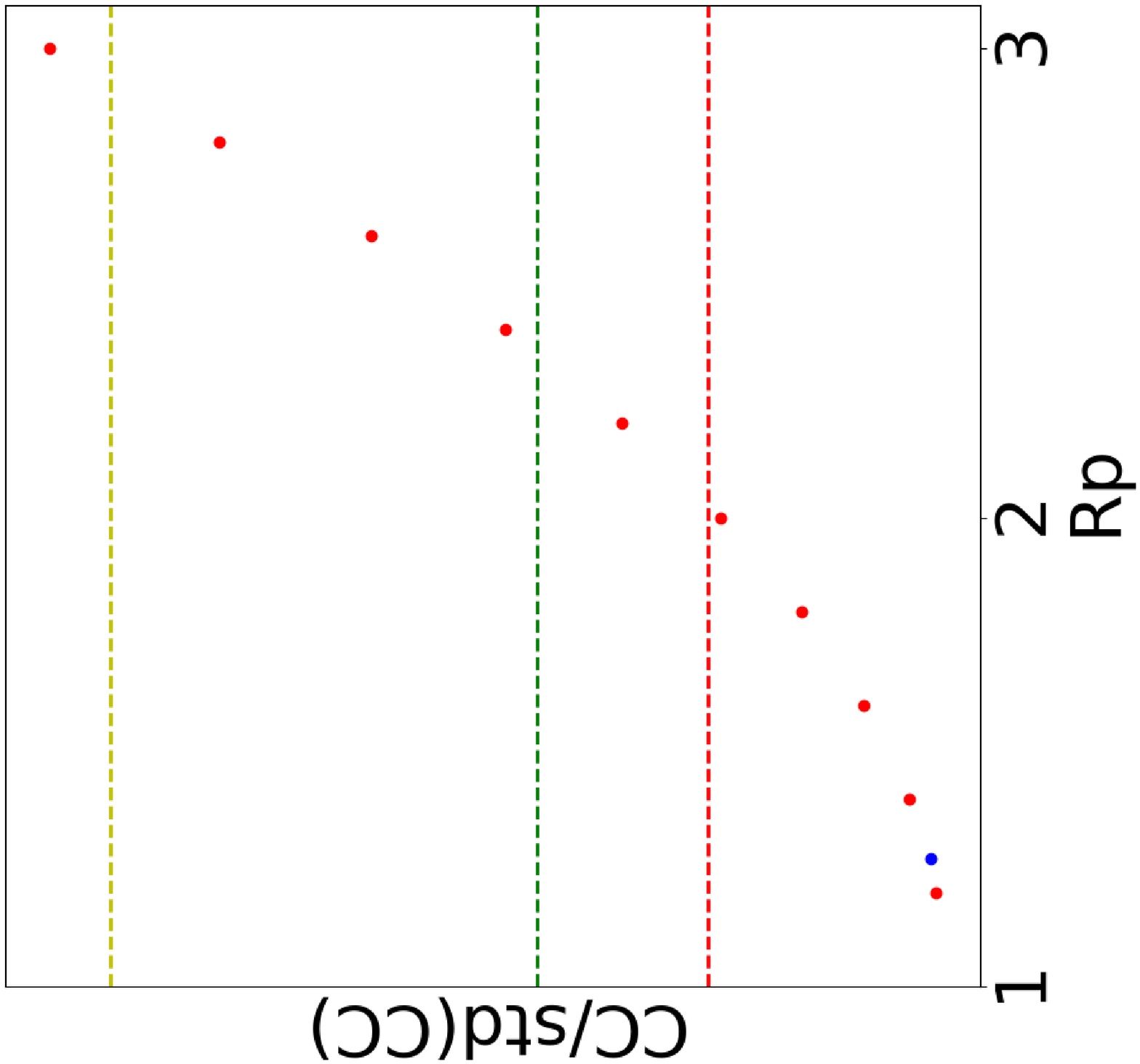}}%
  \qquad
  \subfloat{\includegraphics[scale = 0.12, angle =90 ]{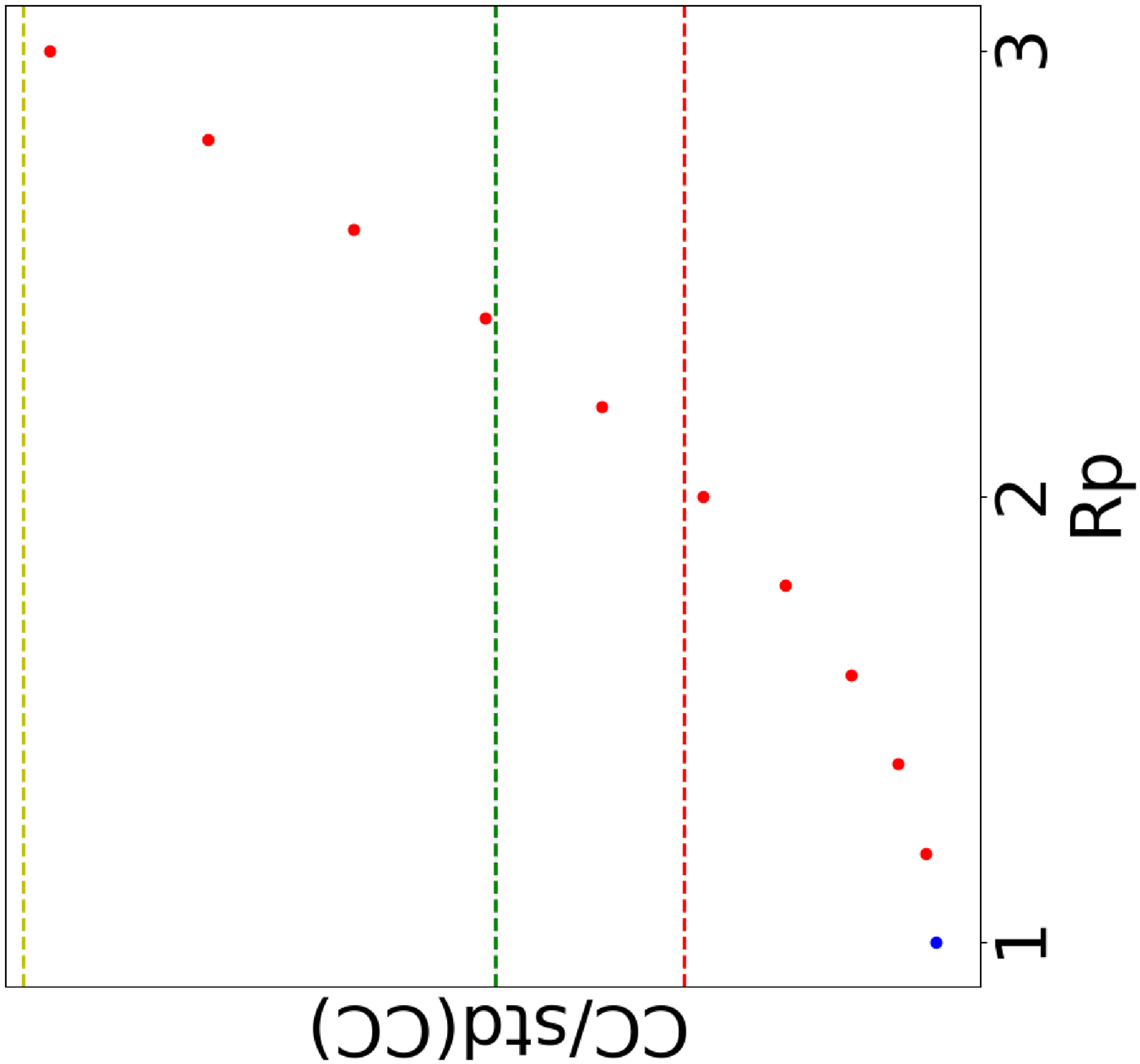}}%
  \qquad
  \subfloat{\includegraphics[scale = 0.12, angle =90 ]{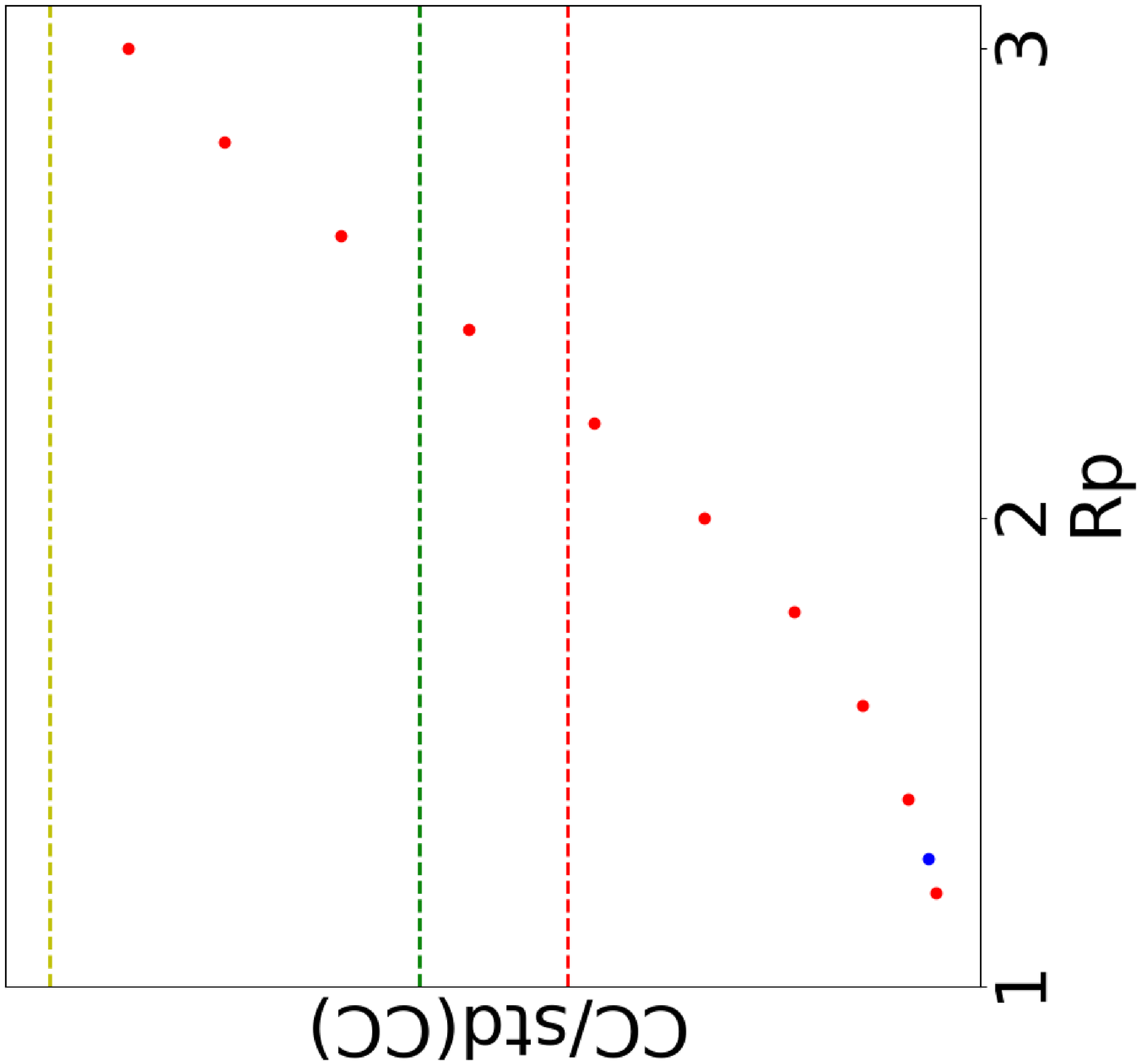}}%
  \qquad
  \subfloat{\includegraphics[scale = 0.12, angle =90 ]{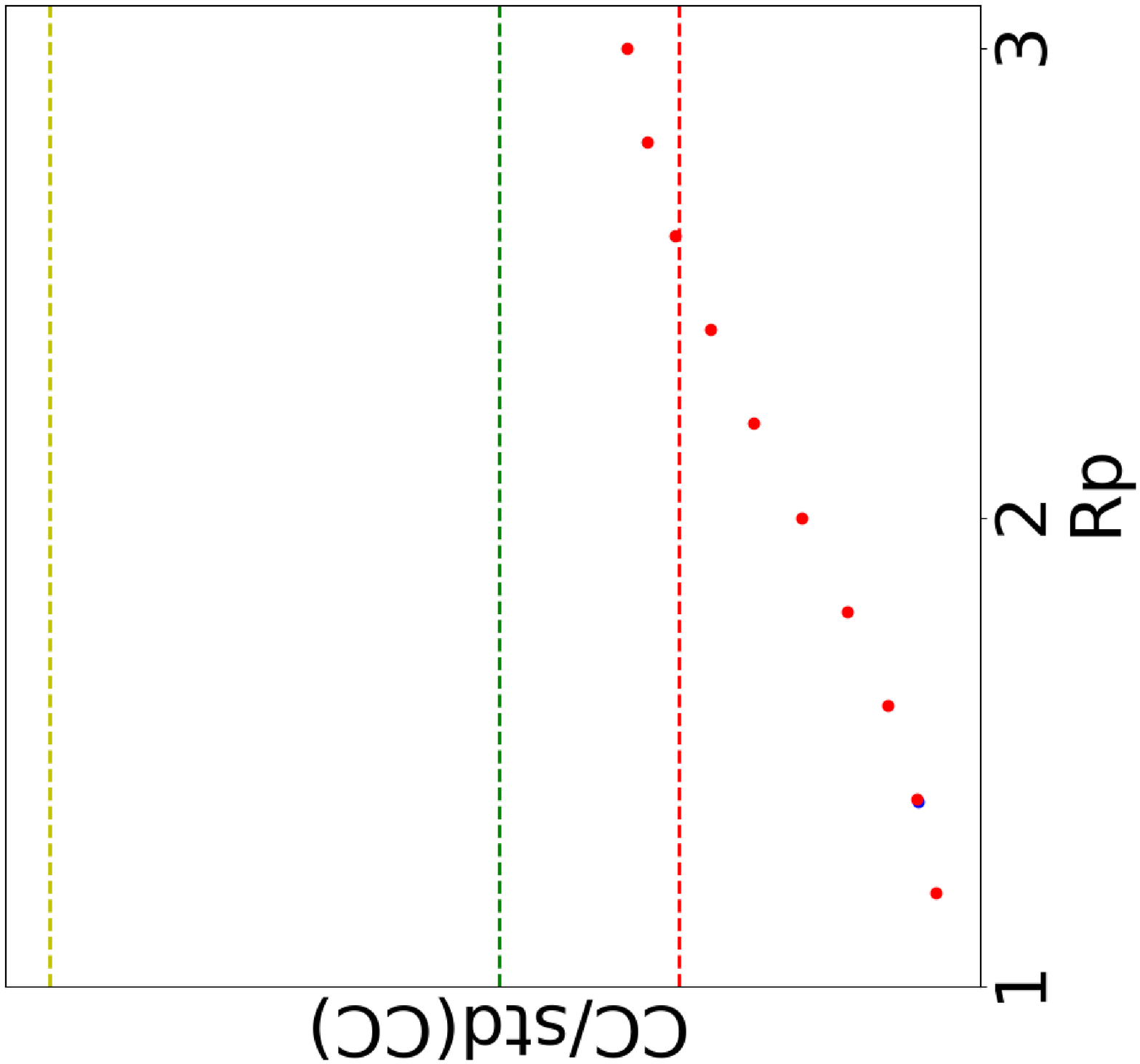}}%
  \qquad
  \subfloat{\includegraphics[scale = 0.12, angle =90 ]{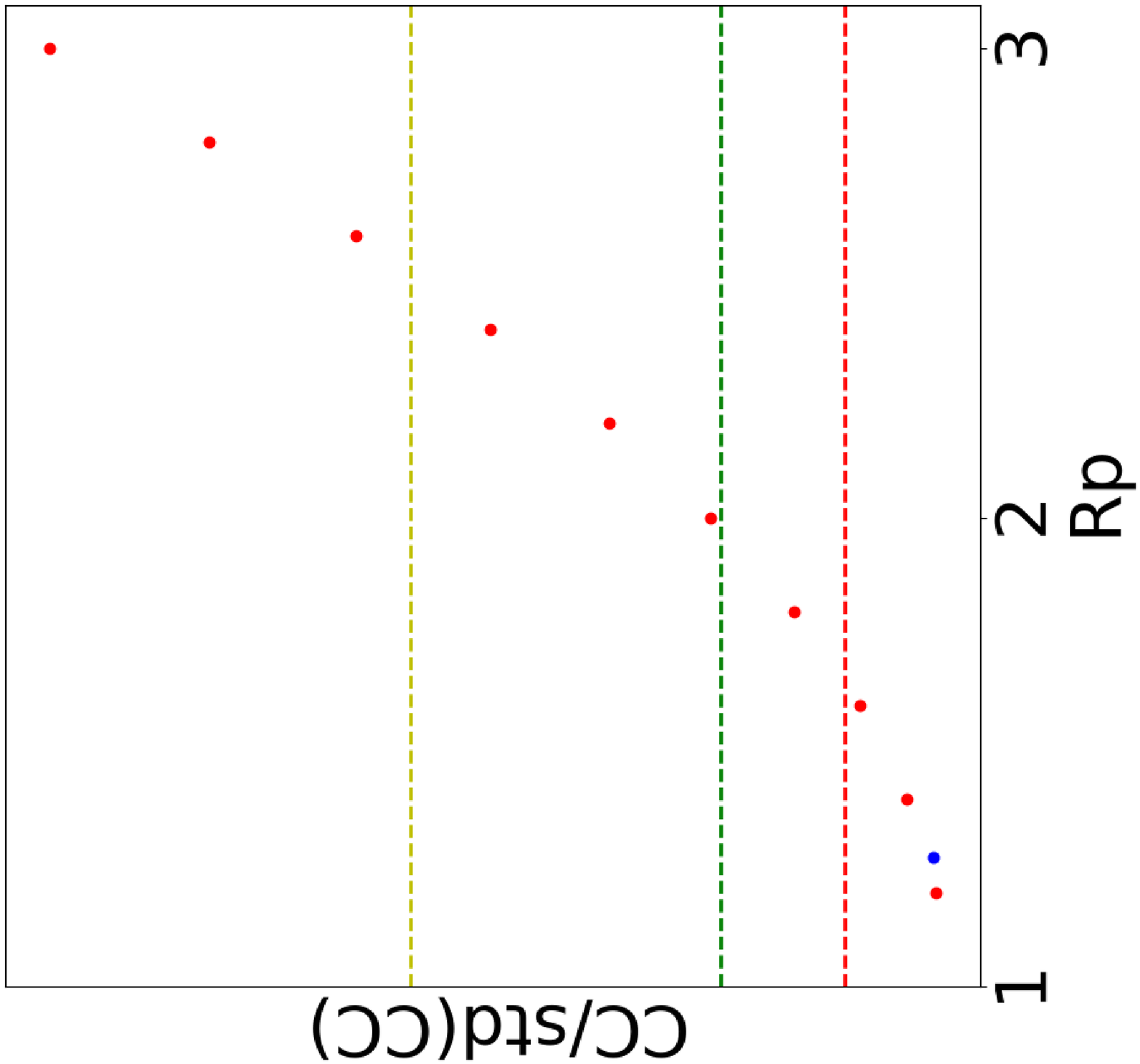}}%
  \\
  \vspace{0.5cm}
  \centering
  \subfloat{\includegraphics[scale = 0.12, angle =90 ]{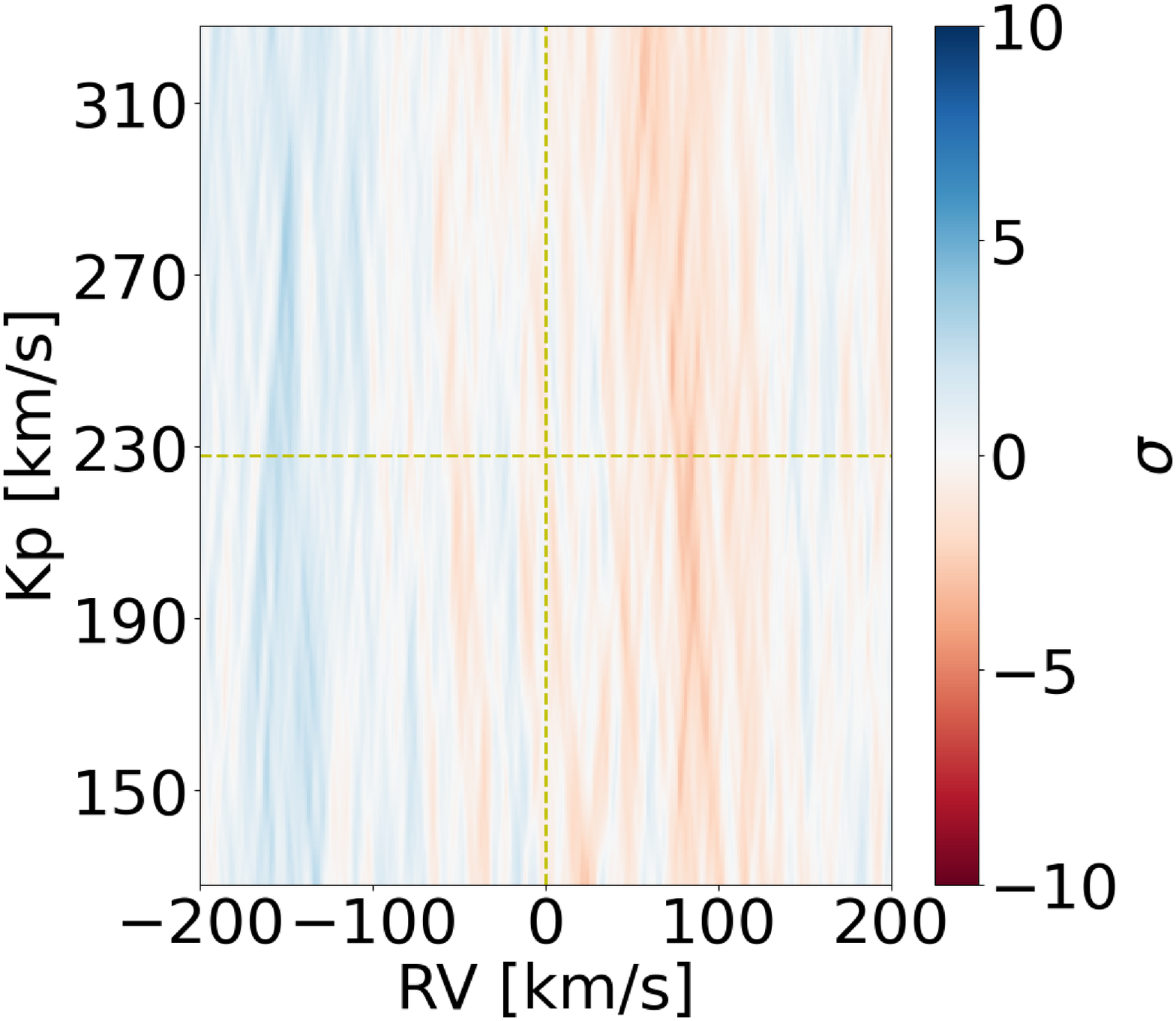}}%
  \qquad
  \subfloat{\includegraphics[scale = 0.12, angle =90 ]{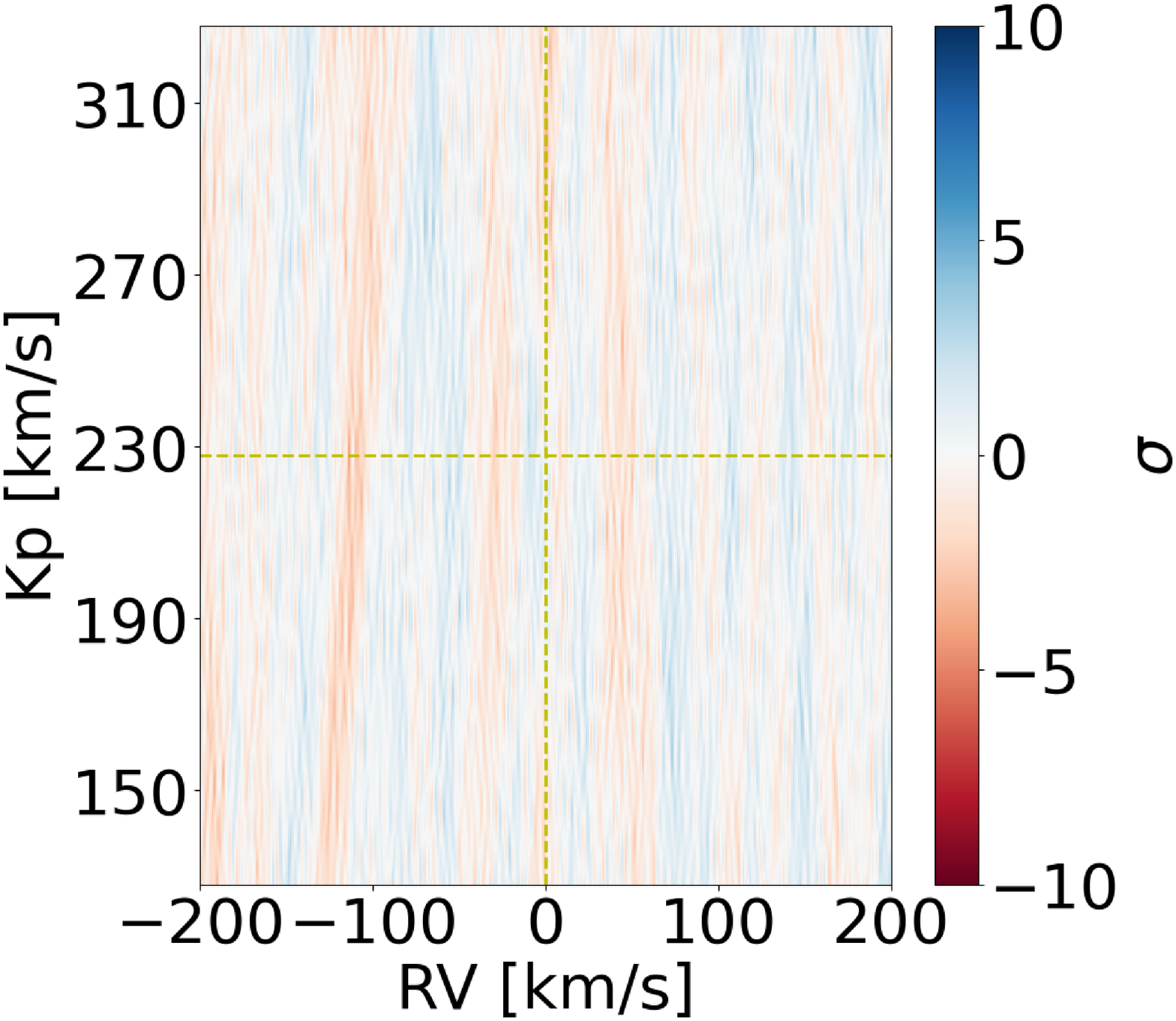}}%
  \qquad
  \subfloat{\includegraphics[scale = 0.12, angle =90 ]{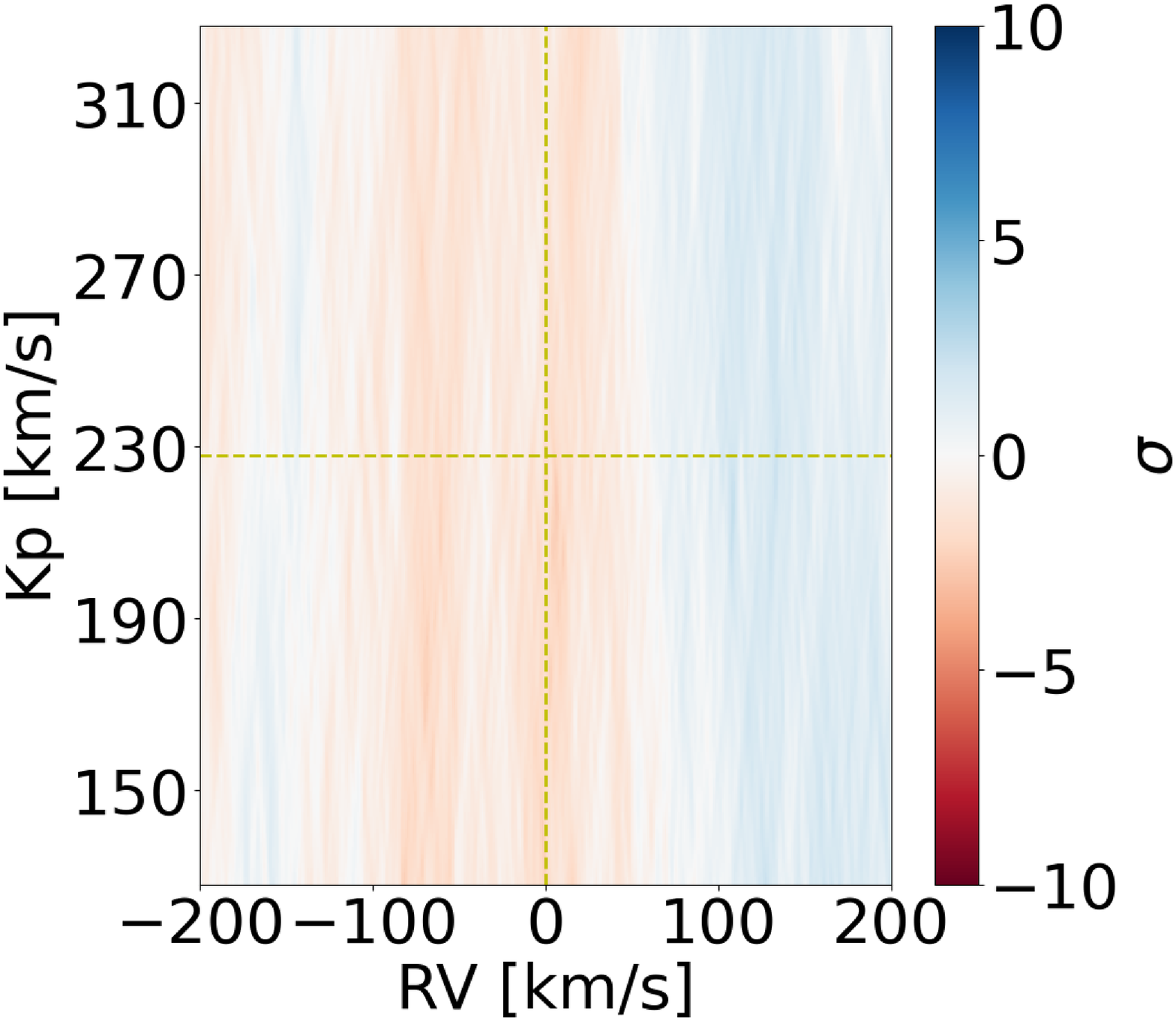}}%
  \qquad
  \subfloat{\includegraphics[scale = 0.12, angle =90 ]{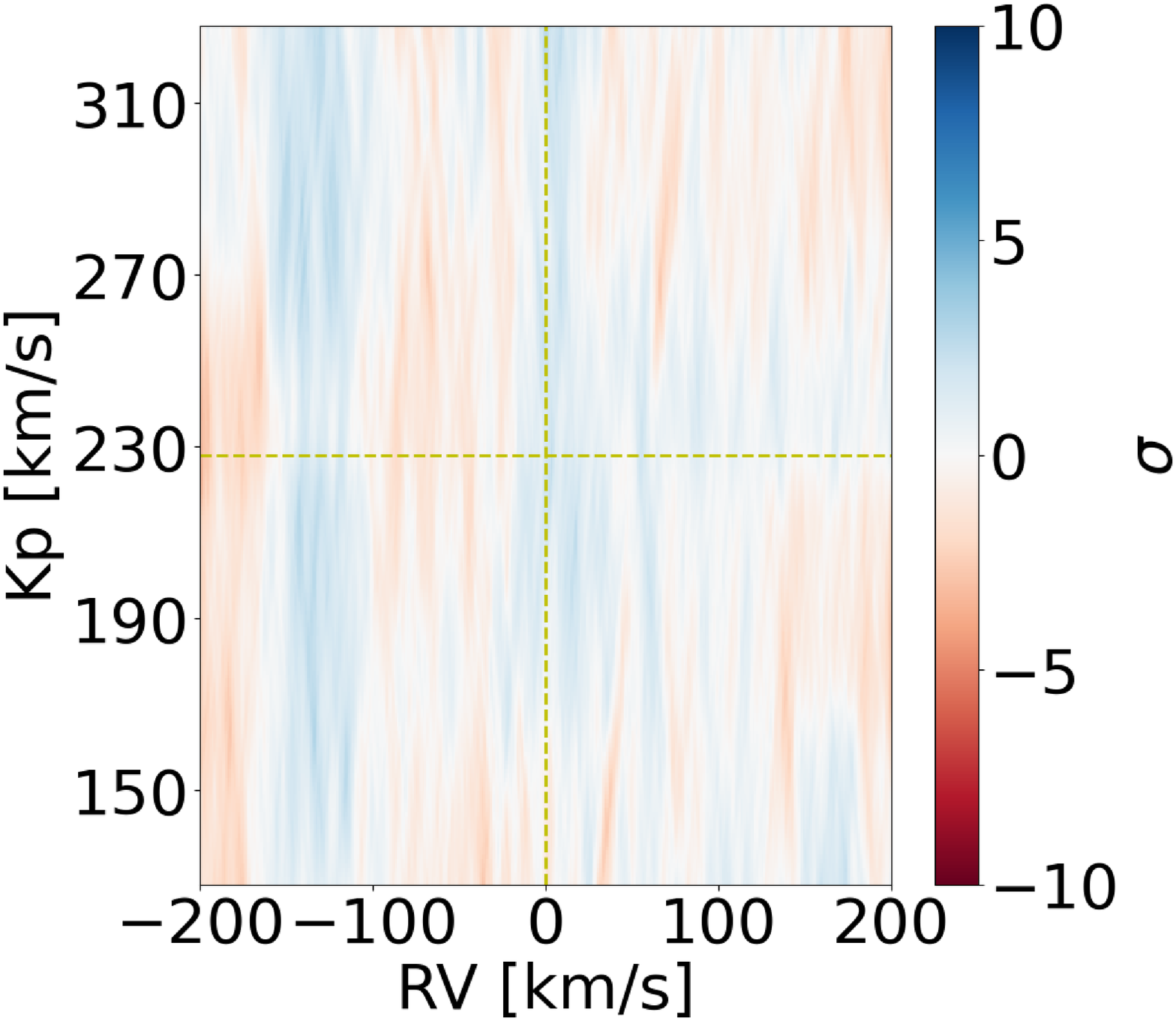}}%
  \qquad
  \subfloat{\includegraphics[scale = 0.12, angle =90 ]{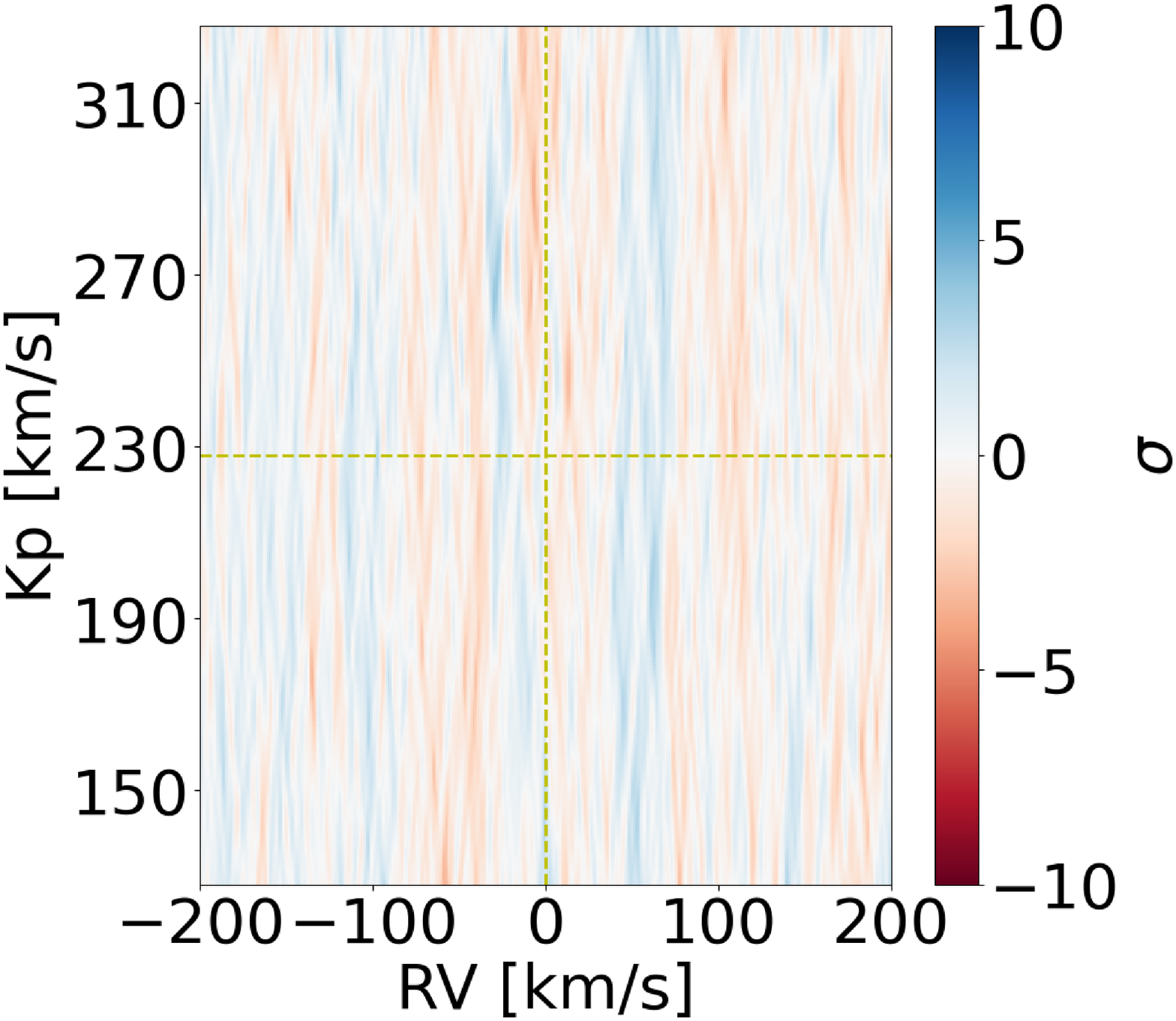}}%
  \\
  \vspace{0.5cm}
  \centering
  \subfloat{\includegraphics[scale = 0.12, angle =90 ]{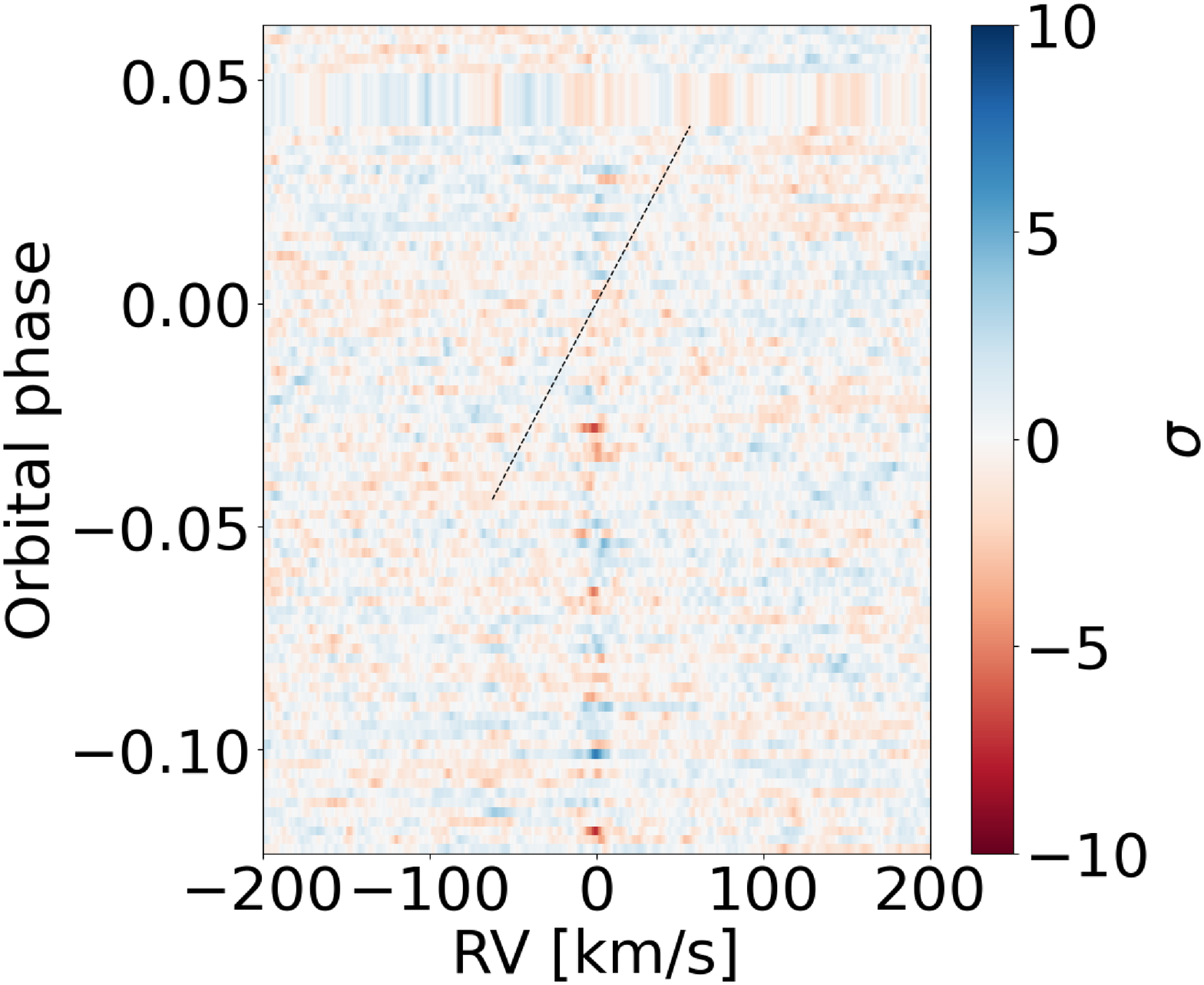}}%
  \qquad
  \subfloat{\includegraphics[scale = 0.12, angle =90 ]{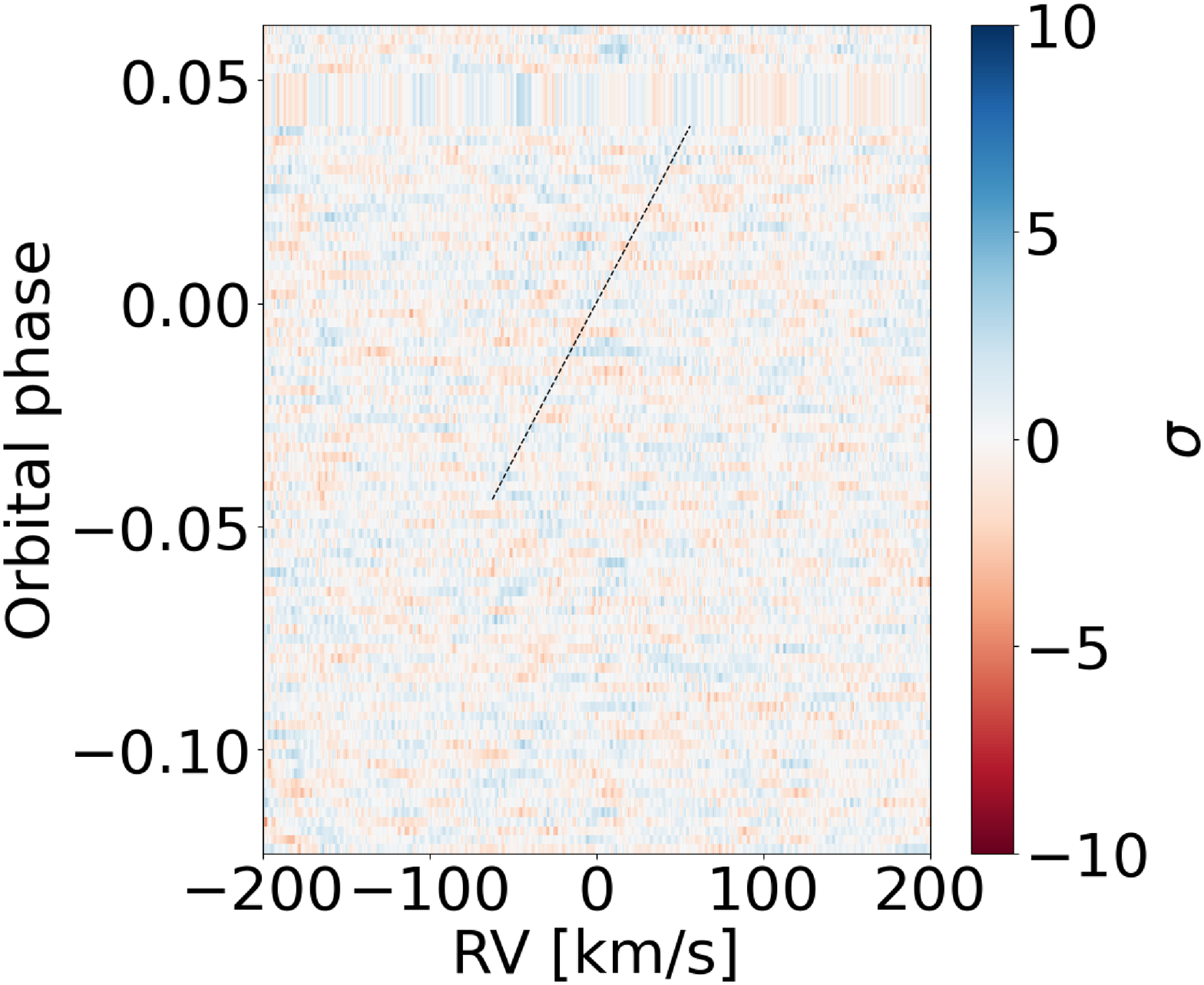}}%
  \qquad
  \subfloat{\includegraphics[scale = 0.12, angle =90 ]{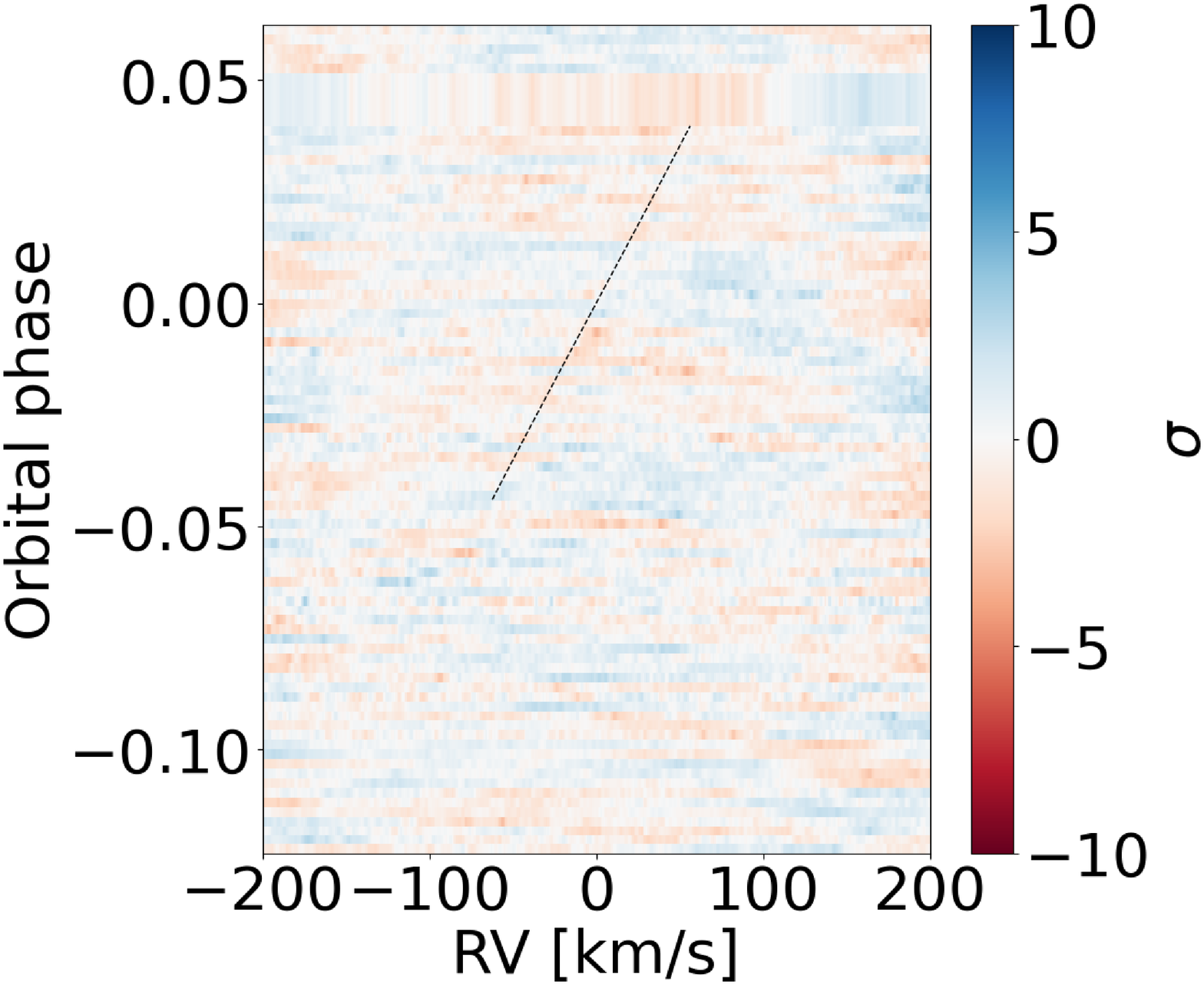}}%
  \qquad
  \subfloat{\includegraphics[scale = 0.12, angle =90 ]{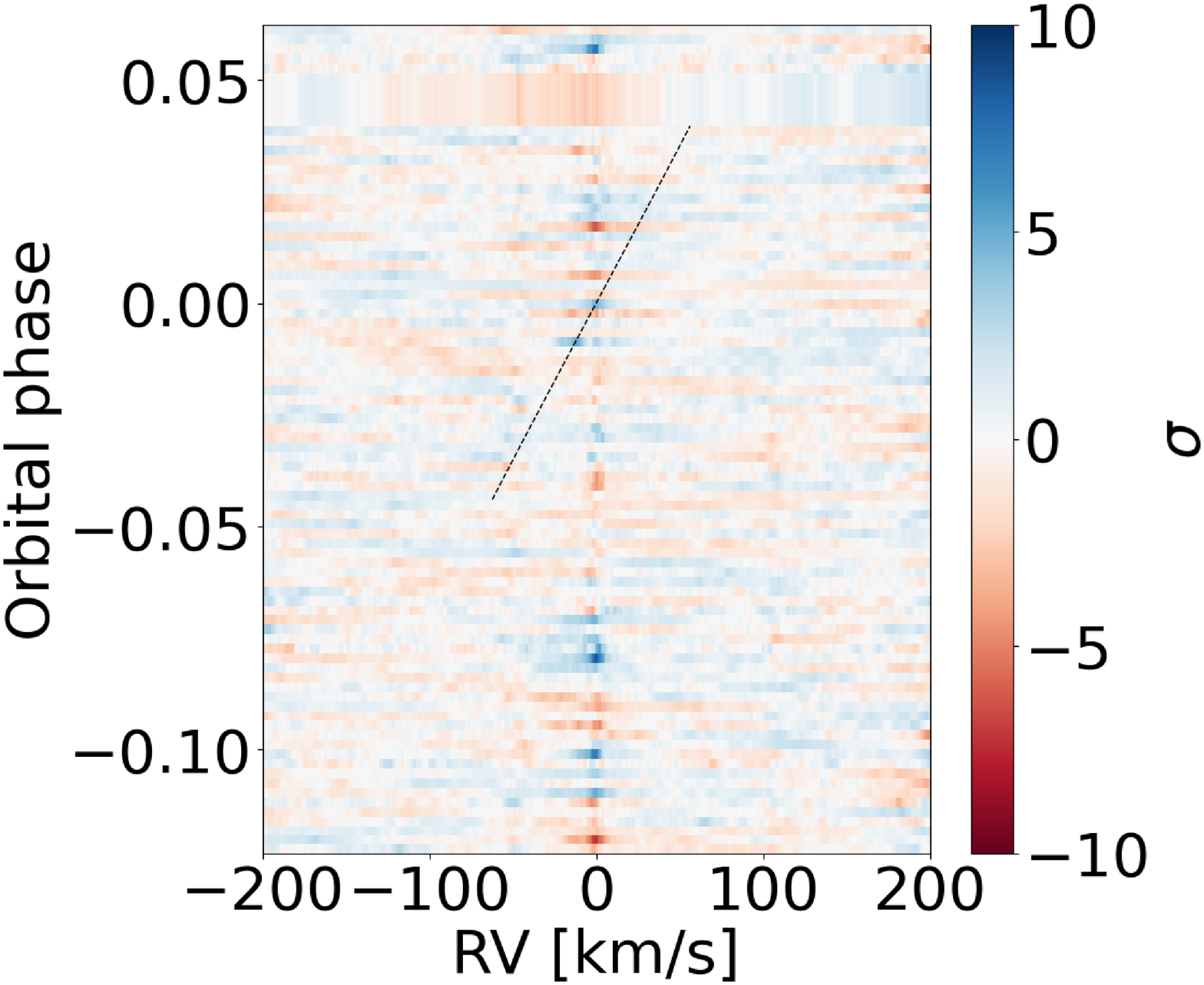}}%
  \qquad
  \subfloat{\includegraphics[scale = 0.12, angle =90 ]{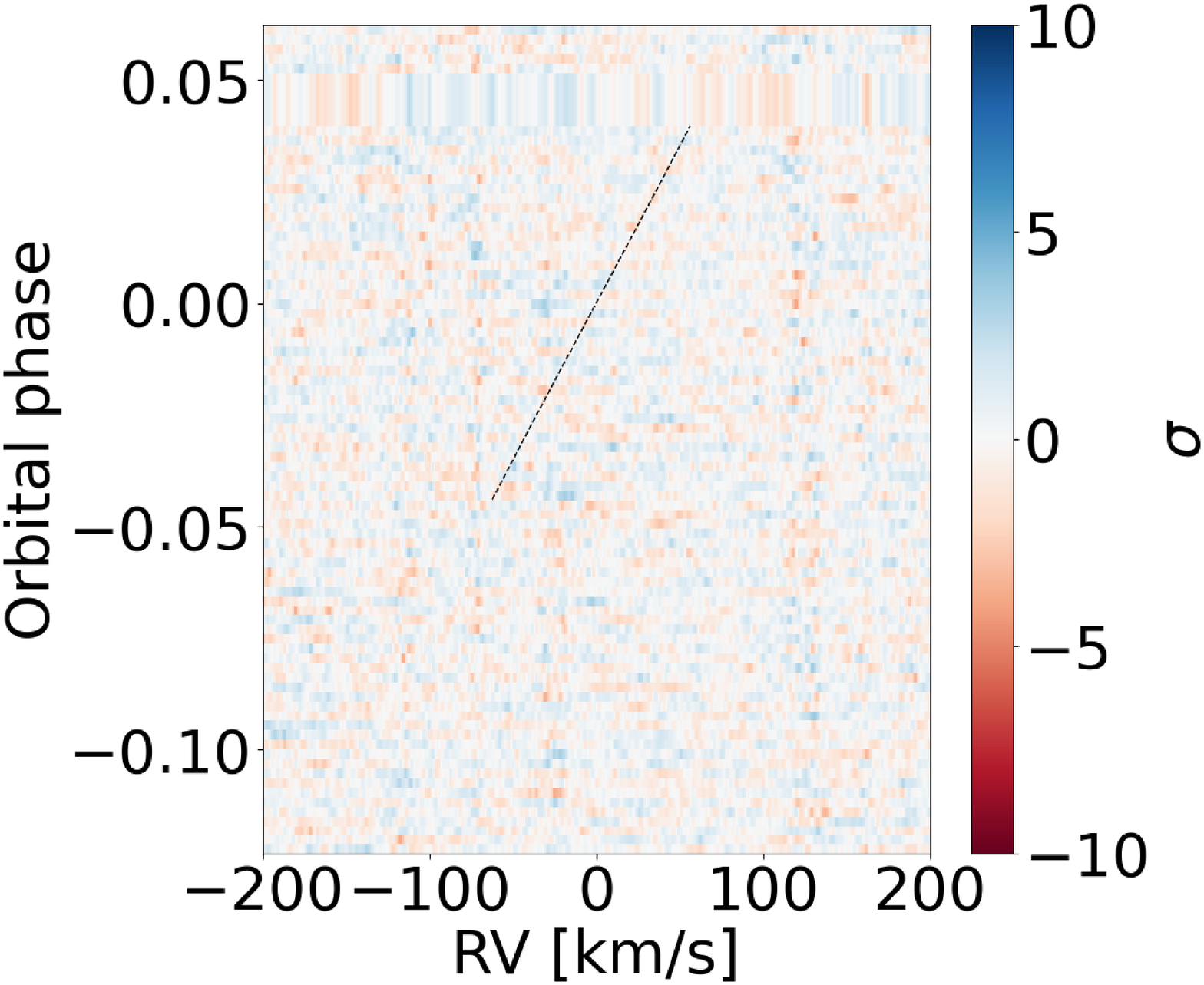}}%
  \\
  \vspace{0.5cm}
  \centering
  \subfloat{\includegraphics[scale = 0.12, angle =90 ]{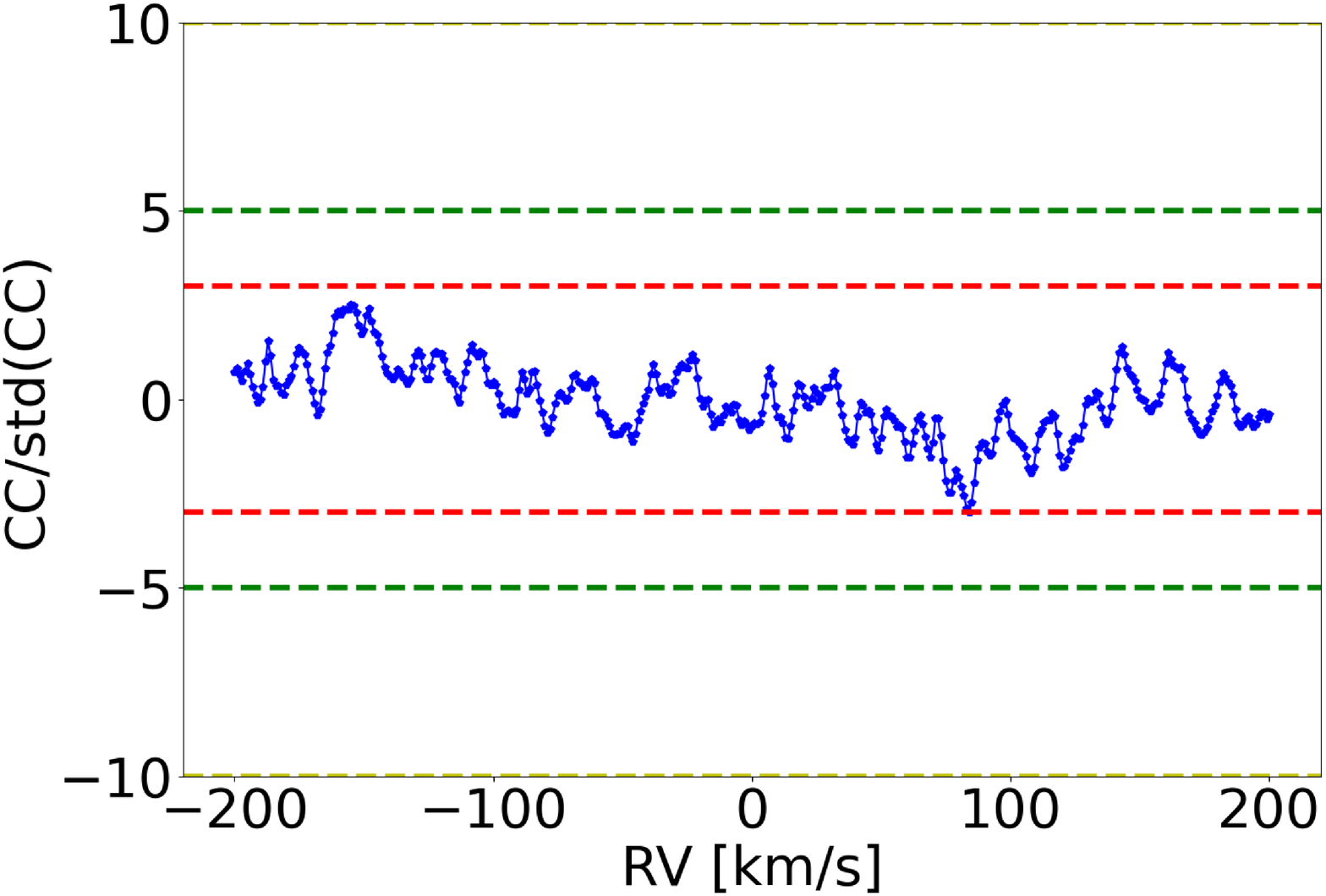}}%
  \qquad
  \subfloat{\includegraphics[scale = 0.12, angle =90 ]{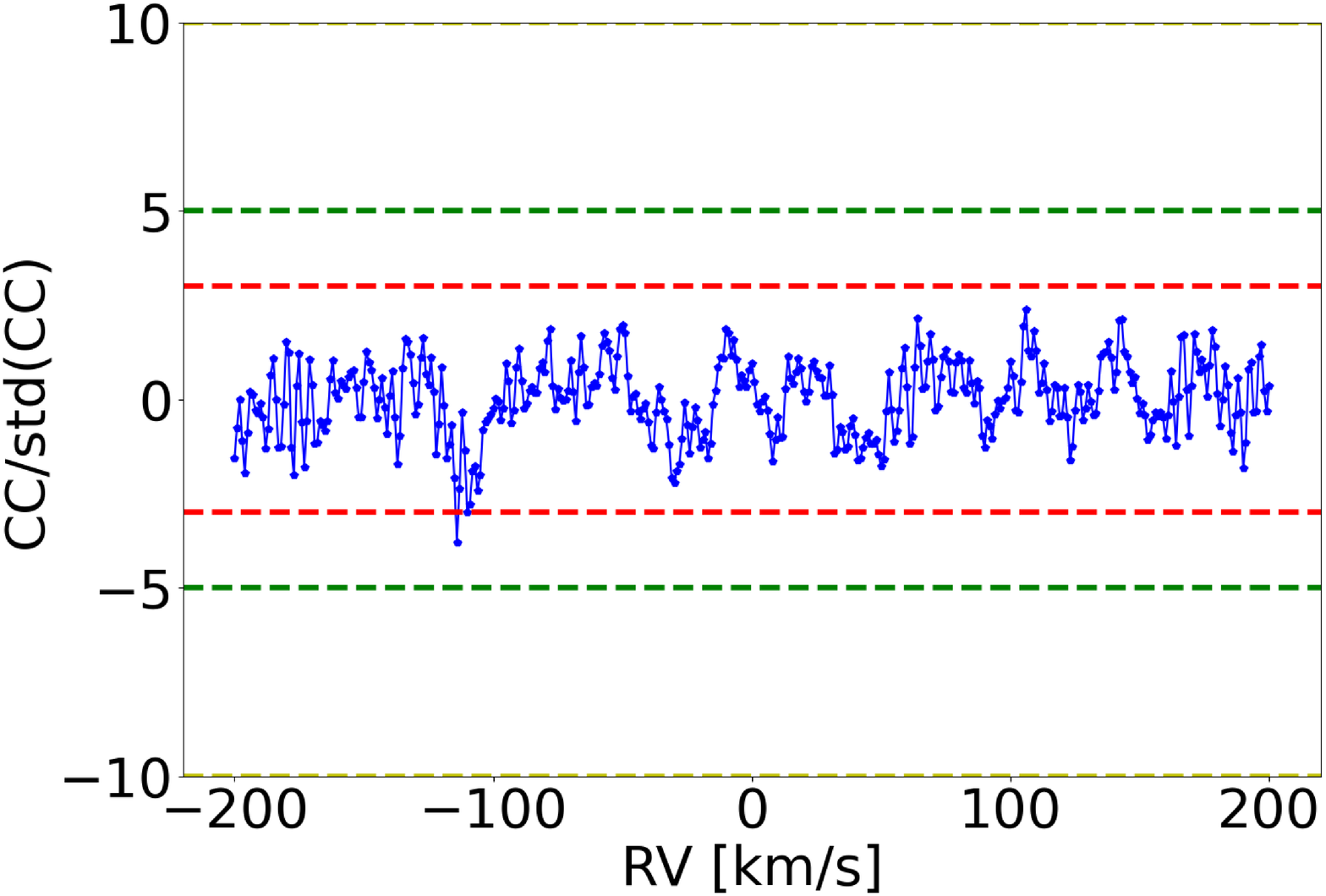}}%
  \qquad
  \subfloat{\includegraphics[scale = 0.12, angle =90 ]{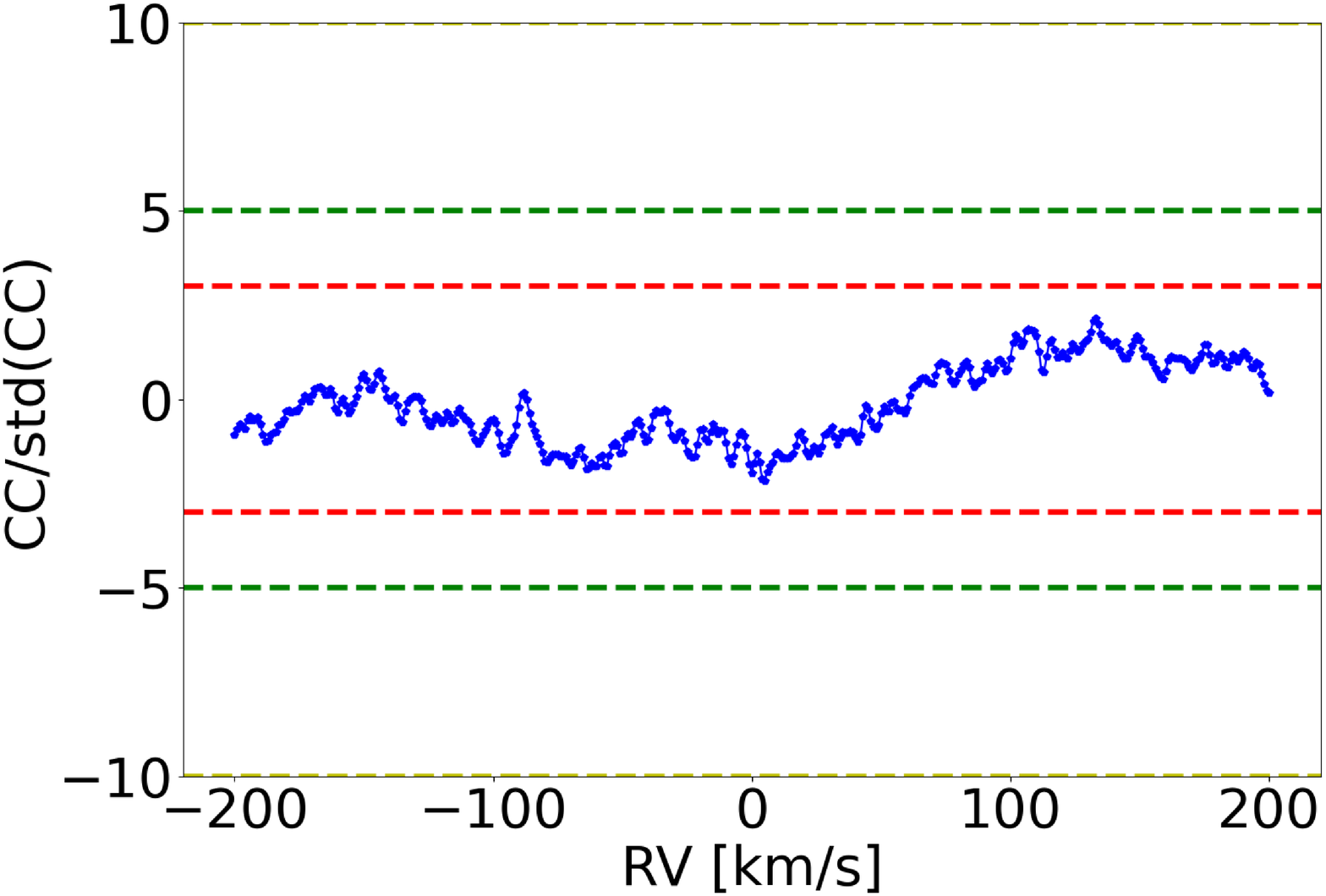}}%
  \qquad
  \subfloat{\includegraphics[scale = 0.12, angle =90 ]{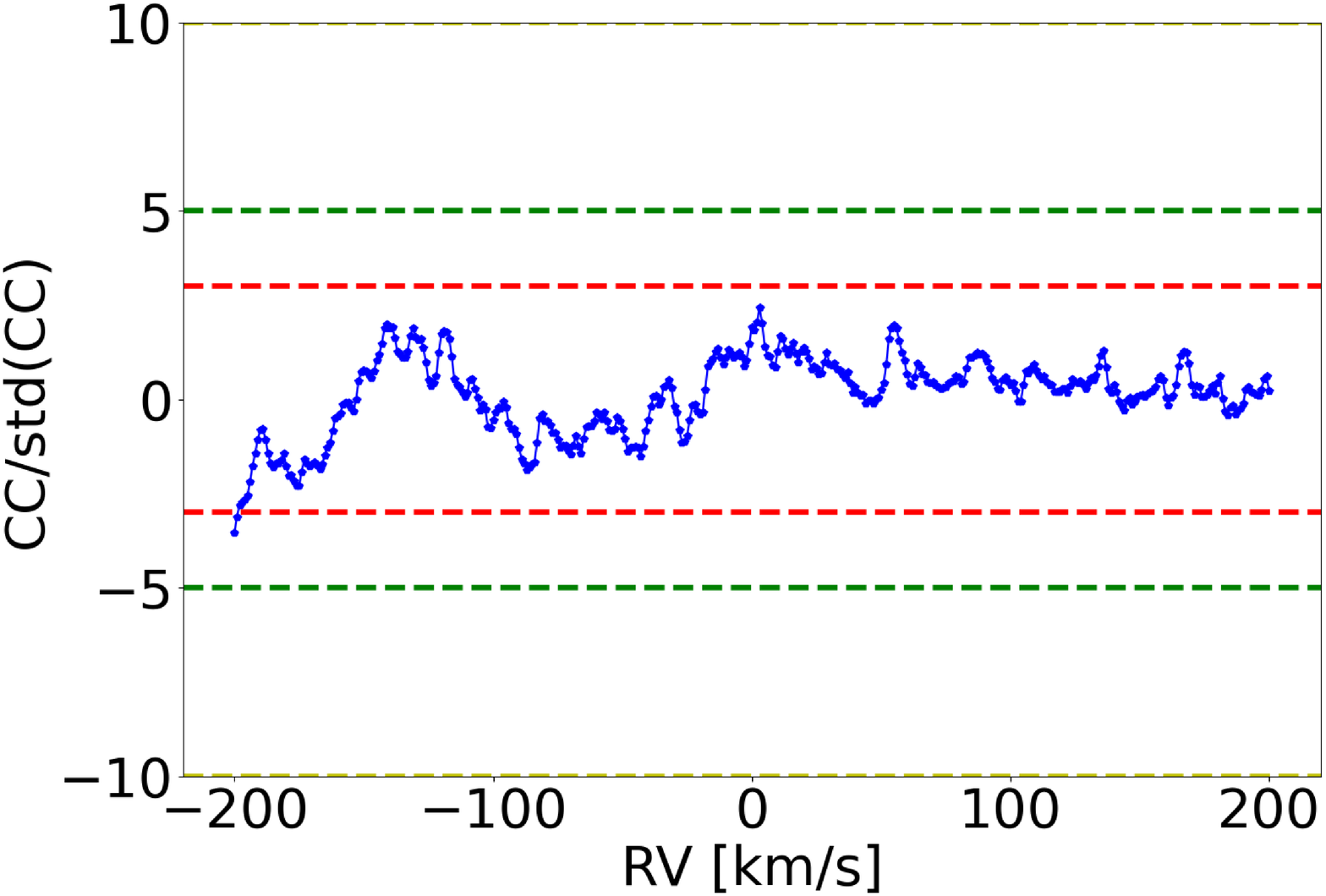}}%
  \qquad
  \subfloat{\includegraphics[scale = 0.12, angle =90 ]{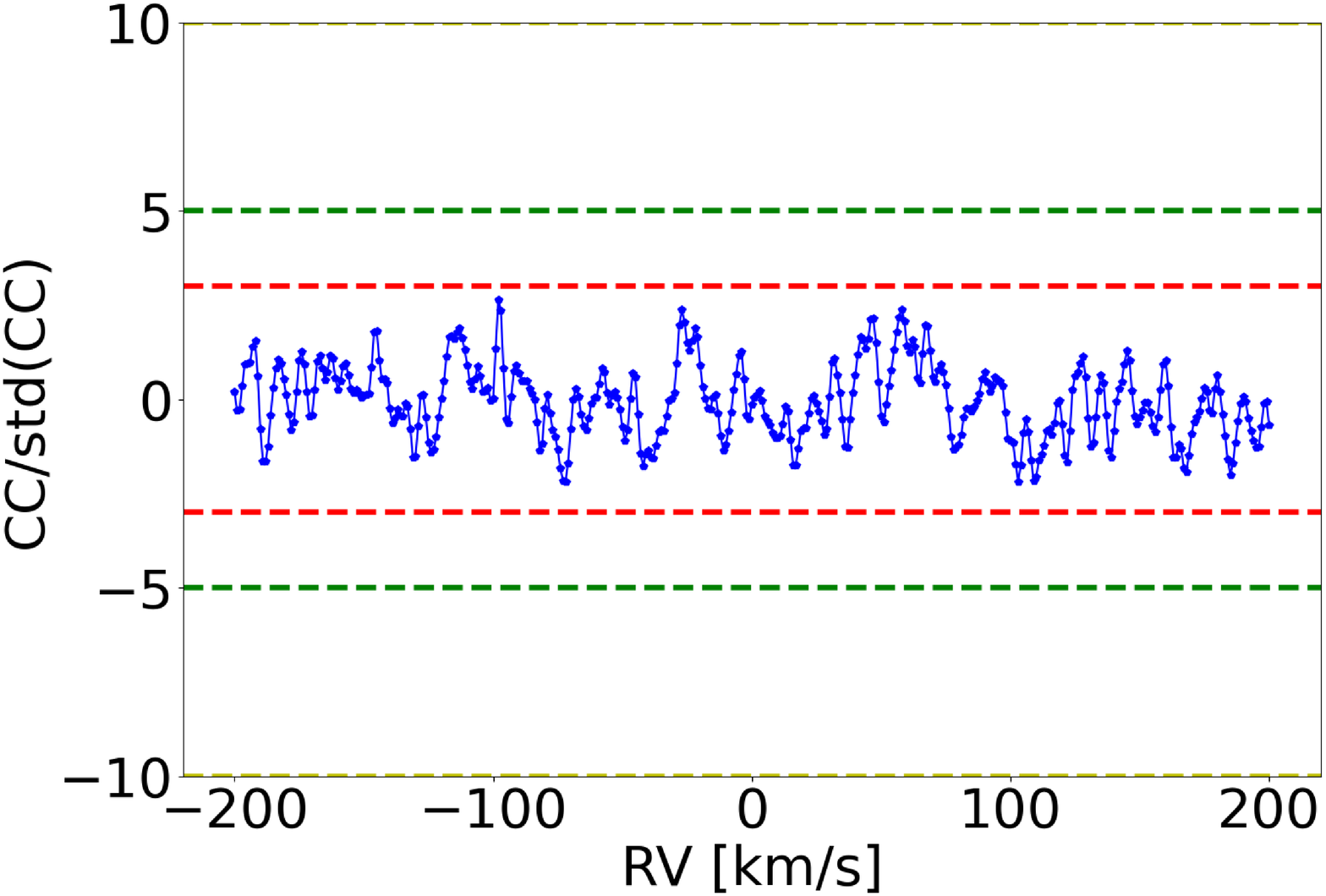}}%
  \\
  \vspace{0.5cm}
  \centering
  \subfloat{\includegraphics[scale = 0.12, angle =90 ]{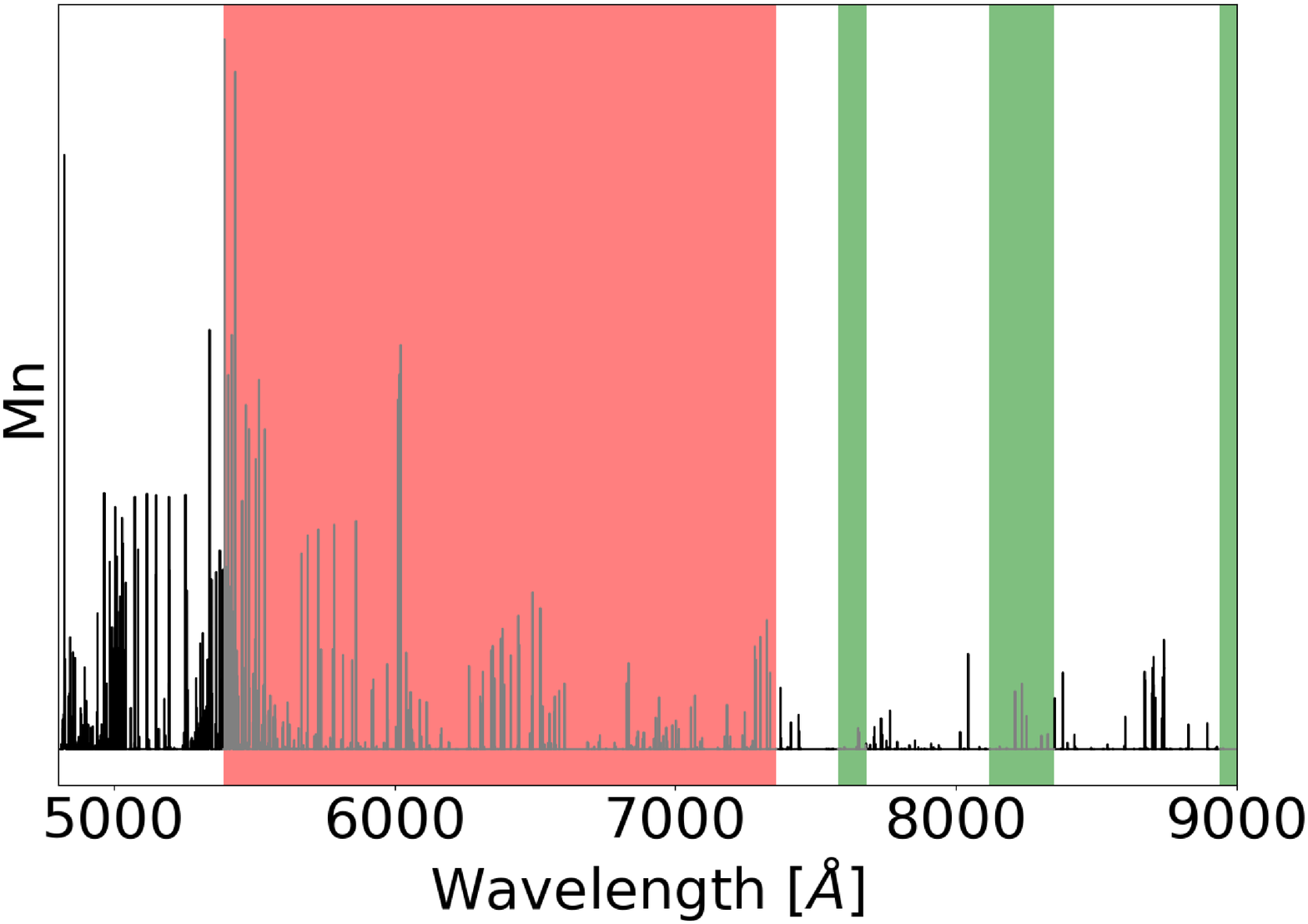}}%
  \qquad
  \subfloat{\includegraphics[scale = 0.12, angle =90 ]{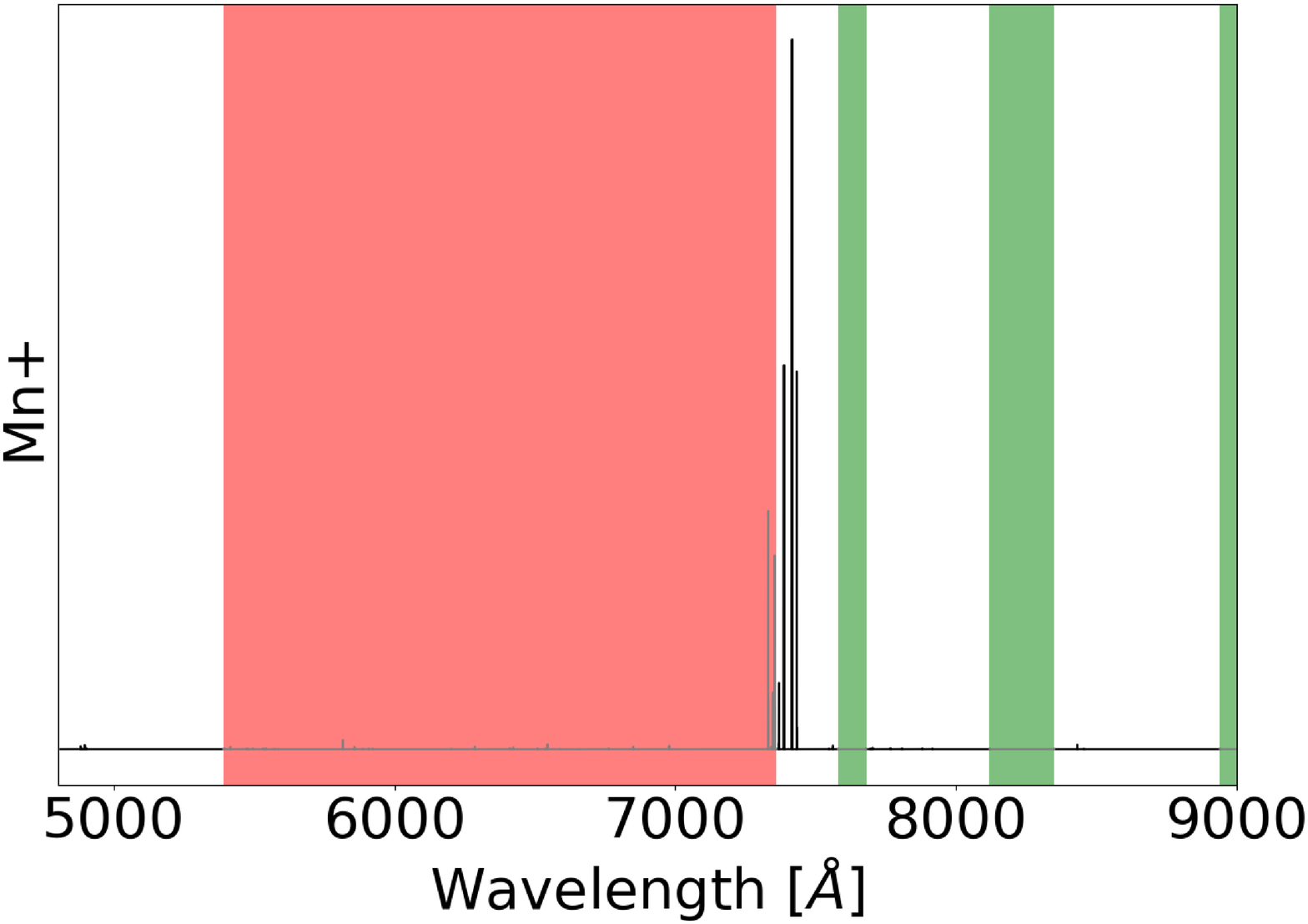}}%
  \qquad
  \subfloat{\includegraphics[scale = 0.12, angle =90 ]{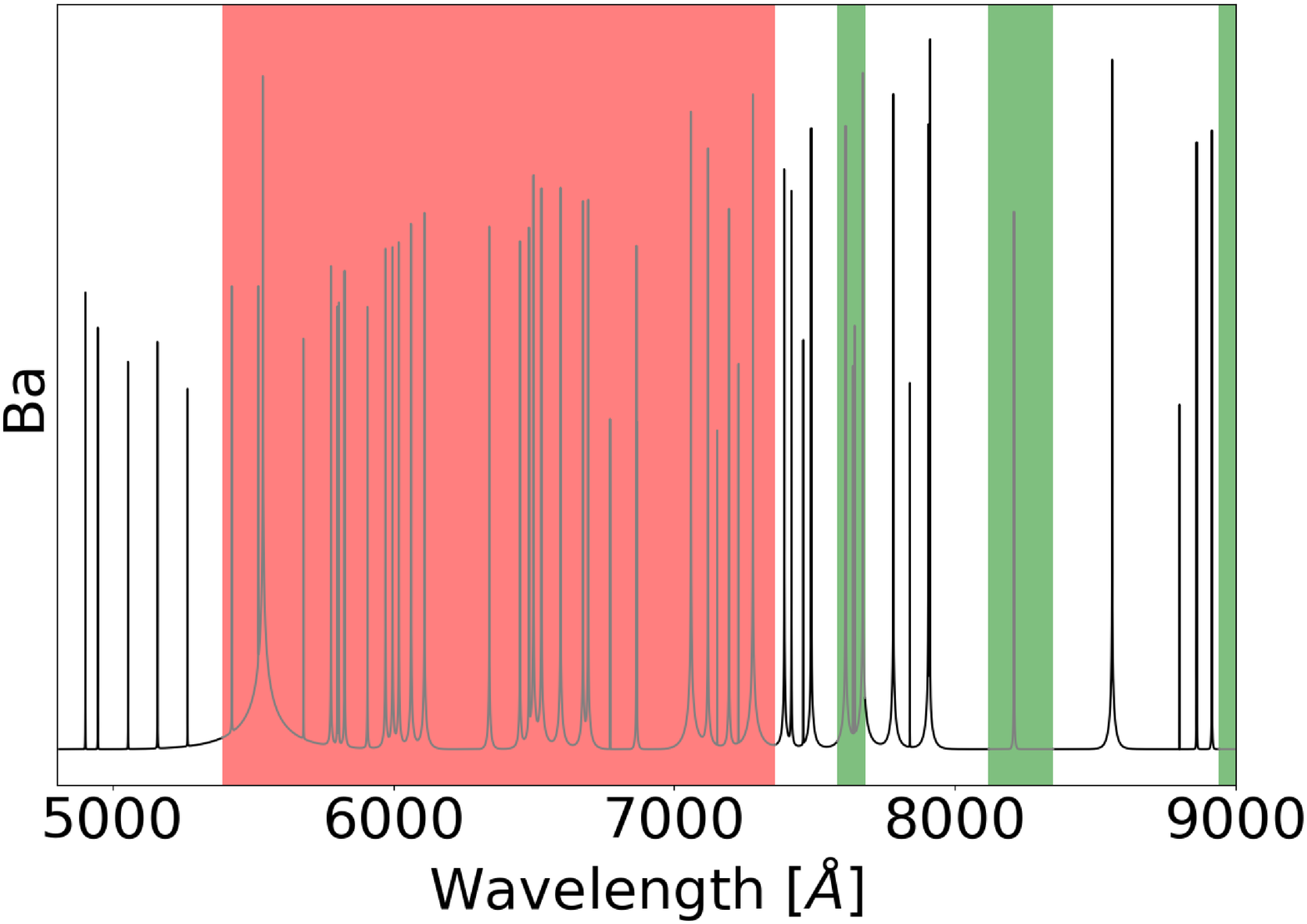}}%
  \qquad
  \subfloat{\includegraphics[scale = 0.12, angle =90 ]{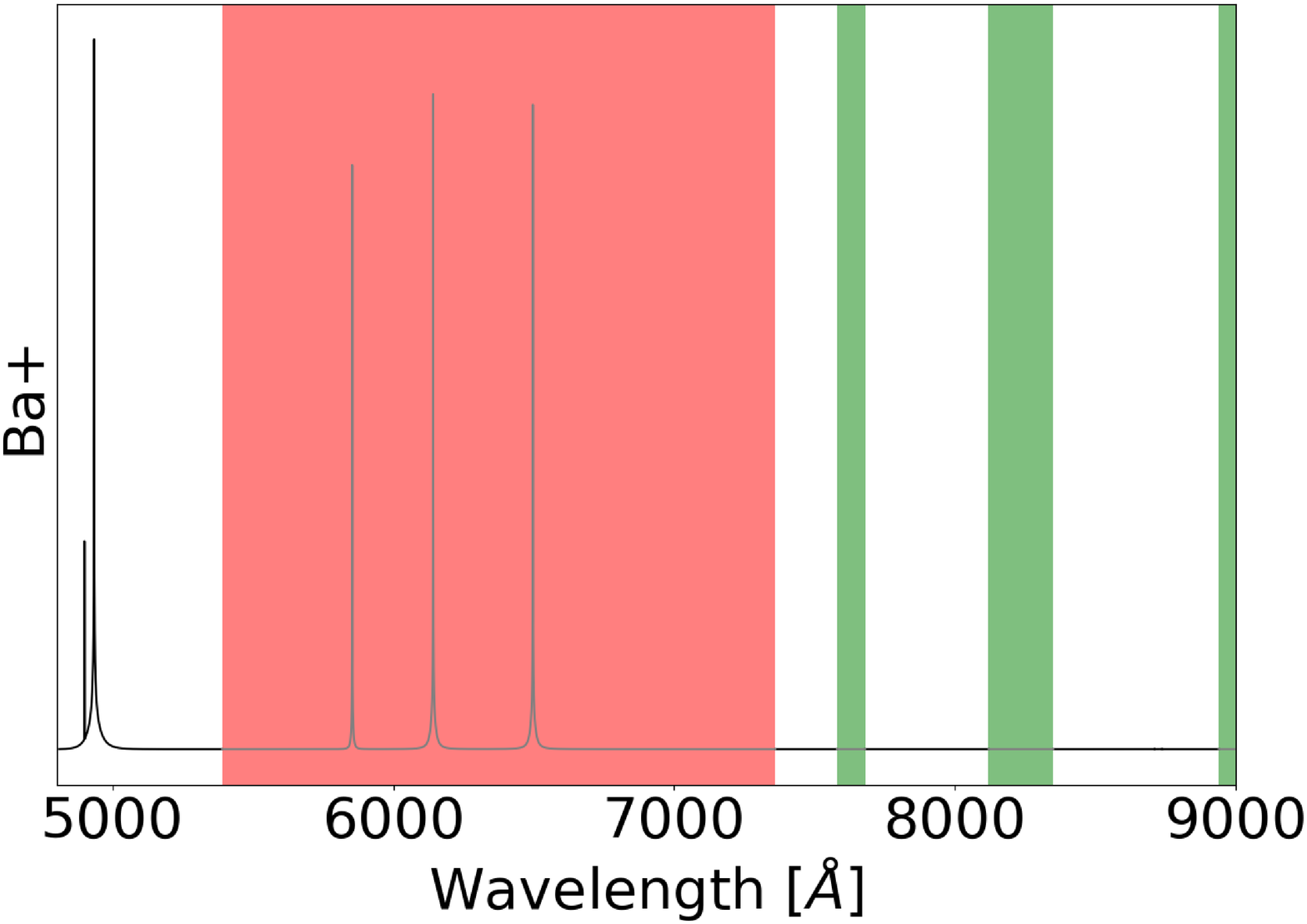}}%
  \qquad
  \subfloat{\includegraphics[scale = 0.12, angle =90 ]{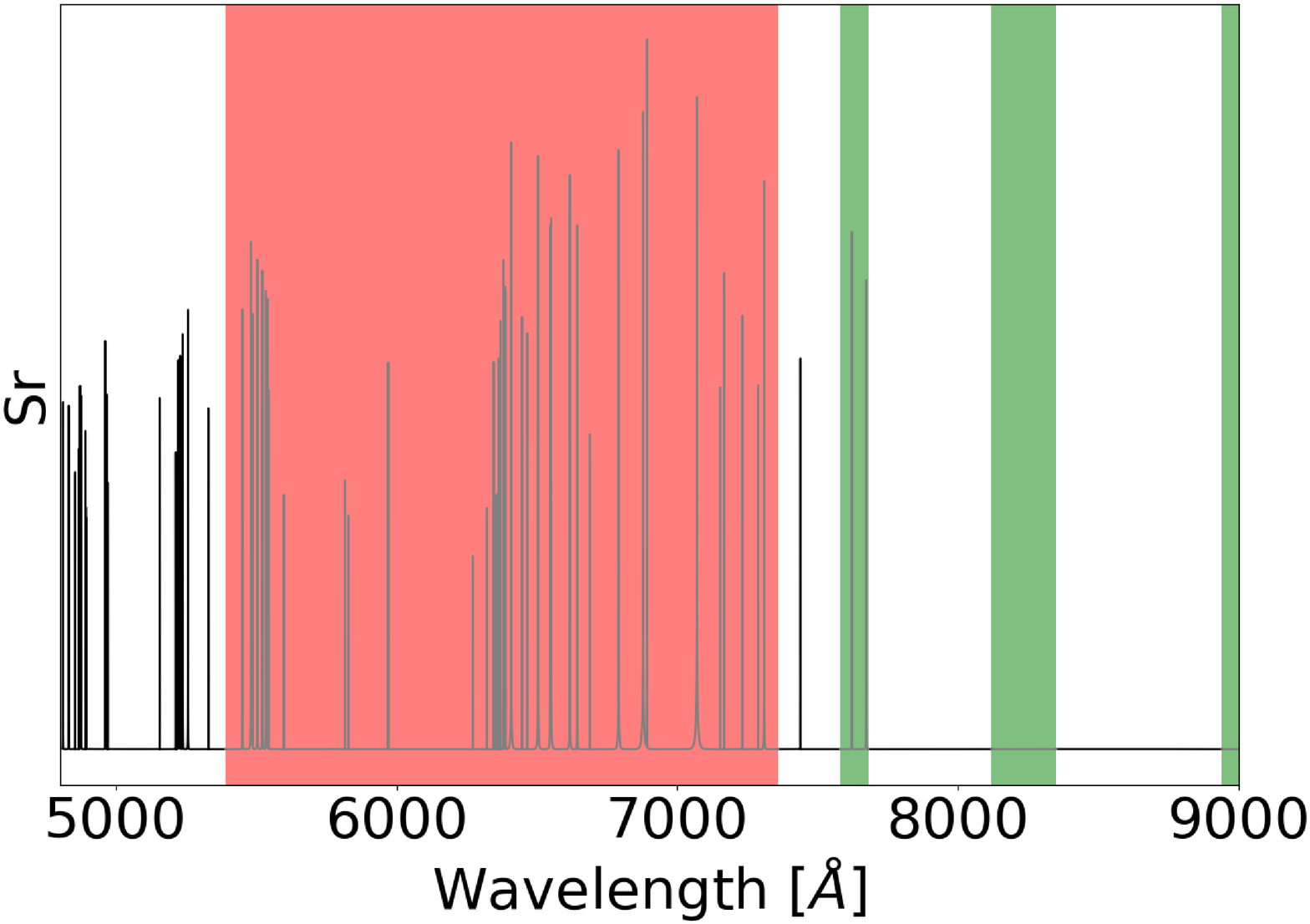}}%
  \caption{Same as Figure~\ref{fig:A1}}%
  \label{fig:A3}
\end{figure*}%

\begin{figure*}%
  \centering
  \subfloat{\includegraphics[scale = 0.12, angle =90 ]{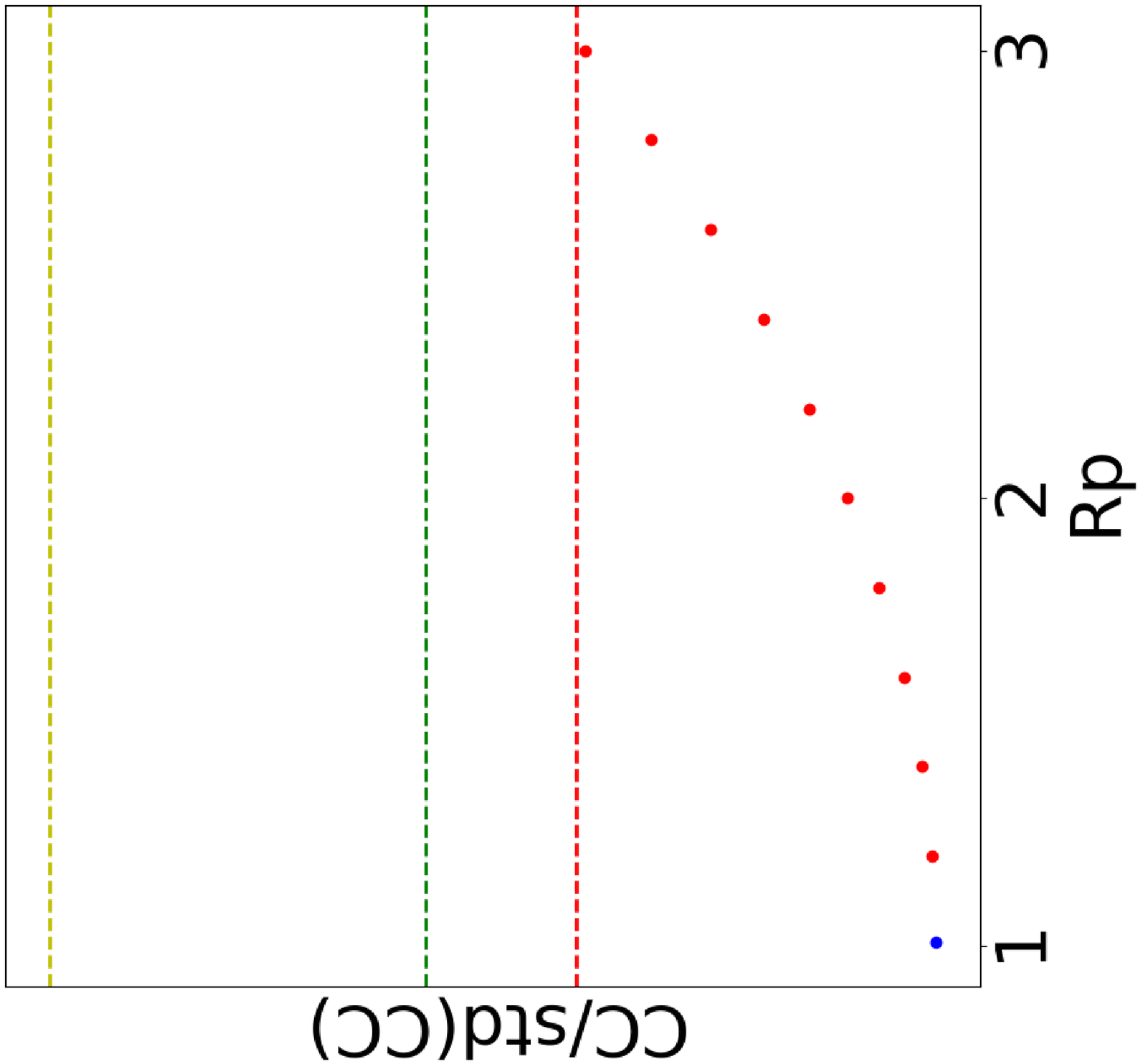}}%
  \qquad
  \subfloat{\includegraphics[scale = 0.12, angle =90 ]{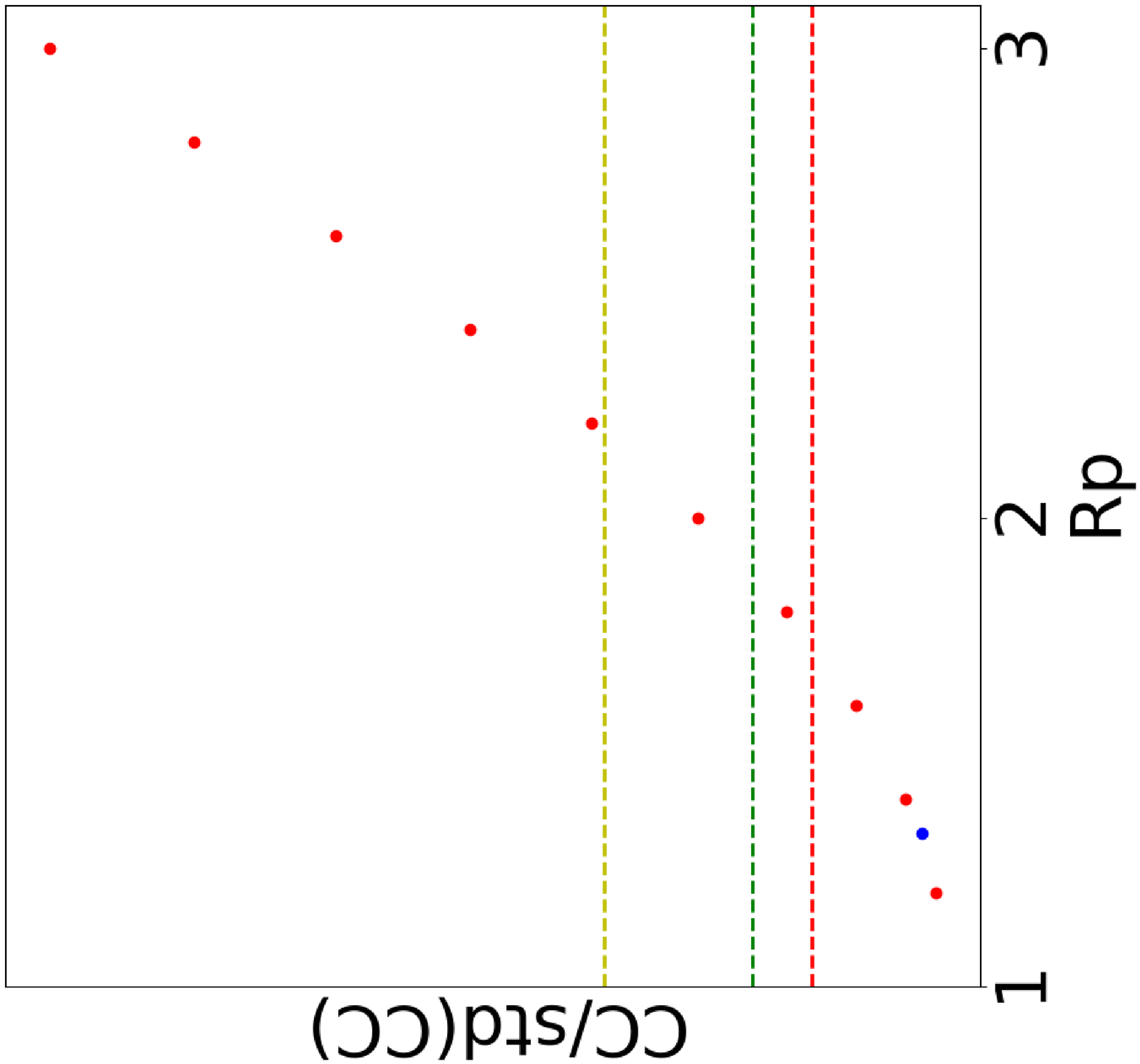}}%
  \qquad
  \subfloat{\includegraphics[scale = 0.12, angle =90 ]{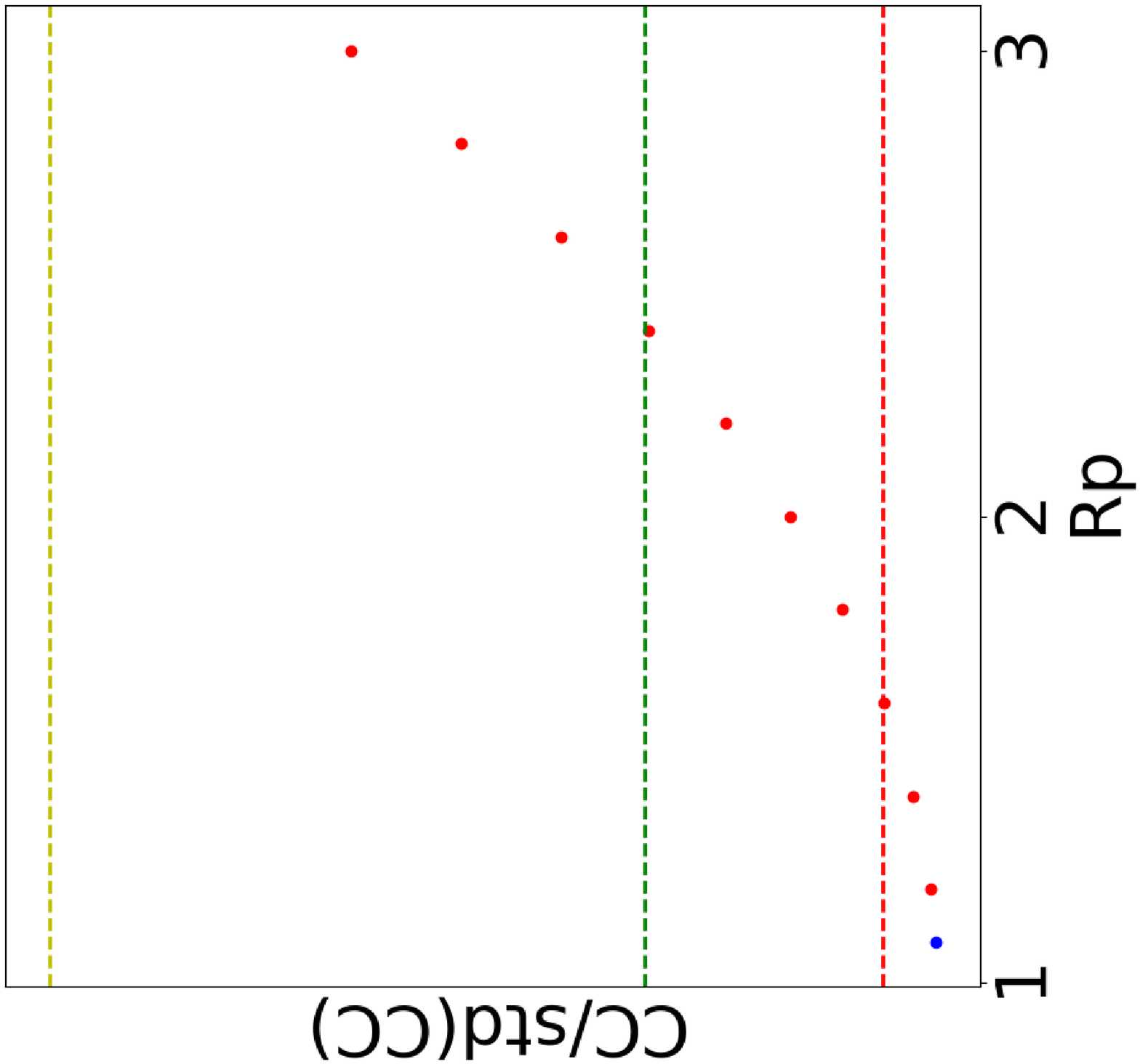}}%
  \qquad
  \subfloat{\includegraphics[scale = 0.12, angle =90 ]{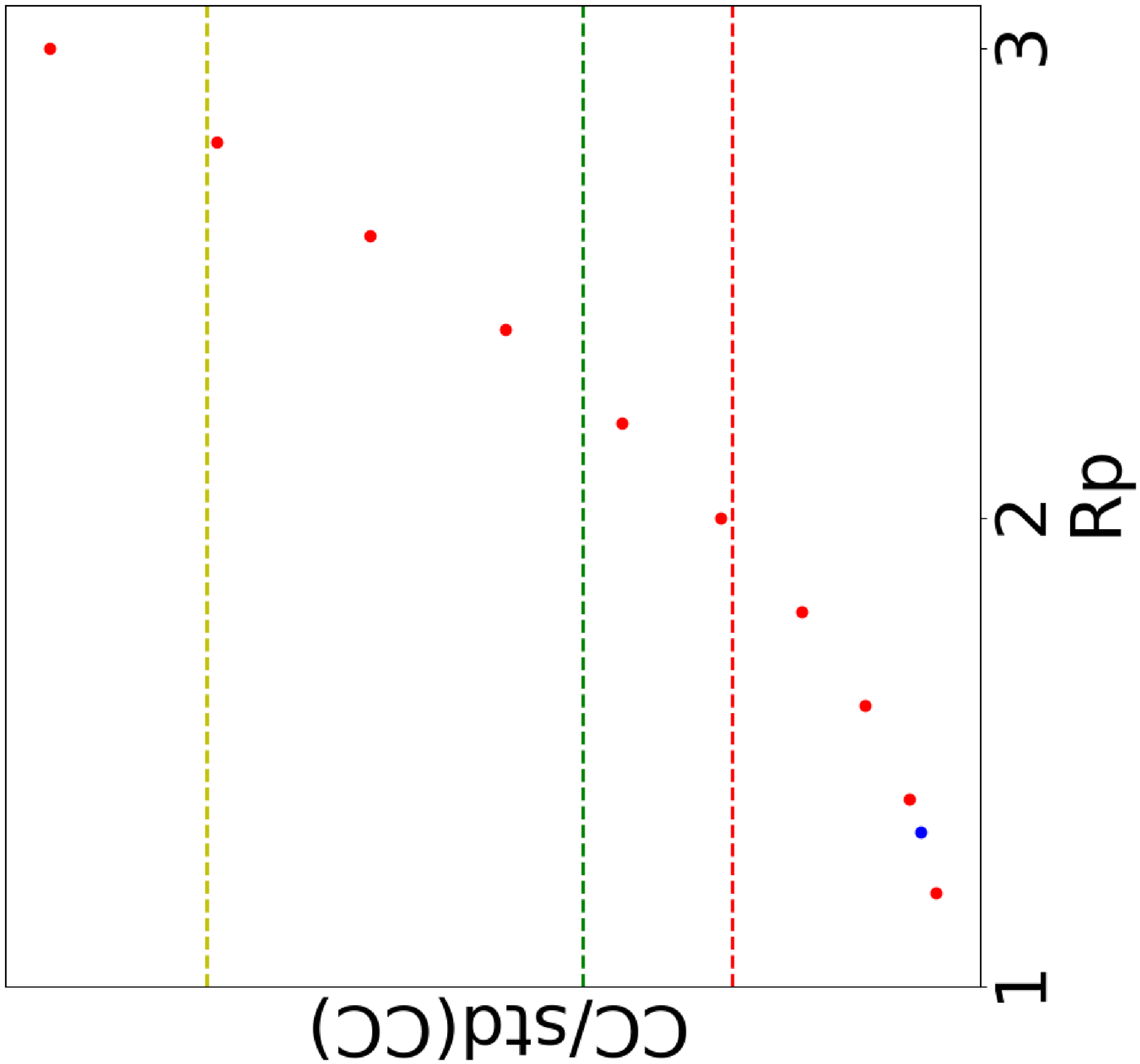}}%
  \qquad
  \subfloat{\includegraphics[scale = 0.12, angle =90 ]{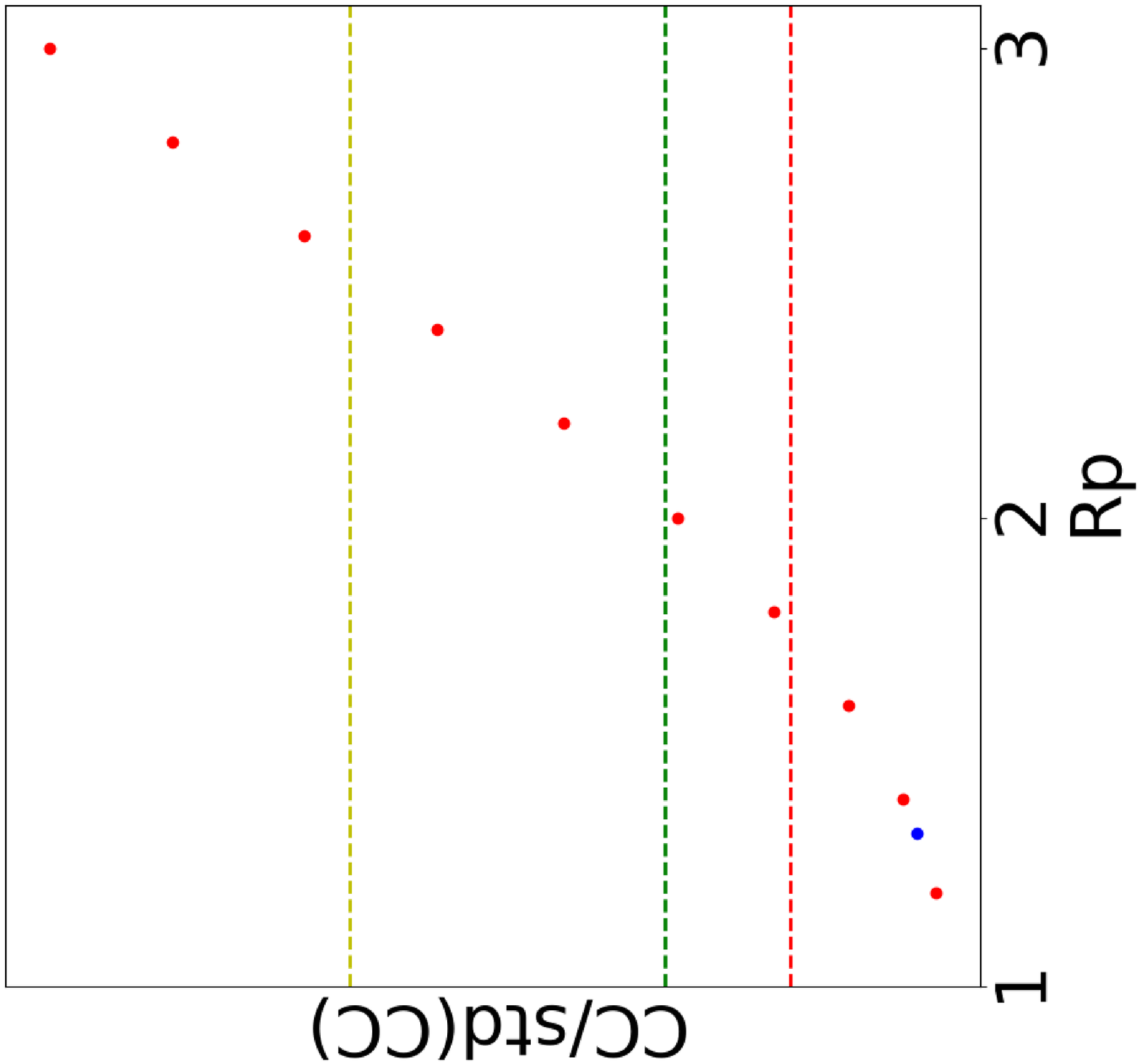}}%
  \\
  \vspace{0.5cm}
  \centering
  \subfloat{\includegraphics[scale = 0.12, angle =90 ]{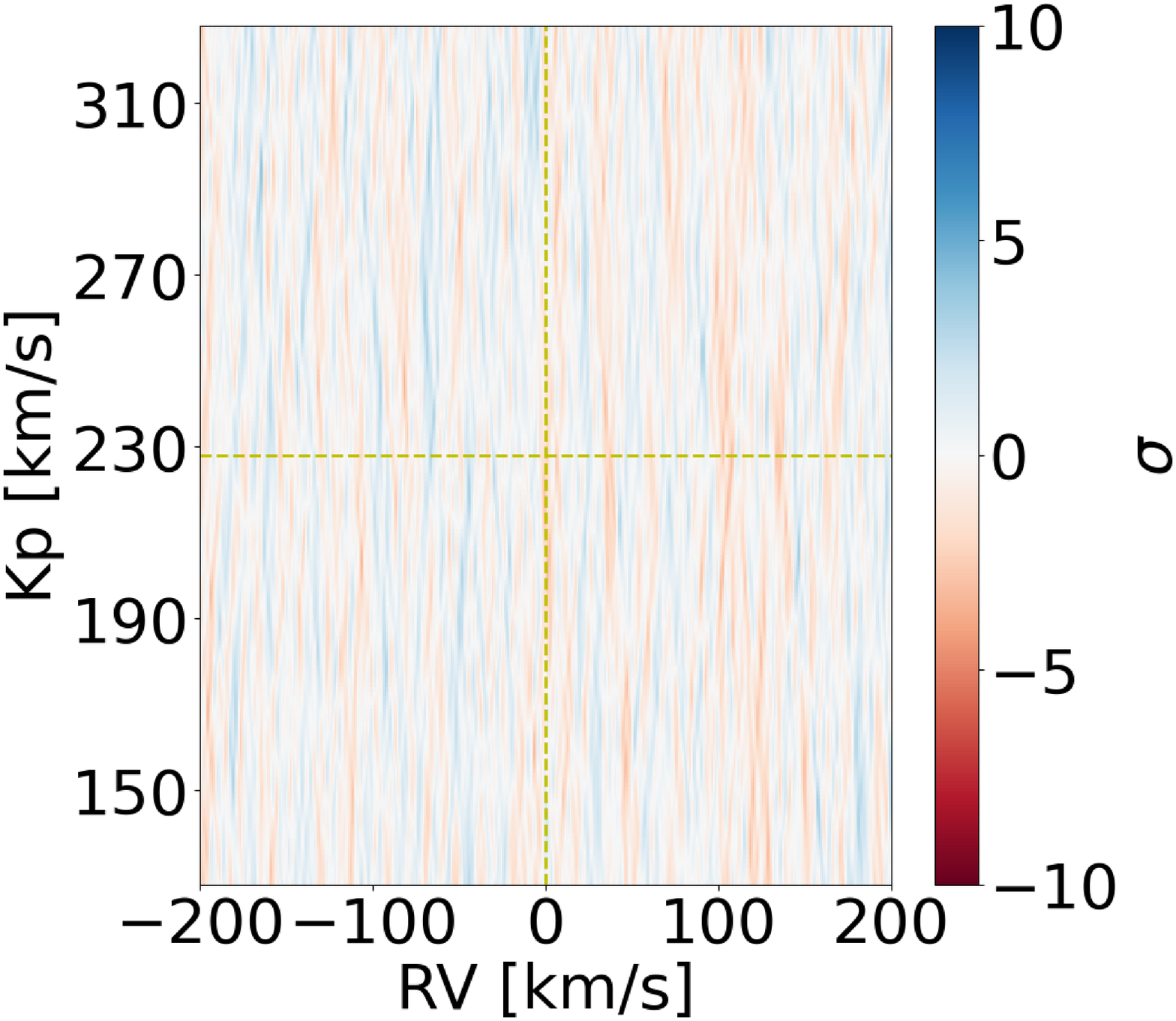}}%
  \qquad
  \subfloat{\includegraphics[scale = 0.12, angle =90 ]{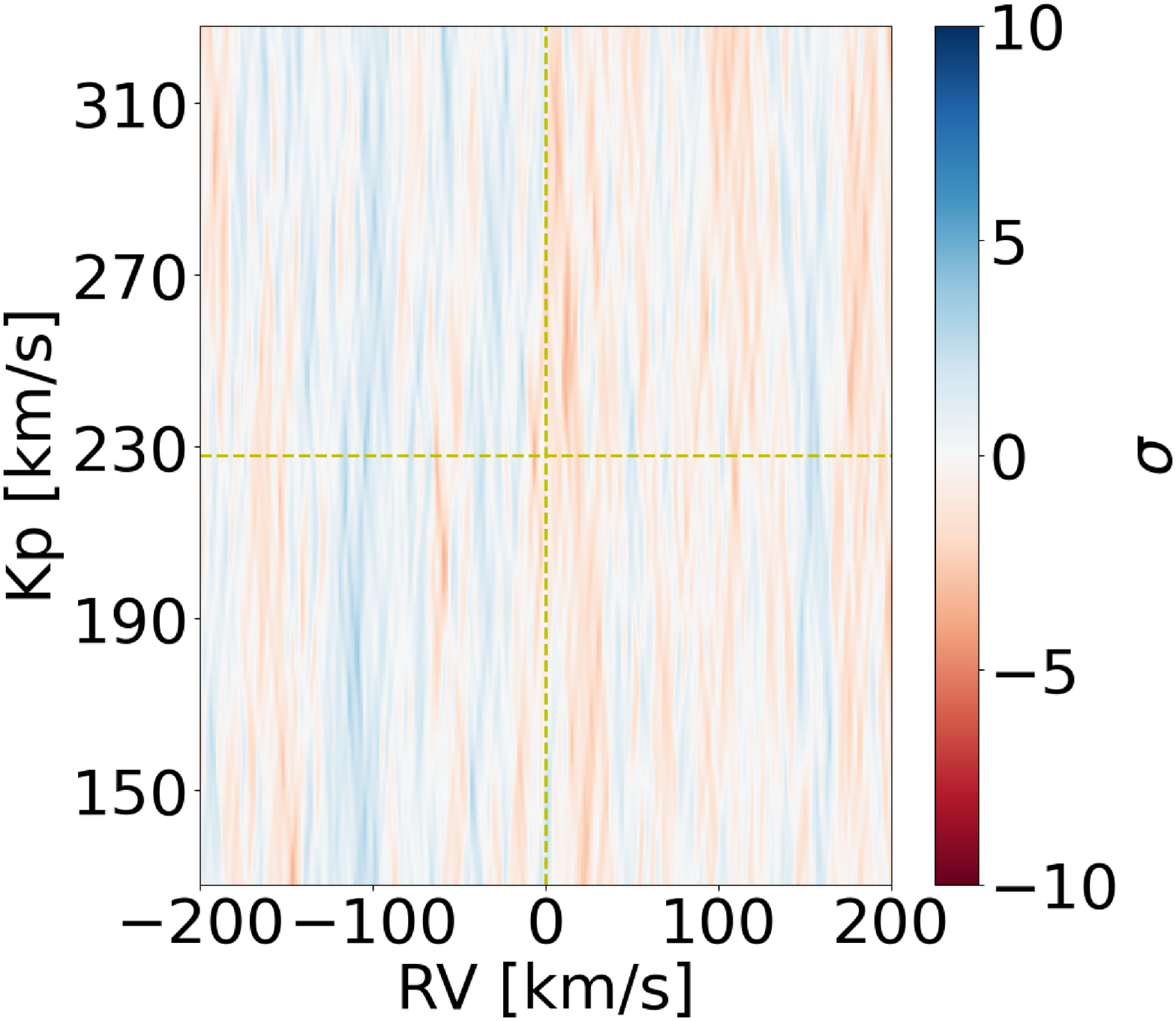}}%
  \qquad
  \subfloat{\includegraphics[scale = 0.12, angle =90 ]{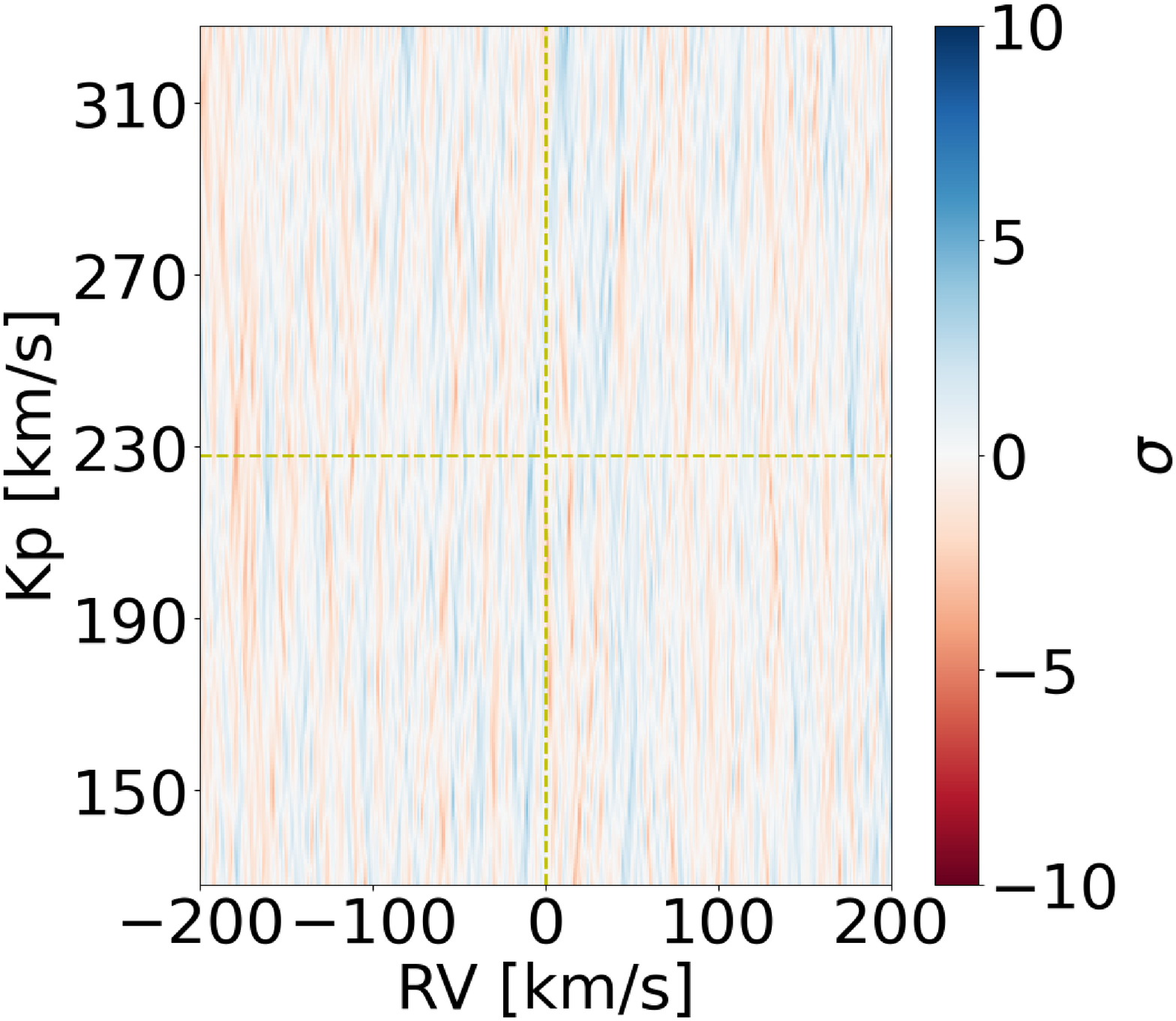}}%
  \qquad
  \subfloat{\includegraphics[scale = 0.12, angle =90 ]{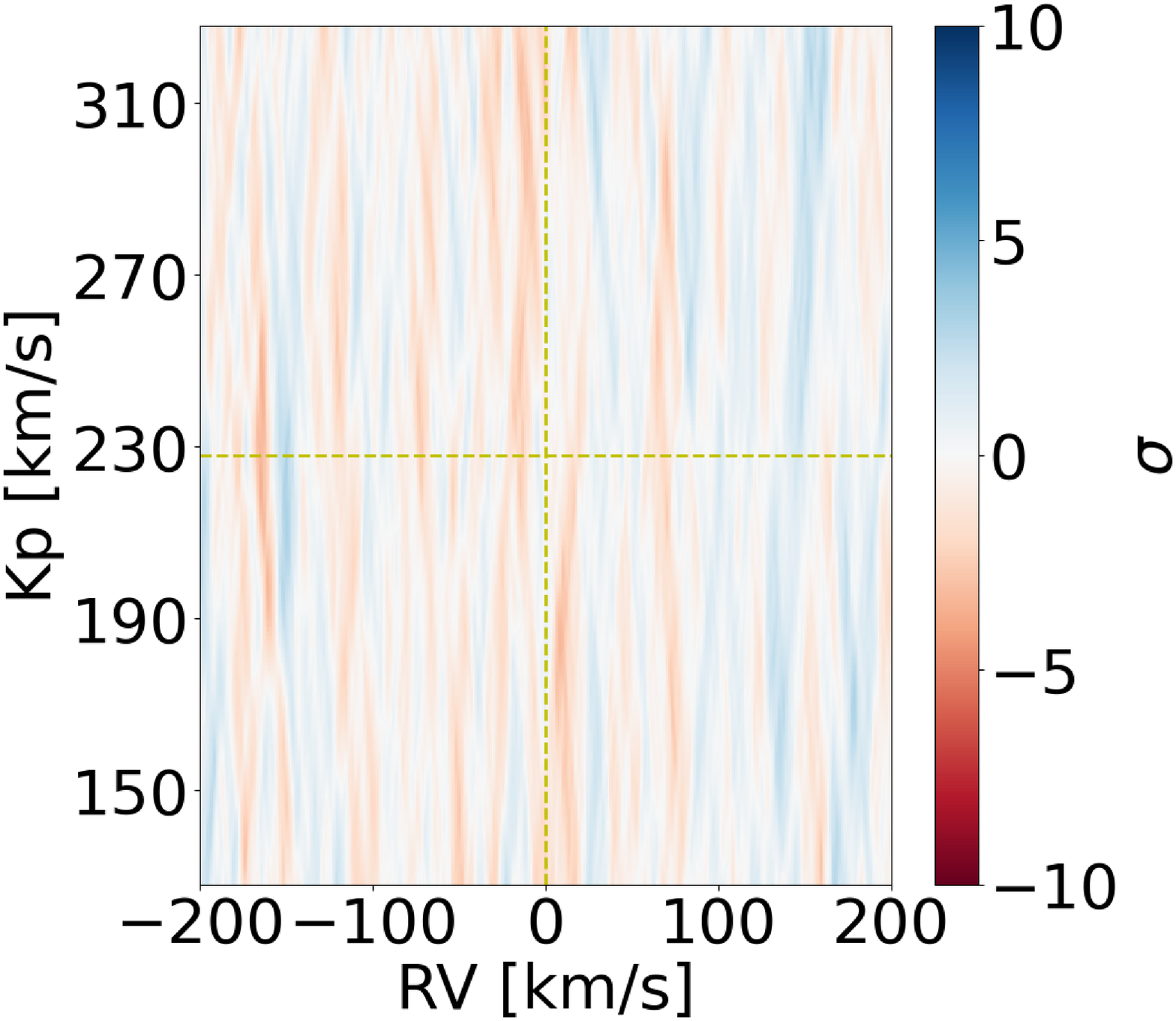}}%
  \qquad
  \subfloat{\includegraphics[scale = 0.12, angle =90 ]{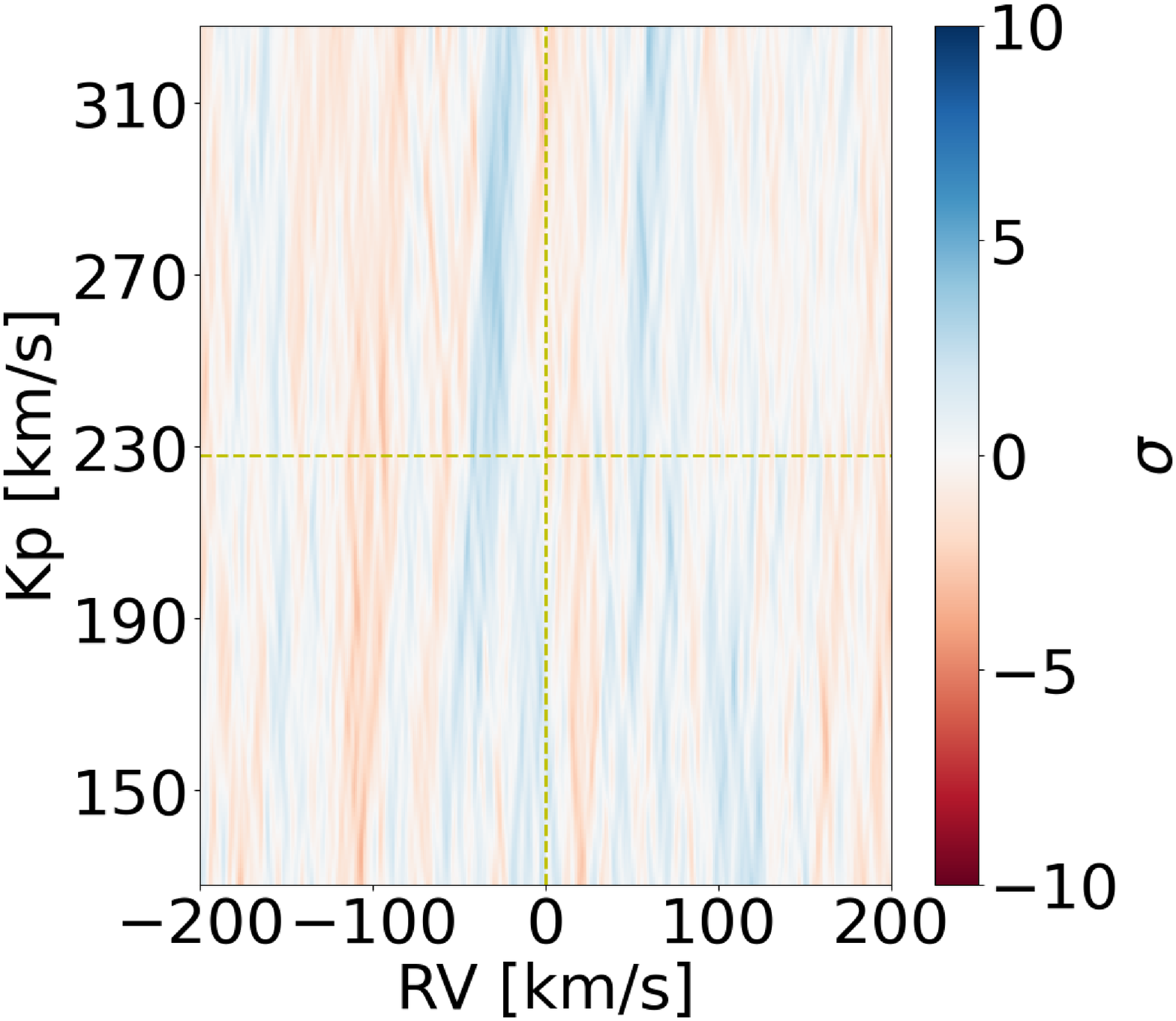}}%
  \\
  \vspace{0.5cm}
  \centering
  \subfloat{\includegraphics[scale = 0.12, angle =90 ]{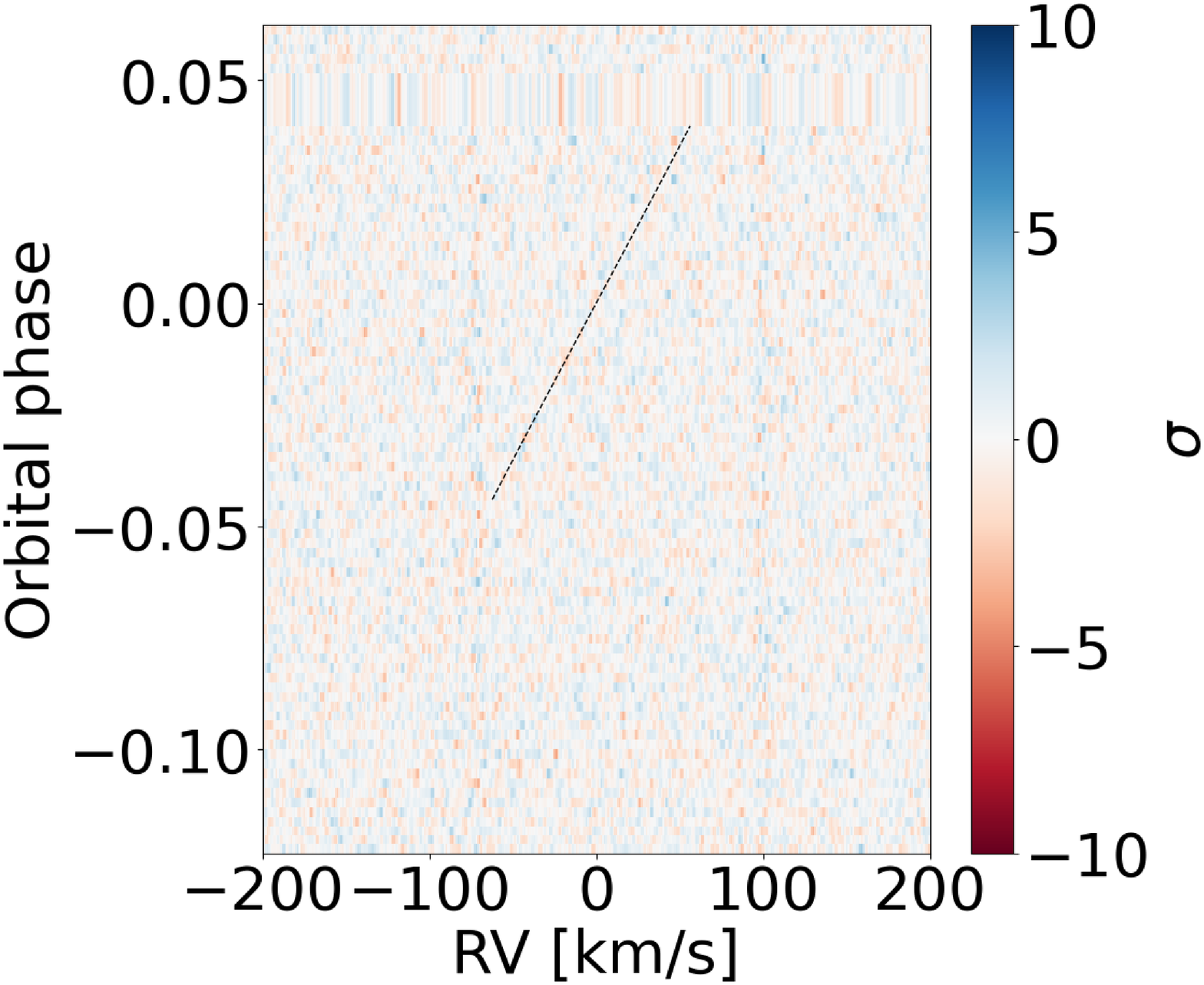}}%
  \qquad
  \subfloat{\includegraphics[scale = 0.12, angle =90 ]{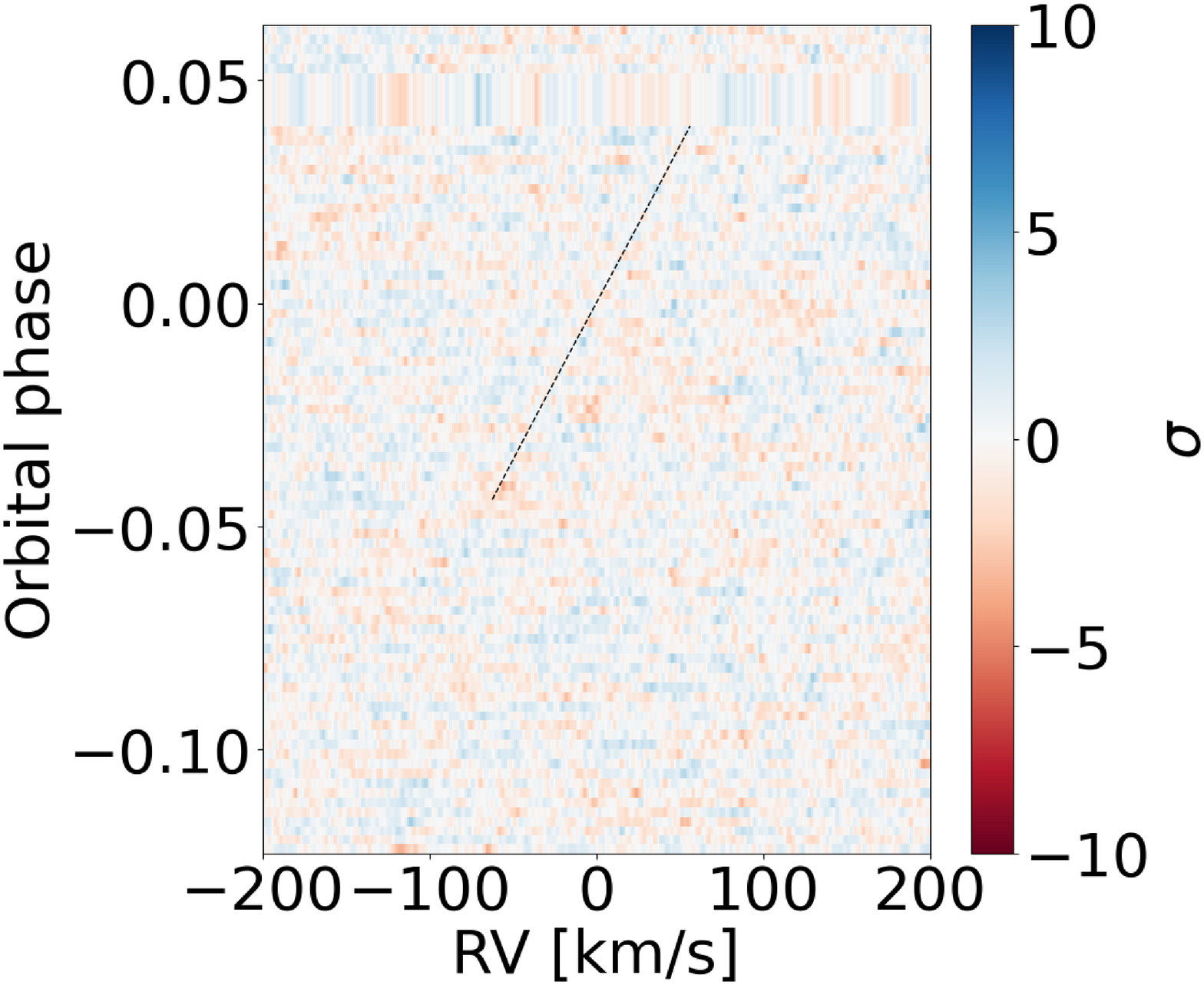}}%
  \qquad
  \subfloat{\includegraphics[scale = 0.12, angle =90 ]{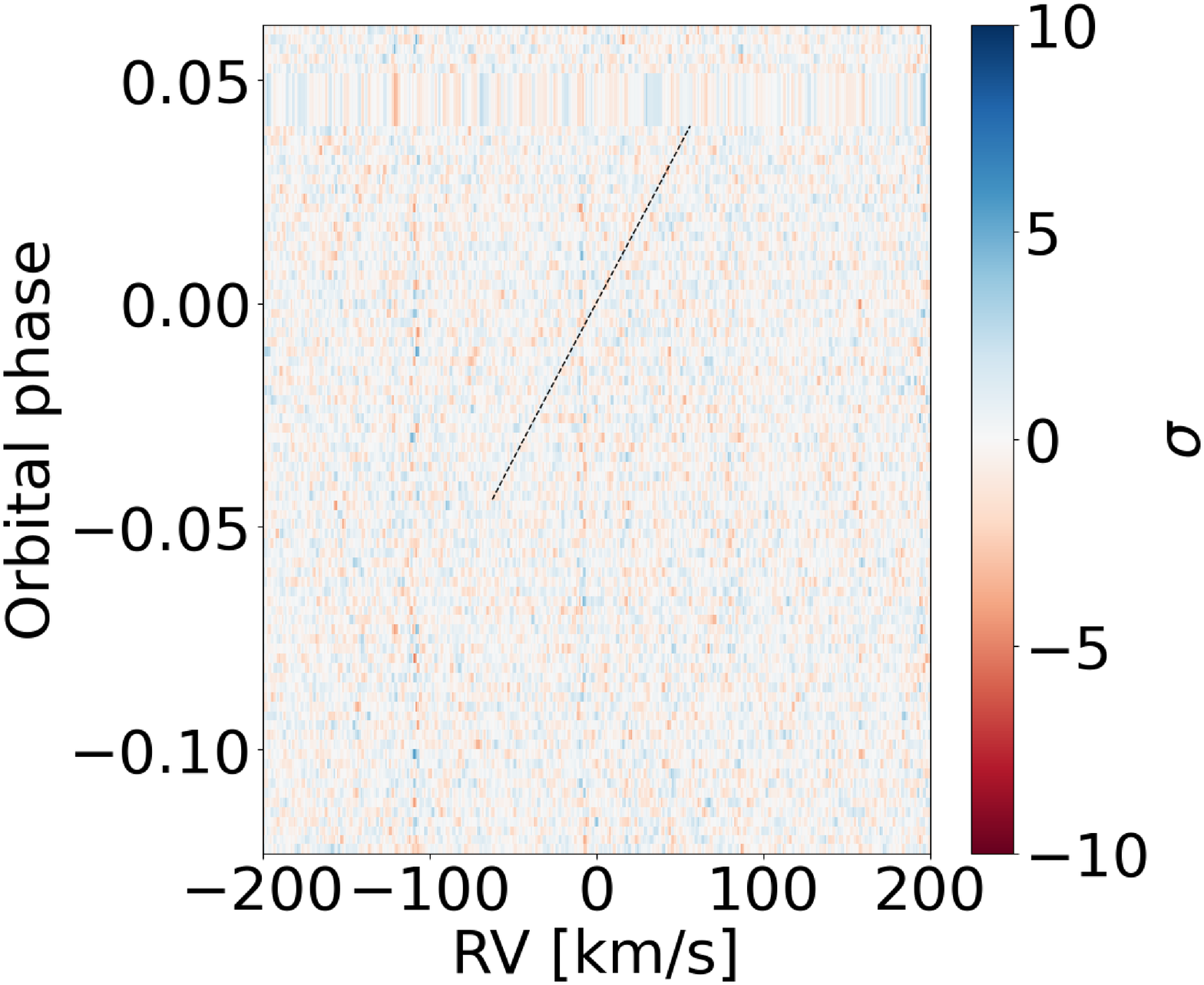}}%
  \qquad
  \subfloat{\includegraphics[scale = 0.12, angle =90 ]{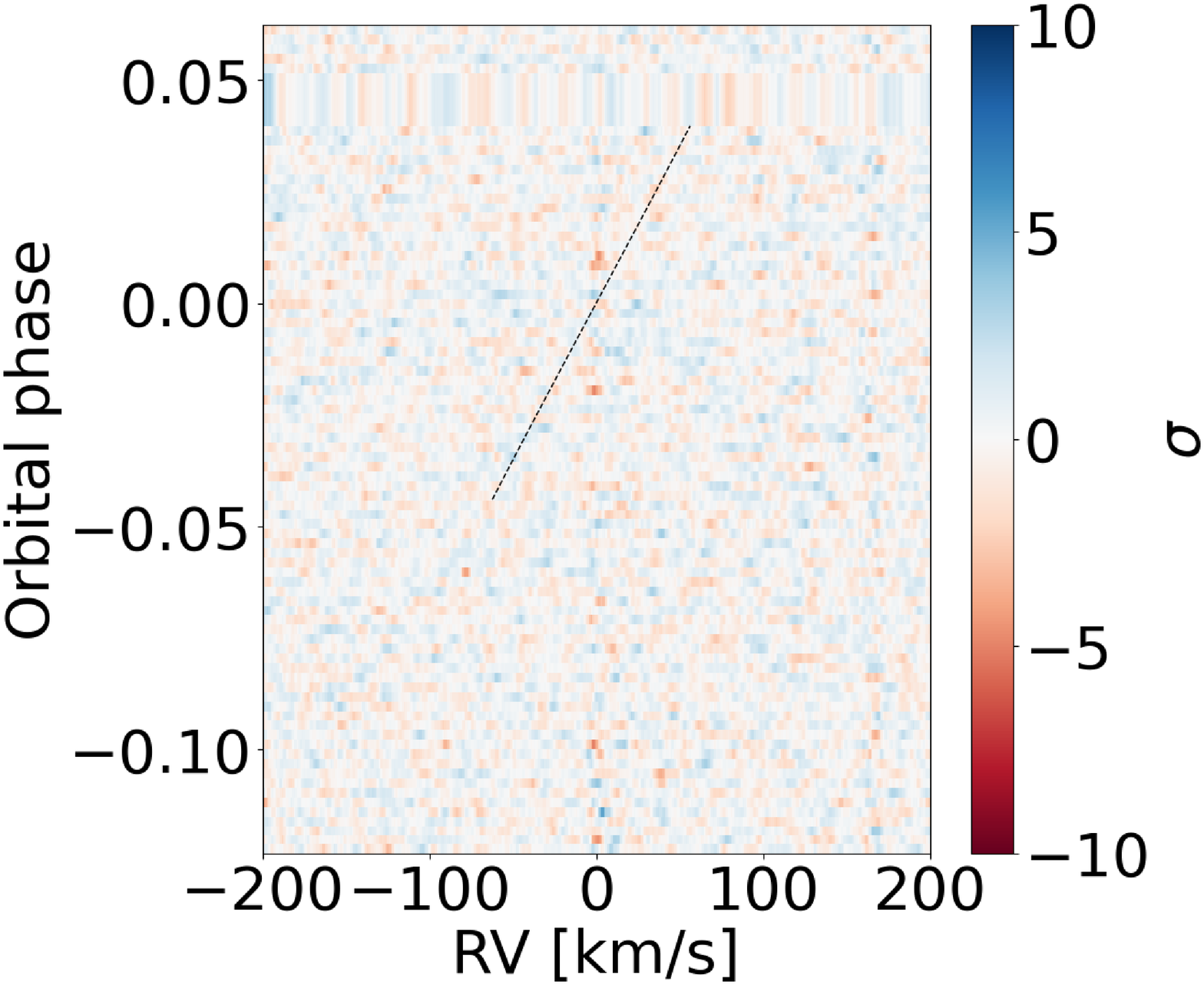}}%
  \qquad
  \subfloat{\includegraphics[scale = 0.12, angle =90 ]{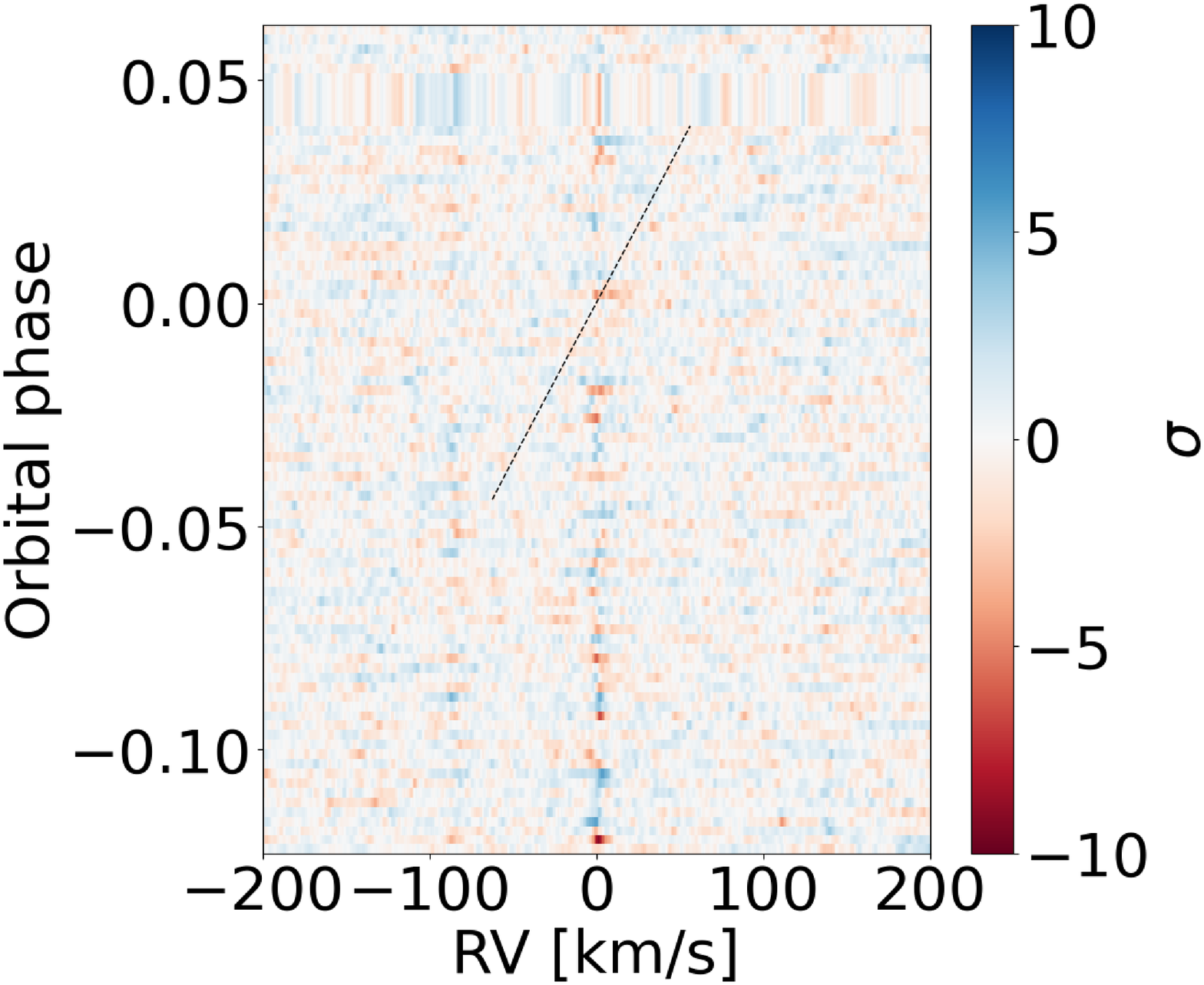}}%
  \\
  \vspace{0.5cm}
  \centering
  \subfloat{\includegraphics[scale = 0.12, angle =90 ]{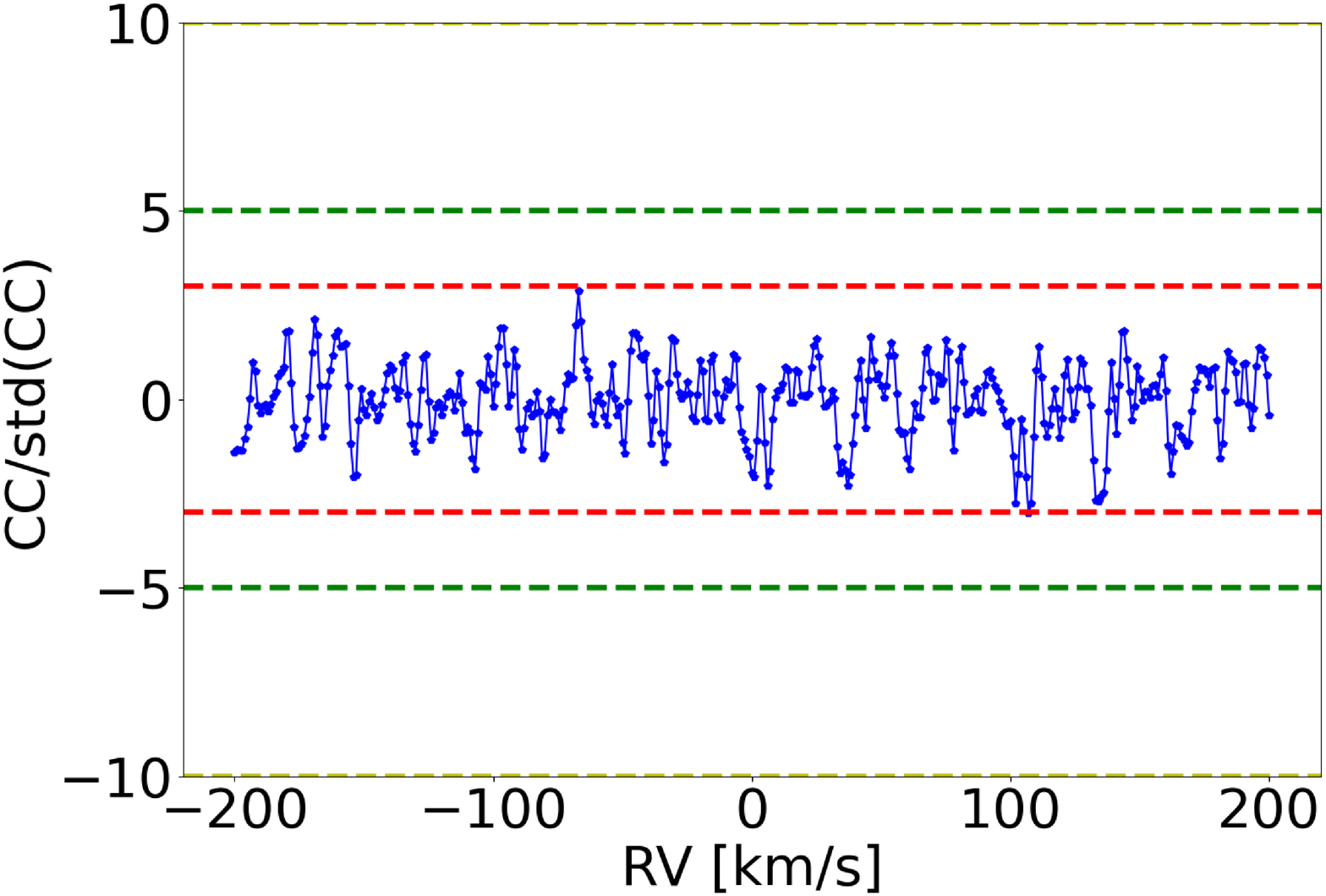}}%
  \qquad
  \subfloat{\includegraphics[scale = 0.12, angle =90 ]{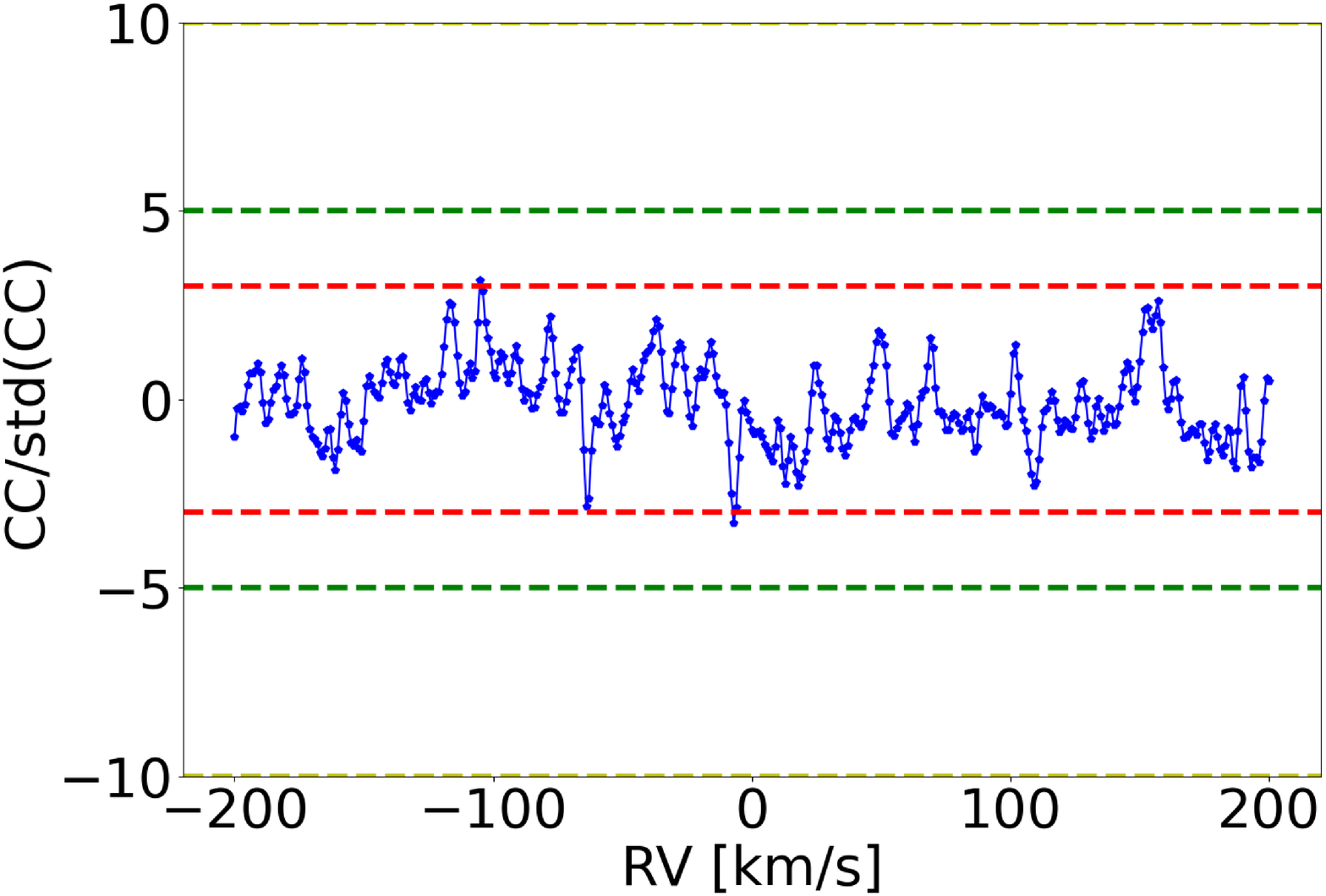}}%
  \qquad
  \subfloat{\includegraphics[scale = 0.12, angle =90 ]{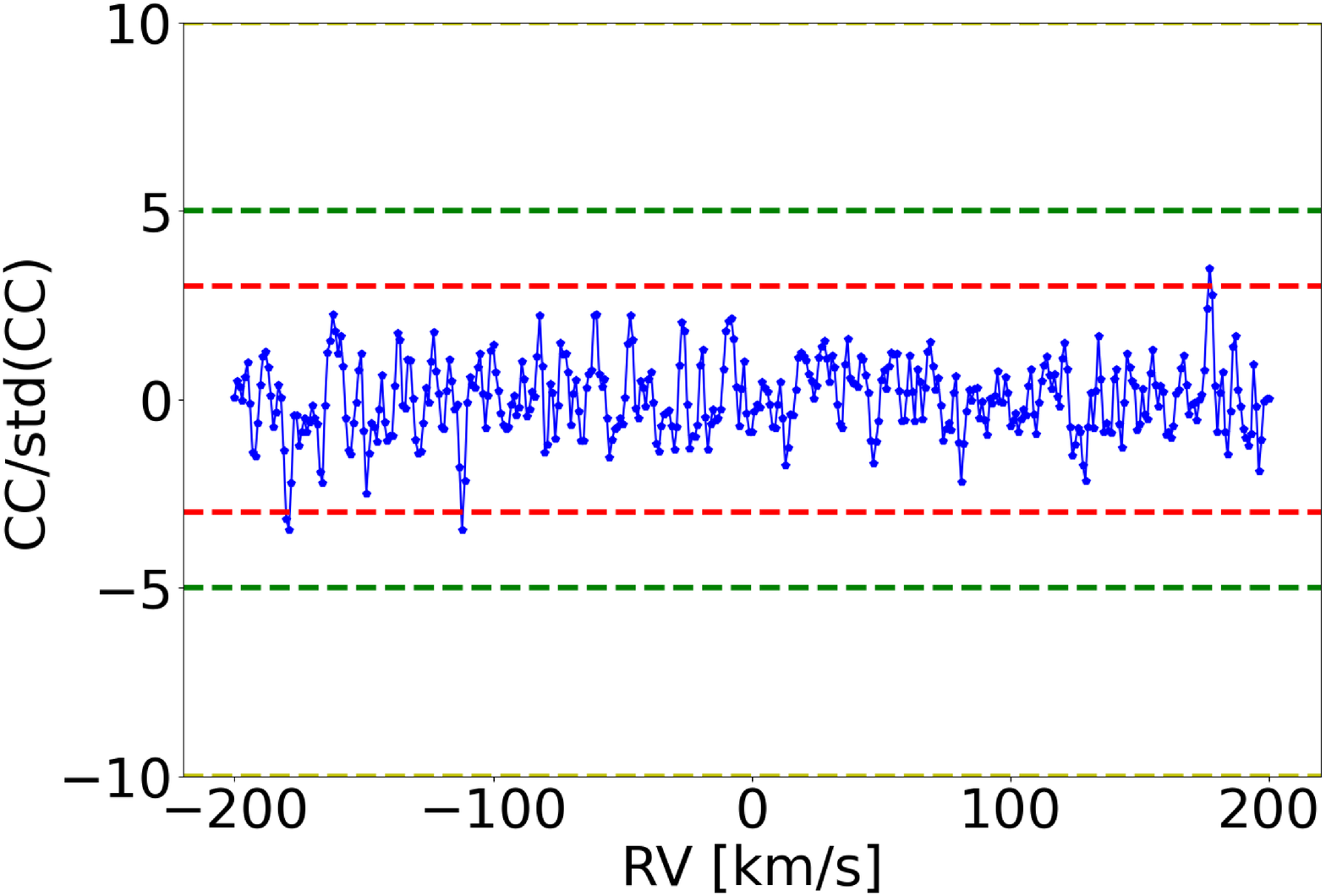}}%
  \qquad
  \subfloat{\includegraphics[scale = 0.12, angle =90 ]{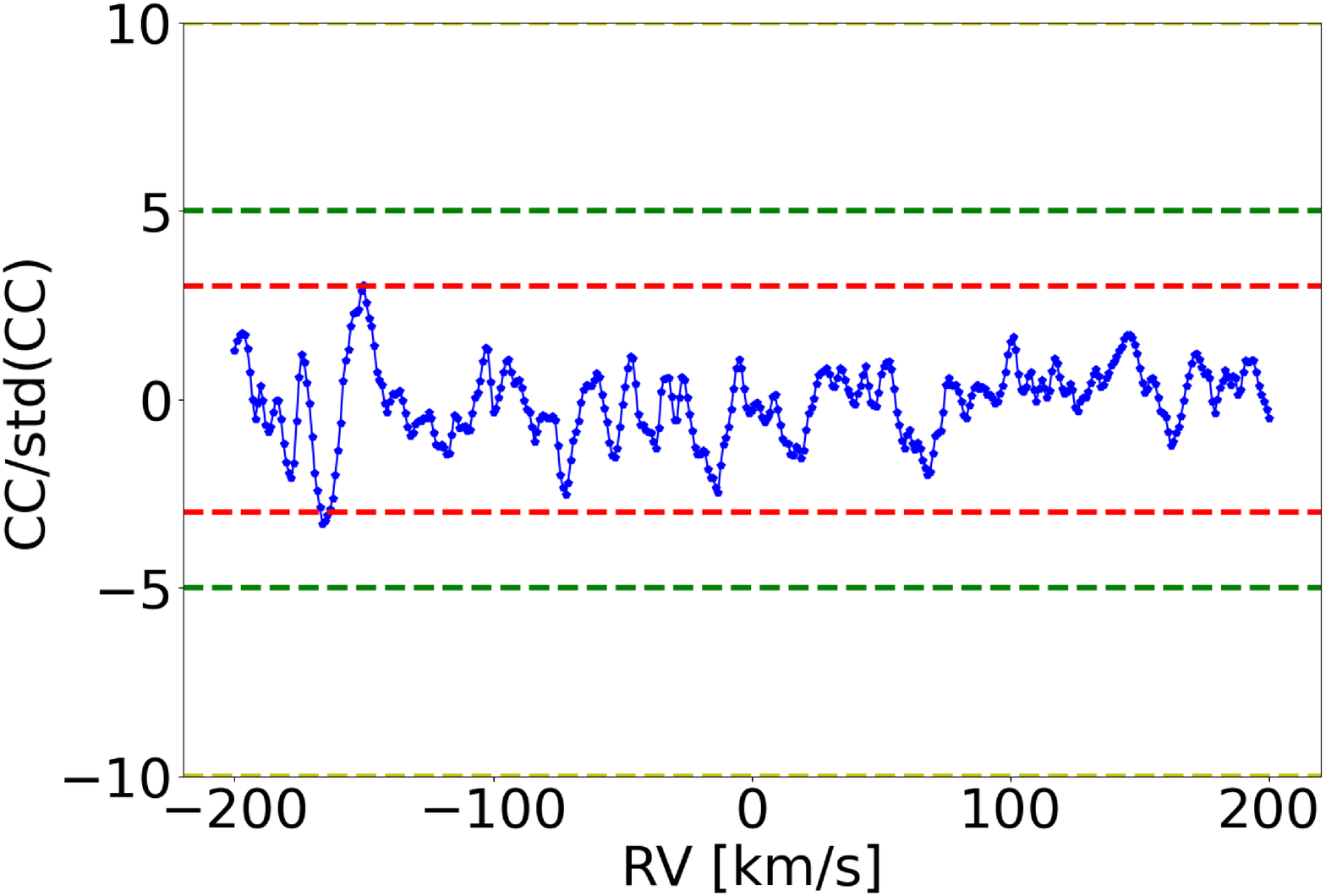}}%
  \qquad
  \subfloat{\includegraphics[scale = 0.12, angle =90 ]{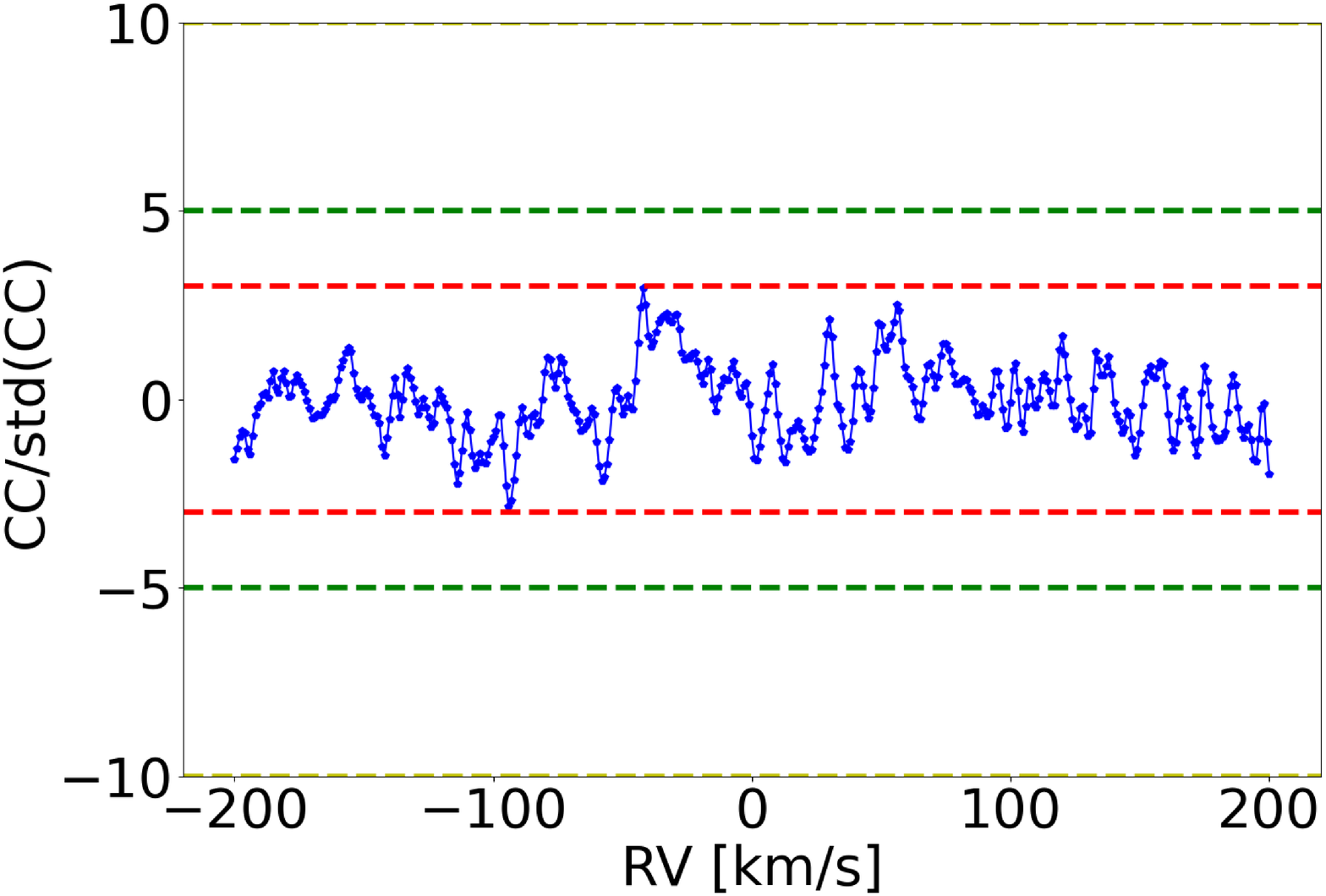}}%
  \\
  \vspace{0.5cm}
  \centering
  \subfloat{\includegraphics[scale = 0.12, angle =90 ]{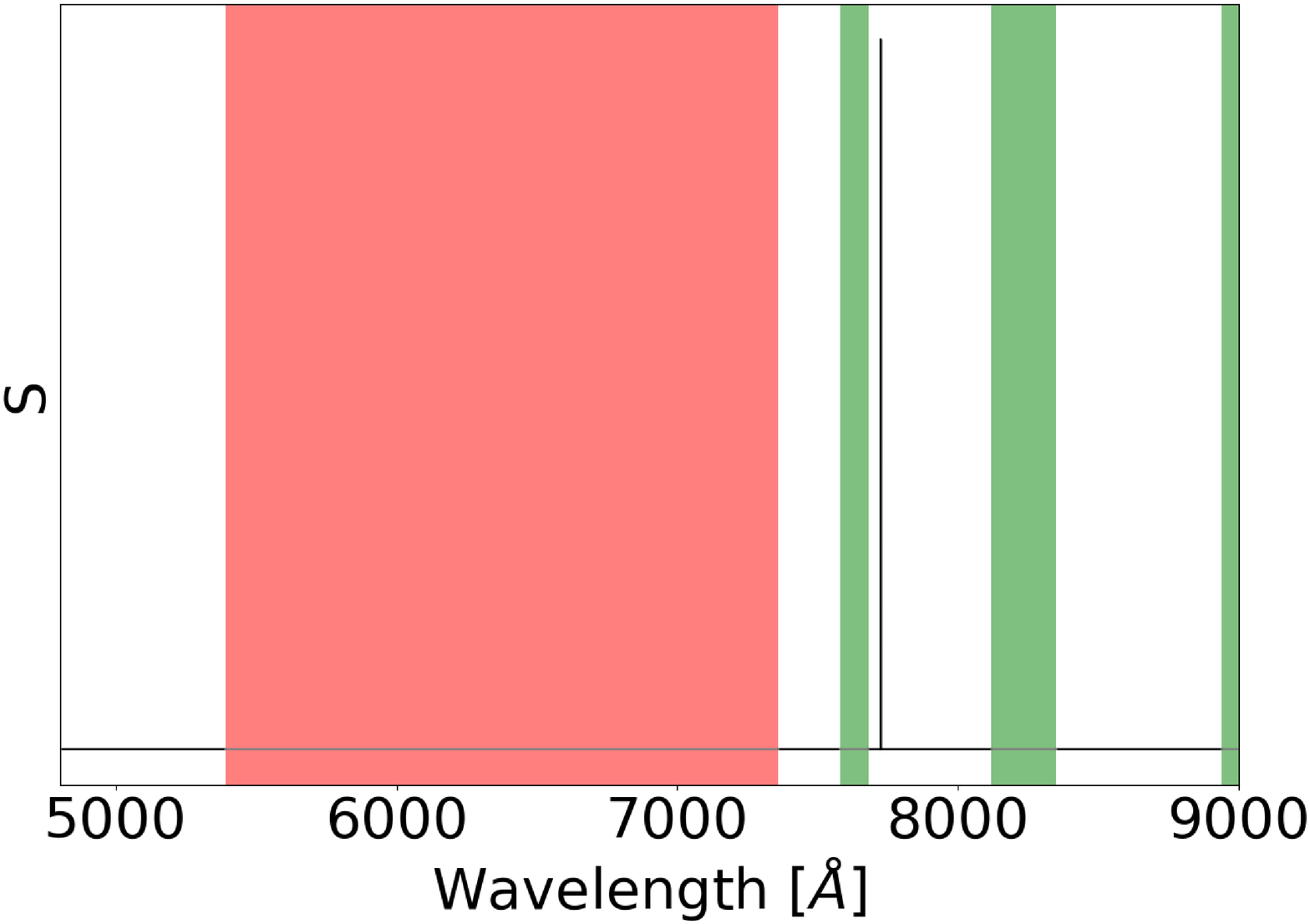}}%
  \qquad
  \subfloat{\includegraphics[scale = 0.12, angle =90 ]{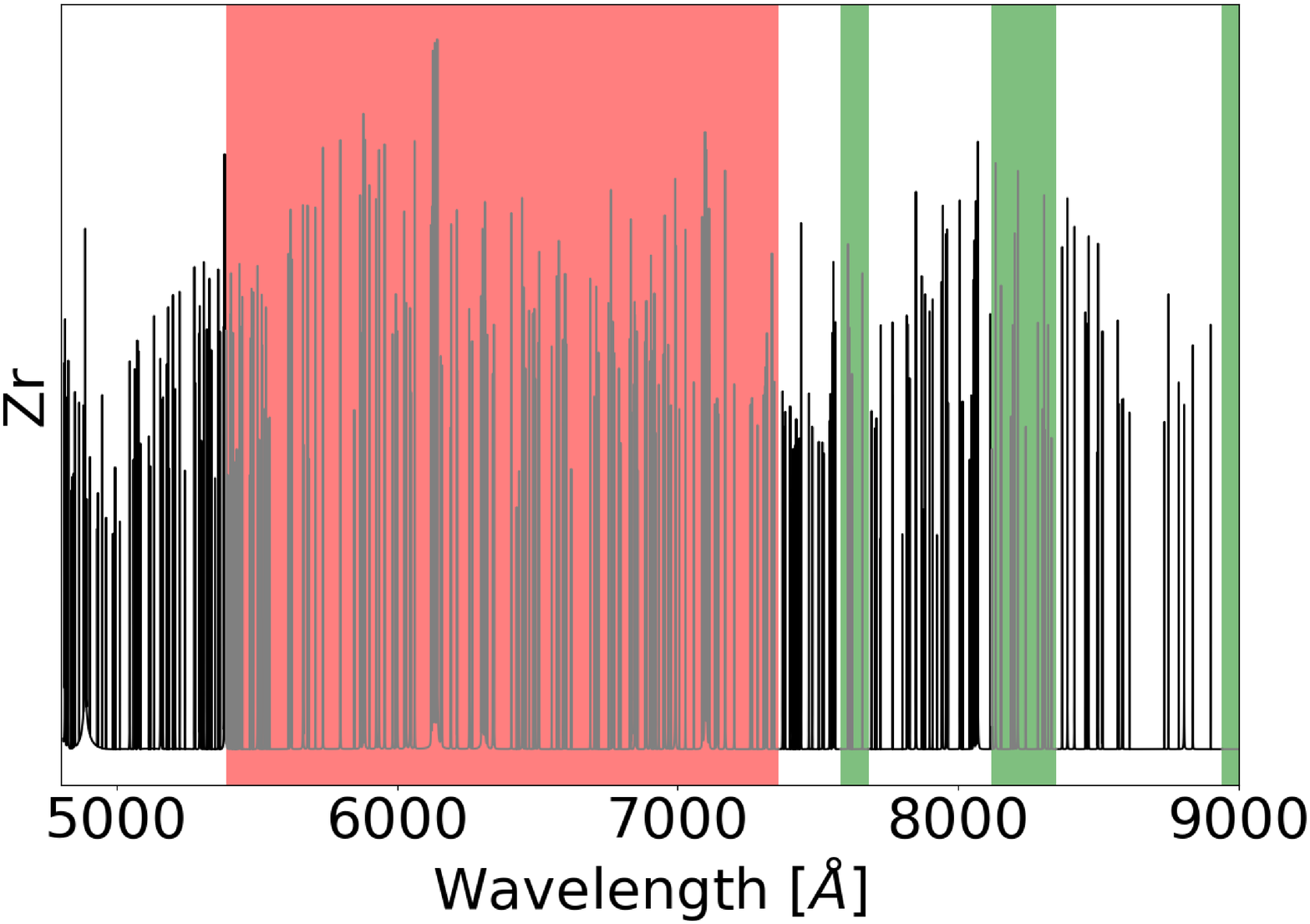}}%
  \qquad
  \subfloat{\includegraphics[scale = 0.12, angle =90 ]{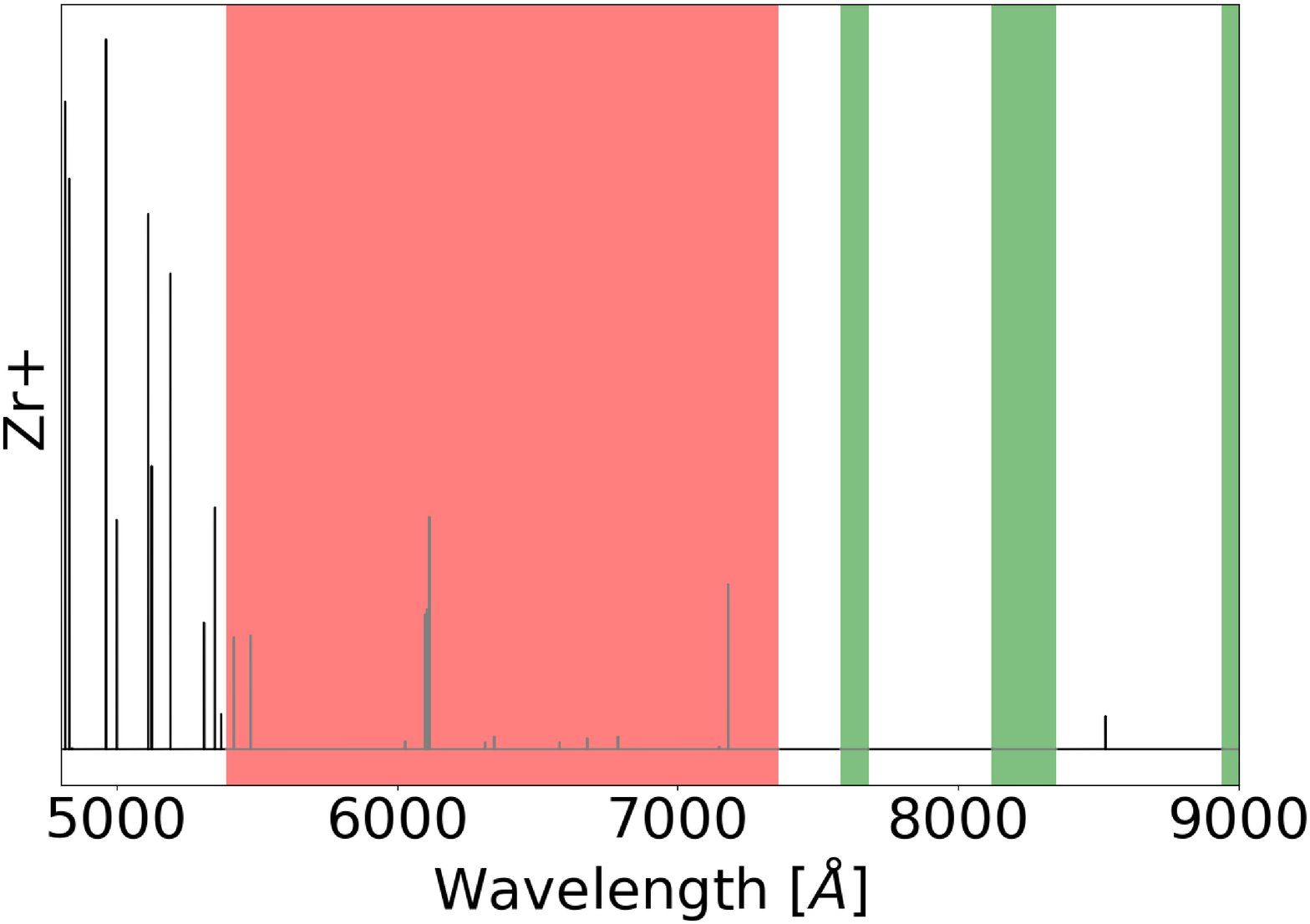}}%
  \qquad
  \subfloat{\includegraphics[scale = 0.12, angle =90 ]{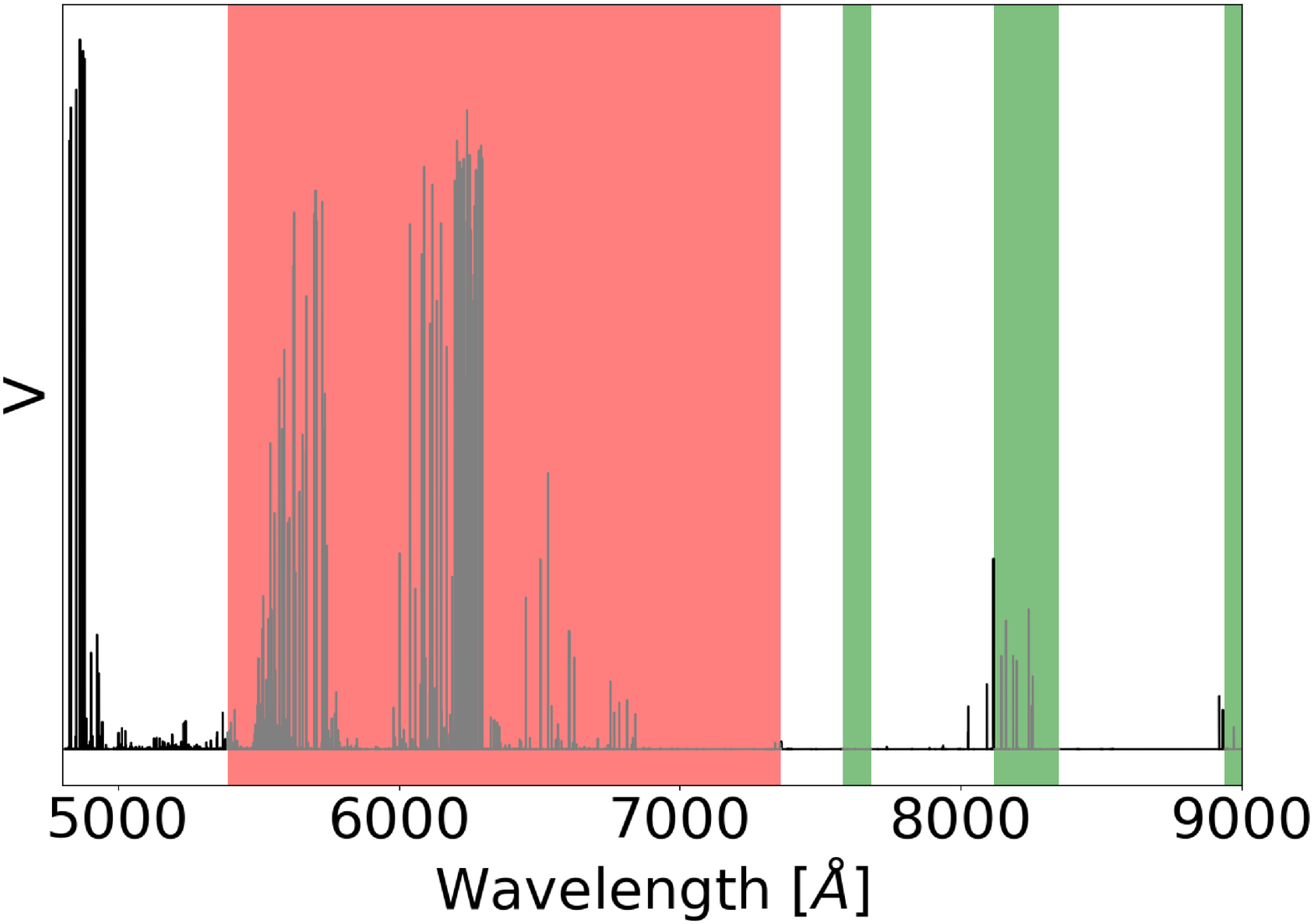}}%
  \qquad
  \subfloat{\includegraphics[scale = 0.12, angle =90 ]{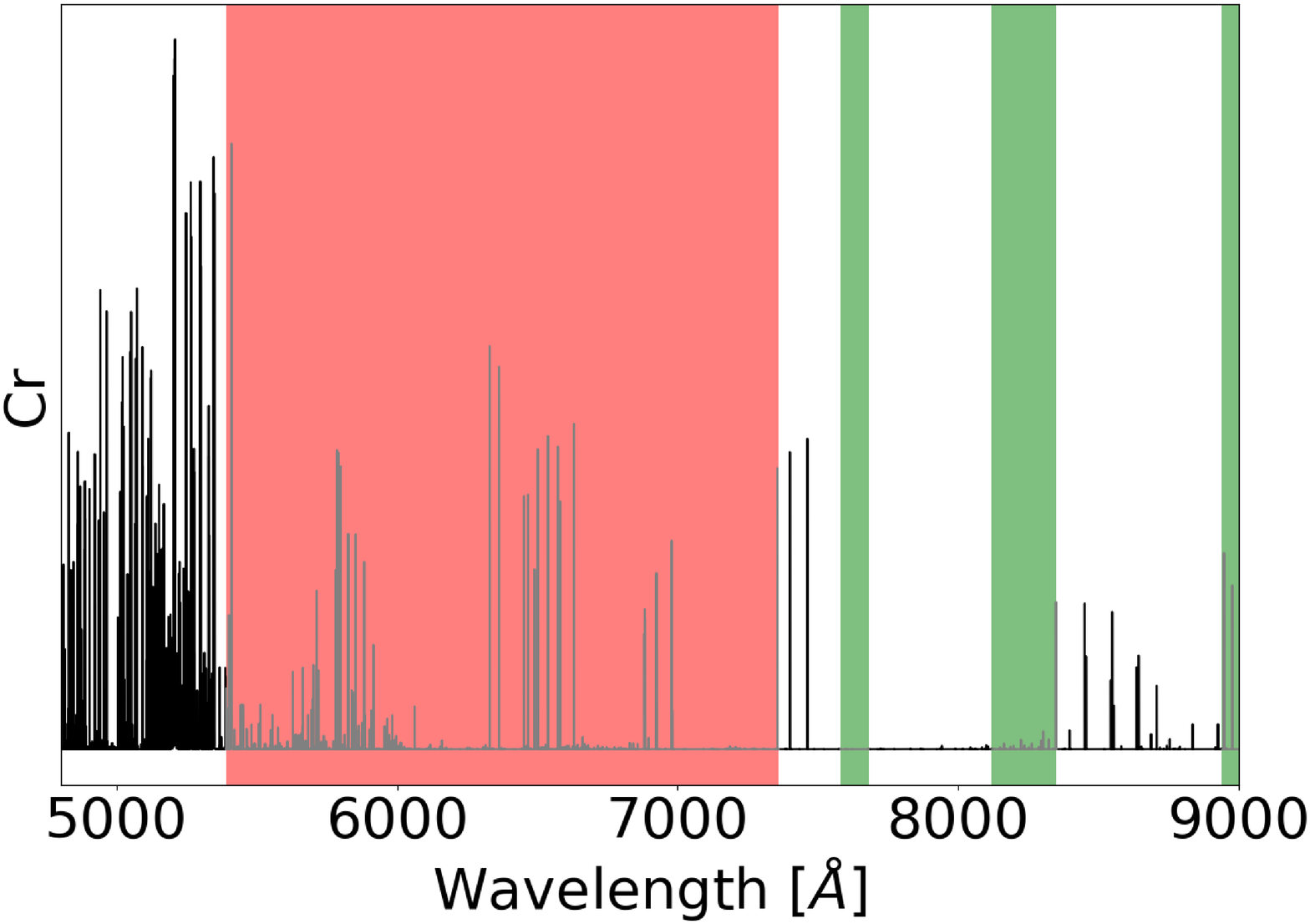}}%
  \caption{Same as Figure~\ref{fig:A1}}%
  \label{fig:A4}
\end{figure*}%


\bsp      
\label{lastpage}
\end{document}